\pdfoutput=1
\documentclass[11pt,aps,a4paper,eqsecnum,amsmath,amssymb,longbibliography,notitlepage,nofootinbib]{revtex4-1}

\usepackage{graphicx}
\usepackage{bbm}
\usepackage{hyperref}
\usepackage{xcolor}
\usepackage{tikz}
\usetikzlibrary{decorations.markings}
\usepackage{mathtools}
\usepackage{float}
\usepackage{subfigure}
\makeatletter
\def\@bibdataout@aps{%
\immediate\write\@bibdataout{%
@CONTROL{%
apsrev41Control%
\longbibliography@sw{%
    ,author="08",editor="1",pages="1",title="0",year="1"%
    }{%
    ,author="08",editor="1",pages="1",title="",year="1"%
    }%
  }%
}%
\if@filesw \immediate \write \@auxout {\string \citation {apsrev41Control}}\fi 
}
\makeatother

\newcommand{\kommu}[2]{\left[#1,#2 \right] }

\newcommand{\tr}[0]{\text{tr}}

\newcommand{\ecelll}[3]{
\draw[<-,thick] (#1,#2+1) -- (#1,#2-1);
\draw[<-,thick] (#1-1,#2) -- (#1+1,#2);
\node[anchor=south west] at (#1+0.1,#2+0.1) {\footnotesize#3};
}
\newcommand{\ecellr}[3]{
\draw[<-,thick] (#1,#2+1) -- (#1,#2-1);
\draw[->,thick] (#1-1,#2) -- (#1+1,#2);
\node[anchor=south west] at (#1+0.1,#2+0.1) {\footnotesize#3};
}

\newcommand{\ecelllscal}[4]{
\draw[<-,thick] (#1,#2+1) -- (#1,#2-1);
\draw[<-,thick] (#1-1,#2) -- (#1+1,#2);
\node[anchor=south west,scale=#4] at (#1+0.1,#2+0.1) {\footnotesize#3};
}
\newcommand{\ecellrscal}[4]{
\draw[<-,thick] (#1,#2+1) -- (#1,#2-1);
\draw[->,thick] (#1-1,#2) -- (#1+1,#2);
\node[anchor=south west,scale=#4] at (#1+0.1,#2+0.1) {\footnotesize#3};
}

\newcommand{\permucellr}[2]{
\draw[->,thick] (#1-1,#2) .. controls (#1,#2) .. (#1,#2+1);
\draw[->,thick] (#1,#2-1) .. controls (#1,#2)  .. (#1+1,#2);
}
\newcommand{\permucelll}[2]{
\draw[<-,thick] (#1-1,#2) .. controls (#1,#2) .. (#1,#2-1);
\draw[->,thick] (#1+1,#2) .. controls (#1,#2)  .. (#1,#2+1);
}

\begin{document}
%\preprint{}

\title{Finite size spectrum of the staggered six-vertex model with \texorpdfstring{$U_q(\mathfrak{sl}(2))$}{Uq(sl(2))}-invariant boundary conditions}

\author{Holger Frahm}
\author{Sascha Gehrmann}
\affiliation{%
Institut f\"ur Theoretische Physik, Leibniz Universit\"at Hannover,
Appelstra\ss{}e 2, 30167 Hannover, Germany}

\date{\today}

\begin{abstract}
The finite size spectrum of the critical $\mathbb{Z}_2$-staggered spin-$1/2$ XXZ model with quantum group invariant boundary conditions is studied.  For a particular (self-dual) choice of the staggering the spectrum of conformal weights of this model has been recently been shown to have a continuous component, similar as in the model with periodic boundary conditions whose continuum limit has been found to be described in terms of the non-compact $SU(2,\mathbb{R})/U(1)$ Euclidean black hole conformal field theory (CFT).  Here we show that the same is true for a range of the staggering parameter.  In addition we find that levels from the discrete part of the spectrum of this CFT emerge as the anisotropy is varied.  The finite size amplitudes of both the continuous and the discrete levels are related to the corresponding eigenvalues of a quasi-momentum operator which commutes with the Hamiltonian and the transfer matrix of the model.
\end{abstract}


\maketitle

%---------------------------------------------------------------------- 
\section{Introduction}
%---------------------------------------------------------------------- 
The analysis of the finite size spectra of two-dimensional vertex models and the corresponding $(1+1)$-dimensional quantum spin chains has long been used e.g.\ to identify the effective field theories describing the low energy behaviour of correlated many-body systems in the presence of strong quantum fluctuations.  Recent studies of (super) spin chains related to network models for quantum Hall transitions, the anti-ferromagnetic Potts model, intersecting loops or two-dimensional polymers have shown that the continuum descriptions may involve conformal field theories (CFTs) with a non-compact target space leading to a continuous component to the spectrum of conformal weights in the thermodynamic limit \cite{EsFS05,IkJS08,FrMa11,IkJS12,VeJS14,FrMa15,FrMa18,FrHM19}.

In this paper we study the $\mathbb{Z}_2$-staggered six-vertex model with anisotropy $0<\gamma<\pi/2$ and staggering parameter $\gamma<\alpha<\pi-\gamma$. At the 'self-dual' point, $\alpha=\pi/2$,  this model is equivalent to the critical anti-ferromagnetic Potts model.  At low energies it can be described effectively in terms of the $SL(2,\mathbb{R})_k/U(1)$ sigma model, a CFT on the two-dimensional Euclidean black hole background \cite{Witten91,HaPT02,RiSc04}, at level $k=\pi/\gamma$ \cite{IkJS12,CaIk13,FrSe14,BaKL21,BKKL21a}: for periodic boundary conditions the observed finite size spectrum of the lattice model and the density of states in the continua emerging in the thermodynamic limit have been found to agree with what is known for this CFT.  Moreover, the quantum number describing the states in the continuum has been related to a conserved quasi-momentum operator in the lattice model.  The construction of this operator relies on the existence of the staggering of the vertex model in the vertical direction: the two-row transfer matrix of the periodic model generating conserved quantities such as the Hamiltonian factorizes into a product of two commuting single-row transfer matrices taken at spectral parameters shifted by the staggering parameter. The quasi-momentum operator, on the other hand, is obtained in an expansion of the ratio of these single-row operators.

More recently, the effect of boundary conditions on the spectrum of this model has been studied: at the self-dual point the staggered six-vertex model has been shown to be related to the $R$-matrix of the $D_2^{(2)}$ affine Lie algebra \cite{RPJS20}.  This has motivated the construction of a $D_2^{(2)}$ spin chain with a particular choice of integrable open boundary conditions which also possesses a continuous spectrum of conformal weights related to the Euclidean black hole CFT \cite{RoJS21}.  Similar to the periodic case, the transfer matrix of this $D_2^{(2)}$ model can be factorized into products of transfer matrices of the six-vertex model \cite{NeRe21a}.  This procedure maps the boundary terms of the $D_2^{(2)}$ chain  to $U_q(\mathfrak{sl}(2))$ quantum group invariant open boundary conditions of the six-vertex model \cite{PaSa90,KuSk91}.  In the latter formulation an integrable model with these boundary conditions can be extended to generic values of the staggering parameter $\alpha$. Furthermore, it allows for the definition of an analog of the quasi-momentum operator for the open boundary model.  This turns out to be particularly useful for the identification of states from the discrete part of the CFT spectrum which are not present in the periodic model (although these states do appear under a twist, see e.g.\ \cite{VeJS14,FrHo17,BKKL21a}).

Below we recall the construction of the double-row transfer matrix of the inhomogeneous six-vertex model with quantum group invariant boundary conditions and its Bethe ansatz solution for $\mathbf{Z}_2$-staggered inhomogeneities $\pm i\alpha/2$.  Introducing the same staggering in the auxiliary direction, we obtain commuting four-row transfer matrices, pairs of which can be related by a duality transformation changing the staggering parameter as $\alpha \to\pi-\alpha$. As for the periodic staggered six-vertex model another family of commuting integrals of motion is generated by a quotient of the double-row transfer matrices.  Representative members of these families are the Hamiltonian and the so-called quasi-momentum operator of the staggered XXZ spin chain constructed in Section~\ref{sec:HamilLimit}.  The Temperley-Lieb representation of the Hamiltonian and its relation to other models in certain limiting cases is shown.  Based on our numerical diagonalization of the Hamiltonian and the quasi-momentum operator for small lattice sizes, we identify the solutions of the Bethe equations relevant for the low energy part of the spectrum. Using the root density formalism \cite{YaYa69} the ground state of the system in the thermodynamic limit is characterized.  For the analysis of the finite size spectrum, we solve the Bethe equations numerically for large system sizes.  This uncovers the role of the quasi-momentum in the characterization of the continuous part of the conformal spectrum and the emergence of discrete states as the anisotropy $\gamma$ is varied.  The paper ends with a summary of our findings.

\section{Definition of the model}
We use the following convention for the $R$-matrix for the XXZ-model,
\begin{align}
    R(u)=\begin{pmatrix}
        \sinh{(u+i\gamma)}&0&0&0\\
        0&\sinh{(u)}&\sinh{(i\gamma)}&0\\
        0&\sinh{(i\gamma)}&\sinh{(u)}&0\\
        0&0&0&\sinh{(u+i \gamma)}\\
        \end{pmatrix}\label{R-Matrix}
\end{align}
This four by four matrix can be interpreted as an operator on $V_1\otimes V_2$, $V_j\sim \mathbb{C}^2$. 
The parameter $\gamma$ measures the anisotropy of the model. 
The $R$-matrix has the following properties
\begin{subequations}
\label{R-properties}
\begin{align}
    R(0)&=\sinh (i\gamma ) P \label{permuR}\,,\\
    PR(u)P&=R(u)\label{PRP-Symmetry}\,,\\
    R^T(u)&=R(u)\label{Symmetrisch}\,,\\
    R(u)R(-u)&=\frac{1}{2}(\cos (2\gamma)-\cosh (2 u))\,\mathbbm{1}=:\rho(u)\,\mathbbm{1}\label{uni}\,,\\
    R^{t_1}(u)R^{t_1}(-u-2i\gamma)&=\frac{1}{2} (\cos (2 \gamma )-\cosh (2 u+2 i \gamma ))\,\mathbbm{1}=\rho(u+i\gamma)\,\mathbbm{1}\label{Crossing-Symmetry}\,,
\end{align}
\end{subequations}
where $P$ denotes the permutation operator.
In addition, the $R$-matrix satisfies the Yang-Baxter equation (YBE)
\begin{align}
R_{2,3}(v)R_{1,3}(u)R_{1,2}(u-v)=R_{1,2}(u-v)R_{1,3}(u)R_{2,3}(v),  \label{Yang-Baxter-Equation}
\end{align}
where the subscripts indicate the factors of $V_1\otimes V_2\otimes V_3$ where the $R$-matrices act non-trivially.
Based on the $R$-matrix, we can construct a monodromy matrix $T\left(u,\{\delta_j\}\right)$ depending on the spectral parameter $u$ and a set of parameters $\{\delta_j\}$, which are called the inhomogeneities:
\begin{align}
        T_0(u,\{\delta_j\})=&R_{0,2L}(u-\delta_{2L})R_{0,2L-1}(u-\delta_{2L-1})...R_{0,1}(u-\delta_{1})\,.
\end{align}
Each $R$-matrix acts on the auxiliary space $V_0\sim\mathbb{C}^2$ and one of the quantum spaces $V_j\sim\mathbb{C}^2$ represented by the second index $j=1,\dots,2L$. The YBE (\ref{Yang-Baxter-Equation}) and the fact that it just depends on the difference of its parameters ensure that the monodromy matrix $T$ is a representation of the Yang-Baxter algebra with commutation relations defined by the $RTT$-relation
\begin{align}
    R_{1,2}(u-v)T_1(u)T_2(v)=T_2(v)T_1(u)R_{1,2}(u-v)\label{RTT}
\end{align}
for arbitrary values of the inhomogeneities $\delta_j$.

\begin{figure}[t]
    \centering
   \begin{tikzpicture}
    \node[right] at (-4,1) {$\alpha$}; 
    \node[right] at (-4,-1) {$\beta$}; 
    \draw[thick,<-] (1-5,-1)--(-1-5,0);
    \draw[thick,<-] (-1-5,0)--(1-5,1);
    \node[left] at (-1-5,0) {$u$};
    \node[left] at (-1.65-5,0) {$(K_+)^\beta_\alpha(u)=$};
    
    \node[left] at (4.5,1) {$\beta$}; 
    \node[left] at (4.5,-1) {$\alpha$}; 
    \draw[thick,->] (4.5,-1)--(6.5,0);
    \draw[thick,->] (6.5,0)--(4.5,1);
    \node[right] at (6.5,0) {$u$};
    \node[left] at (3.65+0.5,0) {,\,\,$(K_-)^\beta_\alpha(u)=$};

    \node[left] at (-1.5,0) {,\,\,$R^{\gamma\delta}_{\alpha\beta}(u)=$}; 
    \node[above] at (0.25,0) {$u$}; 
    \node[left] at (-1,0) {$\alpha$}; 
    \draw[->,thick] (-1,0)--(1,0);
    \node[right] at (1,0) {$\gamma$}; 
    \node[below] at (0,-1) {$\beta$}; 
    \draw[->,thick] (0,-1)--(0,1);
    \node[above] at (0,1) {$\delta$}; 
    \end{tikzpicture}
    \caption{$R$-matrix and $K$-matrices in graphical notation}
    \label{Graphical Notation of R-Matrix}
\end{figure}
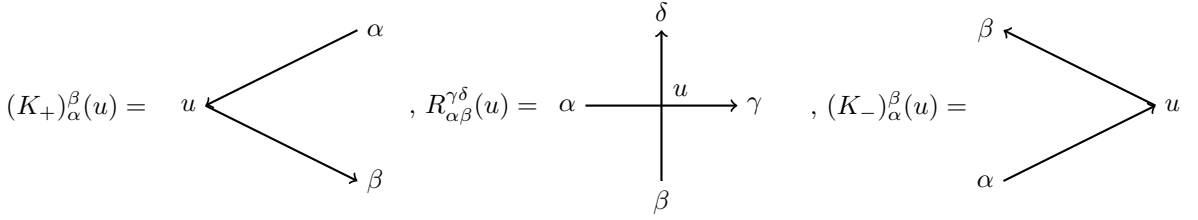

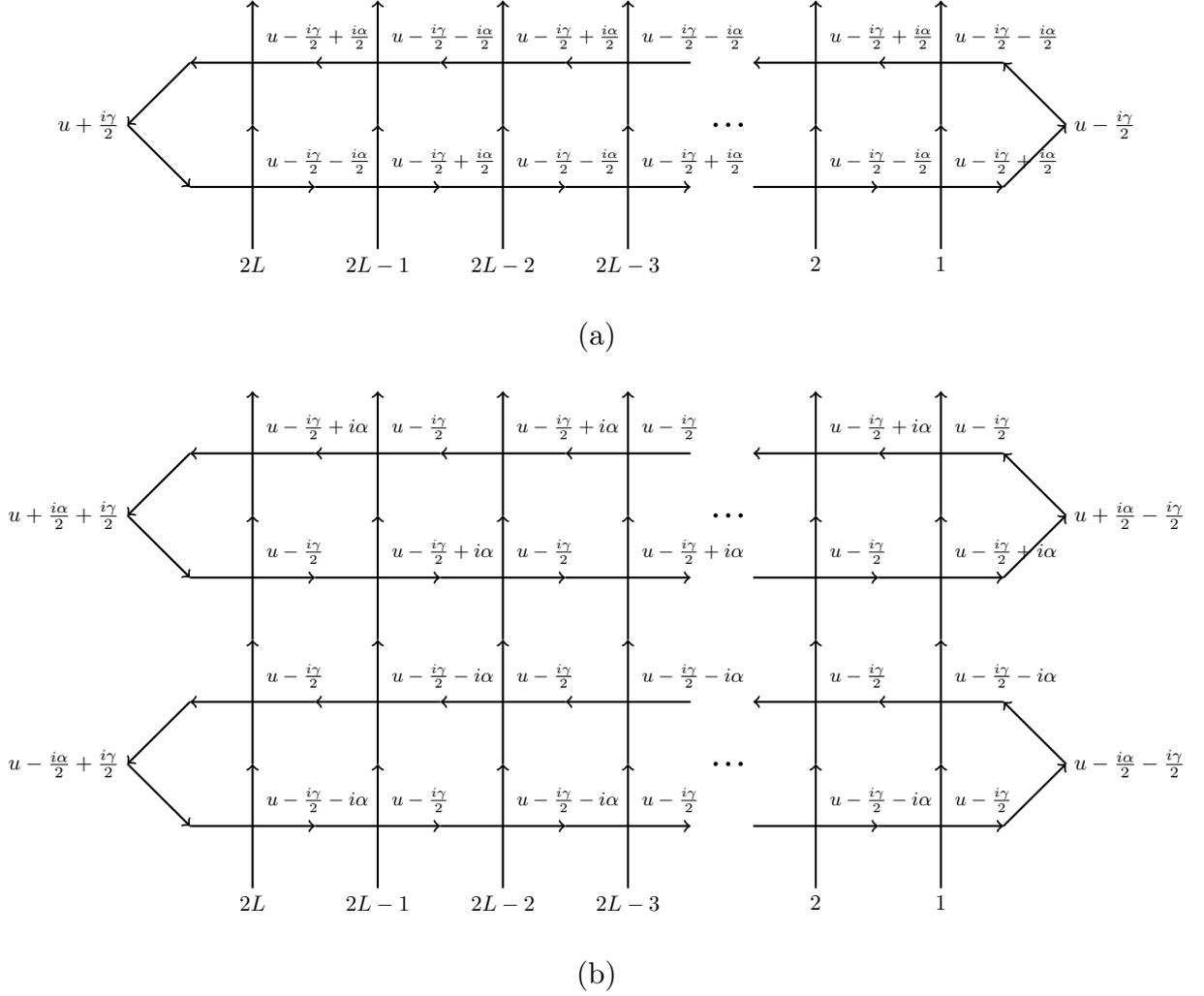
\begin{figure}[t]
    \centering
    \subfigure{
    \begin{tikzpicture}[scale=0.85, transform shape]
\ecellr{0}{0}{$u-\frac{i\gamma}{2}-\frac{i\alpha}{2}$}
\ecellr{2}{0}{$u-\frac{i\gamma}{2}+\frac{i\alpha}{2}$}
\ecellr{4}{0}{$u-\frac{i\gamma}{2}-\frac{i\alpha}{2}$}
\ecellr{6}{0}{$u-\frac{i\gamma}{2}+\frac{i\alpha}{2}$}
\draw[->,thick] (-1,2)--(-2,1) node[left] {$u+\frac{i\gamma}{2}$};
\draw[->,thick] (-2,1)--(-1,0);
\ecelll{0}{2}{$u-\frac{i\gamma}{2}+\frac{i\alpha}{2}$}
\ecelll{2}{2}{$u-\frac{i\gamma}{2}-\frac{i\alpha}{2}$}
\ecelll{4}{2}{$u-\frac{i\gamma}{2}+\frac{i\alpha}{2}$}
\ecelll{6}{2}{$u-\frac{i\gamma}{2}-\frac{i\alpha}{2}$}
\fill (7.4,1) circle (1pt);
\fill (7.6,1) circle (1pt);
\fill (7.8,1) circle (1pt);
\ecellr{9}{0}{$u-\frac{i\gamma}{2}-\frac{i\alpha}{2}$}
\ecellr{11}{0}{$u-\frac{i\gamma}{2}+\frac{i\alpha}{2}$}
\ecelll{9}{2}{$u-\frac{i\gamma}{2}+\frac{i\alpha}{2}$}
\ecelll{11}{2}{$u-\frac{i\gamma}{2}-\frac{i\alpha}{2}$}
\draw[<-,thick] (12,2)--(13,1) node[right] {$u-\frac{i\gamma}{2}$};
\draw[<-,thick] (13,1)--(12,0);
\node[below] at (0,-1) {$2L$};
\node[below] at (2,-1) {$2L-1$};
\node[below] at (4,-1) {$2L-2$};
\node[below] at (6,-1) {$2L-3$};
\node[below] at (9,-1) {$2$};
\node[below] at (11,-1) {$1$};
\node[below,scale=1.38] at (5.5,-2) {(a)};
\end{tikzpicture}
\label{Double_Row_Transfermatrix_Graphic}   
    }
    
    \subfigure{
\begin{tikzpicture}[scale=0.85, transform shape]
\ecellr{0}{0}{$u-\frac{i\gamma}{2}-i\alpha$}
\ecellr{2}{0}{$u-\frac{i\gamma}{2}$}
\ecellr{4}{0}{$u-\frac{i\gamma}{2}-i\alpha$}
\ecellr{6}{0}{$u-\frac{i\gamma}{2}$}
\draw[->,thick] (-1,2)--(-2,1) node[left] {$u-\frac{i\alpha}{2}+\frac{i\gamma}{2}$};
\draw[->,thick] (-2,1)--(-1,0);
\ecelll{0}{2}{$u-\frac{i\gamma}{2}$}
\ecelll{2}{2}{$u-\frac{i\gamma}{2}-i\alpha$}
\ecelll{4}{2}{$u-\frac{i\gamma}{2}$}
\ecelll{6}{2}{$u-\frac{i\gamma}{2}-i\alpha$}
\fill (7.4,1) circle (1pt);
\fill (7.6,1) circle (1pt);
\fill (7.8,1) circle (1pt);
\ecellr{9}{0}{$u-\frac{i\gamma}{2}-i\alpha$}
\ecellr{11}{0}{$u-\frac{i\gamma}{2}$}
\ecelll{9}{2}{$u-\frac{i\gamma}{2}$}
\ecelll{11}{2}{$u-\frac{i\gamma}{2}-i\alpha$}
\draw[<-,thick] (12,2)--(13,1) node[right] {$u-\frac{i\alpha}{2}-\frac{i\gamma}{2}$};
\draw[<-,thick] (13,1)--(12,0);

\ecellr{0}{4}{$u-\frac{i\gamma}{2}$}
\ecellr{2}{4}{$u-\frac{i\gamma}{2}+i\alpha$}
\ecellr{4}{4}{$u-\frac{i\gamma}{2}$}
\ecellr{6}{4}{$u-\frac{i\gamma}{2}+i\alpha$}
\draw[->,thick] (-1,6)--(-2,5) node[left] {$u+\frac{i\alpha}{2}+\frac{i\gamma}{2}$};
\draw[->,thick] (-2,5)--(-1,4);
\ecelll{0}{6}{$u-\frac{i\gamma}{2}+i\alpha$}
\ecelll{2}{6}{$u-\frac{i\gamma}{2}$}
\ecelll{4}{6}{$u-\frac{i\gamma}{2}+i\alpha$}
\ecelll{6}{6}{$u-\frac{i\gamma}{2}$}
\fill (7.4,5) circle (1pt);
\fill (7.6,5) circle (1pt);
\fill (7.8,5) circle (1pt);
\ecellr{9}{4}{$u-\frac{i\gamma}{2}$}
\ecellr{11}{4}{$u-\frac{i\gamma}{2}+i\alpha$}
\ecelll{9}{6}{$u-\frac{i\gamma}{2}+i\alpha$}
\ecelll{11}{6}{$u-\frac{i\gamma}{2}$}
\draw[<-,thick] (12,6)--(13,5) node[right] {$u+\frac{i\alpha}{2}-\frac{i\gamma}{2}$};
\draw[<-,thick] (13,5)--(12,4);

\node[below] at (0,-1) {$2L$};
\node[below] at (2,-1) {$2L-1$};
\node[below] at (4,-1) {$2L-2$};
\node[below] at (6,-1) {$2L-3$};
\node[below] at (9,-1) {$2$};
\node[below] at (11,-1) {$1$};
\node[below,scale=1.38] at (5.5,-2) {(b)};
\end{tikzpicture}
\label{Four_Row_Transfermatrix_Graphic}
}
    \caption{ (a) The double row transfer matrix $\tau(u)$. The horizontal lines correspond to the auxiliary space $V_0$, while the vertical lines represent the quantum spaces $V_j$ as labeled underneath.  (b) Graphical representation of the four-row transfer matrix $\mathcal{T}(u)$ with two auxiliary spaces. Observe the staggering in the horizontal and vertical direction.}
\end{figure}

While the above is enough to define an integrable model with periodic boundary conditions, to study open boundary conditions, one needs two $K$-matrices related to the $R$-matrix of the periodic model via the so-called reflection algebras \cite{Sk88}
\begin{equation}
\label{reflection-algebra}
\begin{aligned}
    R_{1,2}(u-v)K_{1,-}(u)R_{1,2}(u+v)K_{2,-}(v)&=K_{2,-}(v)R_{1,2}(u+v)K_{1,-}(u)R_{1,2}(u-v)\,,\\
    R_{1,2}(-u+v)K^{t_1}_{1,+}(u)R_{1,2}(-u-v-2i\gamma)K^{t_2}_{2,+}(v)&=K^{t_2}_{2,+}(v)R_{1,2}(-u-v-2i\gamma)K^{t_1}_{1,+}(u)R_{1,2}(-u+v)\,.
\end{aligned}
\end{equation}
Here we take the reflection matrix $K_-$ to be 
\begin{align}
    K_-(u)=\begin{pmatrix}
            e^{u}&0\\
            0&e^{-u}
        \end{pmatrix}\label{Kminus},
\end{align}
and use the $X$-isomorphism \cite{Sk88} of the reflection algebras to construct the second one as $K_{-}(-u-i \gamma)$. 
For notational reasons, we define the matrix 
\begin{align}
        K_{+}(u)=\begin{pmatrix}
            e^{-u}&0\\
            0&e^{u}
        \end{pmatrix}\label{Kplus}.
\end{align}
The $R$-matrix and the $K$-matrices allow the construction of a family of commuting operators for arbitrary inhomogeneities $\delta_j$\cite{Sk88}
\begin{equation}
\begin{aligned}
    \tau(u)&=\tr_0\left(K_{0,+}\left(u+\frac{i\gamma}{2}\right)T_0\left(u-\frac{i\gamma}{2},\{\delta_j\}\right)K_{0,-}\left(u-\frac{i\gamma}{2}\right)T^{-1}_0\left(-\left(u-\frac{i\gamma}{2}\right),\{\delta_j\}\right) \right)\,,\\
    0&=\kommu{\tau(u)}{\tau(v)}\,, \qquad \forall u,v\in \mathbb{C}
\end{aligned}
\end{equation}
which act on the Hilbert space $\mathcal{H}=\otimes_{j=1}^{2L} V_j$. 
We will refer to $\tau(u)$ as the double-row transfer matrix depending on the spectral parameter $u$. 
Due to the boundary matrices' particular choice, the double-row transfer matrix commutes with the generators of the algebra $U_q(\mathfrak{sl}(2))$ 
on $\mathcal{H}$ \cite{KuSk91}: 
\begin{align}
    S_z=\frac{1}{2}\Big(\sigma^z_1+...+\sigma^z_{2L}\Big)\,,\qquad X^{\pm}=\sum^{2L}_{j=1}e^{\pm \delta_j}e^{\frac{i\gamma}{2} \sum^{j-1}_{i=1}\sigma^z_i}\sigma^\pm_j e^{-\frac{i\gamma}{2}\sum^{2L}_{i=j+1}\sigma^z_i}\label{Uqsl2_Gen}
\end{align}In this study, we will restrict to a $\mathbb{Z}_2$-staggering, meaning that 
\begin{align}
    \delta_{2j-1}=\frac{i \alpha}{2}=-\delta_{2j}\,, \qquad j=1,...,L \,,
\end{align}
where we call $\alpha$ the staggering parameter. 
For the $\mathbb{Z}_2$-staggering, the double-row transfer matrix $\tau(u)$ becomes in terms of the $R$-matrices
\begin{equation}
\label{transfermatrix}
\begin{aligned}
    \tau(u)=c_{\tau}\,\tr_0&\left(K_{0,+}(u+\frac{i\gamma}{2})R_{0,2L}(u+\frac{i\alpha}{2}-\frac{i\gamma}{2})\dots R_{0,1}(u-\frac{i\alpha}{2}-\frac{i\gamma}{2})\right.\\
    &\quad \times \left.K_{0,-}(u-\frac{i\gamma}{2})R_{1,0}(u+\frac{i\alpha}{2}-\frac{i\gamma}{2})\dots R_{2L,0}(u-\frac{i\alpha}{2}-\frac{i\gamma}{2})\right),
\end{aligned}
\end{equation}
where we picked up the factor 
\begin{align}
   c_{\tau}= \left({\rho\left(-u+\frac{i\gamma}{2}+\frac{i\alpha}{2}\right) \rho\left(-u+\frac{i\gamma}{2}-\frac{i\alpha}{2}\right)}\right)^{-L},
\end{align}
due to the evaluation of the inverse monodromy matrix using the unitarity relation (\ref{uni}). Introducing the pictorial representations of the $K$-matrices and the $R$-matrix displayed in Figure \ref{Graphical Notation of R-Matrix},
we can represent the double-row transfer matrix graphically in Figure \ref{Double_Row_Transfermatrix_Graphic}.

The double-row transfer matrix (\ref{transfermatrix}) can be diagonalized by means of the algebraic Bethe ansatz. The Bethe equations and the eigenvalues of the transfer matrix have been derived in \cite{KuSk91}. Note that our normalization of the transfer matrix (\ref{transfermatrix}) differs from the one used in \cite{KuSk91}, where it is multiplied by the quantum determinant of the monodromy matrix
\begin{equation}
\label{Quantendeterminate}
\begin{aligned}
 q\det(T(-u))= &\sinh^L \left(\frac{i \alpha }{2}-\frac{i \gamma }{2}-u\right) \sinh^L \left(\frac{i \alpha }{2}+\frac{3 i \gamma }{2}-u\right)\\&\times \sinh^L \left(-\frac{i \alpha }{2}-\frac{i \gamma }{2}-u\right) \sinh^L \left(-\frac{i \alpha }{2}+\frac{3 i \gamma }{2}-u\right).
\end{aligned}
\end{equation}
Taking this normalization difference into account, the eigenvalues of (\ref{transfermatrix}) in the sector with $S_z=L-M$ read
\begin{equation}
\label{eq:tau_eigenval}
\begin{aligned}
    \Lambda(u)=&\frac{\sinh(2u+i\gamma)}{\sinh(2u)}\sinh^{2L}\left(u+\frac{i(\alpha+\gamma)}{2}\right)\sinh^{2L}\left(u-\frac{i(\alpha-\gamma)}{2}\right)\\&\quad\times\frac{1}{q\det (T(-u))}\prod^{M}_{m=1} \frac{\sinh(u-v_m-i\gamma)\sinh(u+v_m-i\gamma)}{\sinh(u-v_m)\sinh(u+v_m)}\\
    &+\frac{\sinh(2u-i\gamma)}{\sinh(2u)}\sinh^{2L}\left(u+\frac{i(\alpha-\gamma)}{2}\right)\sinh^{2L}\left(u-\frac{i(\alpha+\gamma)}{2}\right)\\&\quad\times\frac{1}{q\det (T(-u))}\prod^{M}_{m=1} \frac{\sinh(u-v_m+i\gamma)\sinh(u+v_m+i\gamma)}{\sinh(u-v_m)\sinh(u+v_m)}
\end{aligned}
\end{equation}
and the Bethe roots $v_m$, $m=1,\dots,M$, satisfy the Bethe equations
\begin{align}
    \left(\frac{\sinh{(v_m+\frac{i\gamma-i\alpha}{2})}\sinh{(v_m+\frac{i\gamma+i\alpha}{2})}}{\sinh{(v_m-\frac{i\gamma-i\alpha}{2})}\sinh{(v_m-\frac{i\gamma+i\alpha}{2})}}\right)^{2L}=\prod_{\substack{k=1{}\\k\neq m}}^{M}\frac{\sinh(v_m-v_k+i\gamma)\sinh(v_m+v_k+i\gamma)}{\sinh(v_m-v_k-i\gamma)\sinh(v_m+v_k-i\gamma)}\,.\label{BAE}
\end{align}
Similar to the periodic case \cite{IkJS08}, we introduce a vertical staggering by multiplying two transfer matrices with arguments differing by $i\alpha$. This procedure leads to the four-row transfer matrix 
\begin{align}
  \mathcal{T}(u)=  \tau\left(u+\frac{i\alpha}{2}\right)\tau\left(u-\frac{i\alpha}{2}\right),\label{TM_END}
\end{align}
shown in Figure \ref{Four_Row_Transfermatrix_Graphic}. By  construction $\tau(u)$ and $\mathcal{T}(u)$ commute for different arguments.  The four-row transfer matrix $\mathcal{T}(u)$ allows a duality transformation regarding the staggering parameter $\alpha$. The spectrum of $\mathcal{T}(u)$ is invariant under the transformation $\mathcal{D}$ which sends $\alpha\to \pi-\alpha$.  Concretely, the $\mathcal{D}$-transformed transfer matrix $\mathcal{D}\left(\mathcal{T}(u)\right)$ is similar to $\mathcal{T}(u)$:
\begin{align}
    \label{duality}
    \mathcal{D}\left(\mathcal{T}(u)\right)=\left(\prod^L_{i=1}\sigma^z_{2j}\right) \mathcal{C}(\alpha)\mathcal{T}(u)\mathcal{C}^{-1}(\alpha)\left(\prod^L_{i=1}\sigma^z_{2j}\right)\,,
\end{align}
where $\mathcal{C}(\alpha)$ is given as a product of local operators
\begin{align}
    \mathcal{C}(\alpha)=\prod^L_{i=1}c_{2i-1,2i}(\alpha) \quad\text{with}\quad c_{i,j}(\alpha)=P_{i,j}R_{i,j}(i\alpha)\,.\label{c-Op}
\end{align}
For the case $\alpha=\pi/2$ the transformation $\mathcal{D}$ becomes the identity and so the transfer matrix $\mathcal{T}(u)$ is invariant under the action of $\left(\prod^L_{i=1}\sigma^z_{2j}\right)\mathcal{C}(\alpha)$. Hence, an additional symmetry arises for this choice of $\alpha$ and we will refer to the parameter subspace $\{(\alpha=\pi/2,\gamma)|\gamma\in (0,\pi/2)\}$ as the self dual line or as the self-dual point regarding the parameter interval of $\alpha$. The general duality (\ref{duality}) can also be seen on the level of the Bethe equations (\ref{BAE}). If we perform the duality transformation and the following redefinition of the Bethe roots
\begin{equation}
\label{BAE_Dual}
\begin{aligned}
    \alpha&\to \pi-\alpha\,,\\
    v_k&\to v_k+\frac{i\pi}{2}\,,
\end{aligned}    
\end{equation}
solutions of (\ref{BAE}) are mapped to solutions of (\ref{BAE}). This can be easily seen by taking into account another symmetry transformation of the Bethe equations: Given a solution $\{v_k\}$, the transformed set  \[v_k\to v_k+i\pi \qquad v_k\in \Omega\] for an arbitrary subset $\Omega$ of the Bethe roots $\{v_k\}$ also solves (\ref{BAE}).
Furthermore, changing the sign of arbitrary many Bethe roots $v_k\to -v_k$ still yields a solution of (\ref{BAE}). Hence, one can restrict the solutions of (\ref{BAE}) to the strip region  $\{v_k\in \mathbb{C}|\text{Re}(v_k)\ge0, -\frac{\pi}{2}<\text{Im}(v_k)\le\frac{\pi}{2}\}$ in the complex plane.

\begin{figure}[ht]
\centering
\subfigure{\begin{tikzpicture}[scale=0.85, transform shape]
\ecellrscal{0}{0}{$-i\alpha$}{1.38}
\permucellr{2}{0}
\ecellrscal{4}{0}{$-i\alpha$}{1.38}
\permucellr{6}{0}
\draw[->,thick] (-1,2)--(-2,1) node[left,scale=1.38] {$-\frac{i\alpha}{2}+i\gamma$};
\draw[->,thick] (-2,1)--(-1,0);
\permucelll{0}{2}
\ecelllscal{2}{2}{$-i\alpha$}{1.38}
\permucelll{4}{2}
\ecelllscal{6}{2}{$-i\alpha$}{1.38}
\fill (7.4,1) circle (1pt);
\fill (7.6,1) circle (1pt);
\fill (7.8,1) circle (1pt);
\ecellrscal{9}{0}{$-i\alpha$}{1.38}
\permucellr{11}{0}
\permucelll{9}{2}
\ecelllscal{11}{2}{$-i\alpha$}{1.38}
\draw[<-,thick] (12,2)--(13,1) node[right,scale=1.38] {$-\frac{i\alpha}{2}$};
\draw[<-,thick] (13,1)--(12,0);

\permucellr{0}{4}
\ecellrscal{2}{4}{$i\alpha$}{1.38}
\permucellr{4}{4}
\ecellrscal{6}{4}{$i\alpha$}{1.38}
\draw[->,thick] (-1,6)--(-2,5) node[left,scale=1.38] {$\frac{i\alpha}{2}+i\gamma$};
\draw[->,thick] (-2,5)--(-1,4);
\ecelll{0}{6}{$i\alpha$}{1.38}
\permucelll{2}{6}
\ecelll{4}{6}{$i\alpha$}{1.38}
\permucelll{6}{6}
\fill (7.4,5) circle (1pt);
\fill (7.6,5) circle (1pt);
\fill (7.8,5) circle (1pt);
\permucellr{9}{4}
\ecellrscal{11}{4}{$i\alpha$}{1.38}
\permucelll{11}{6}
\ecelllscal{9}{6}{$i\alpha$}{1.38}
\draw[<-,thick] (12,6)--(13,5) node[right,scale=1.38] {$\frac{i\alpha}{2}$};
\draw[<-,thick] (13,5)--(12,4);
\node[below,scale=1.15] at (0,-1) {$2L$};
\node[below,scale=1.15] at (2,-1) {$2L-1$};
\node[below,scale=1.15] at (4,-1) {$2L-2$};
\node[below,scale=1.15] at (6,-1) {$2L-3$};
\node[below,scale=1.15] at (9,-1) {$2$};
\node[below,scale=1.15] at (11,-1) {$1$};
\node[below,scale=1.3] at (5.3,-2) {(a)};

\end{tikzpicture}

    \label{transfermatrix_at_u0}
    }
\subfigure{
  \begin{tikzpicture}
    \draw[->,thick] (0,0) .. controls (0,1.5)  .. (1.5,3);
    \draw[<-,thick] (0,3)--(1.5,1.5);
    \draw[<-,thick] (1.5,1.5)--(3,0);
    
    \draw[->,thick] (1.5,0)--(3,1.5);
    \draw[->,thick] (3,1.5)--(4.5,3);
    
    \draw[<-,thick] (3,3)--(4.5,1.5);
    \draw[<-,thick] (4.5,1.5)--(6,0);
    
    \draw[->,thick] (4.5,0)--(6,1.5);
    \draw[->,thick] (6,1.5)--(7.5,3);
    
    \draw[<-,thick] (6,3)--(7.5,1.5);
    \draw[<-,thick] (7.5+2,1.5)--(9+2,0);
    
    \draw[->,thick] (7.5+2,0)--(9+2,1.5);
    \draw[->,thick] (9+2,1.5)--(10.5+2,3);
    \draw[->,thick] (10.5+2,0) .. controls (10.5+2,1.5)  .. (9+2,3);
    
    \node[below] at (0,0) {$2L$};
    \node[below] at (1.5,0) {$2L-1$};
    \node[below] at (3,0) {$2L-2$};
    \node[below] at (4.5,0) {$2L-3$};
    \node[below] at (6,0) {$2L-4$};
    
    \node[below] at (7.5+2,0) {$3$};
    \node[below] at (9+2,0) {$2$};
    \node[below] at (10.5+2,0) {$1$};
\node[below,scale=1.1] at (6.6,-1) {(b)};
    
    \node[below] at (0.75,2) {$-i\alpha$};
    \node[below] at (3.75,2) {$-i\alpha$};
    \node[below] at (6.75,2) {$-i\alpha$};
    \node[below] at (9.75+2,2) {$-i\alpha$};
    
    \node[above] at (2.25,1) {$-i\alpha$};
    \node[above] at (5.25,1) {$-i\alpha$};
    \node[above] at (8.25+2,1) {$-i\alpha$};
    
    \fill (7.7+0.5,2.25) circle (1pt);
    \fill (7.9+0.5,2.25) circle (1pt);
    \fill (8.1+0.5,2.25) circle (1pt);
    \fill (7.7+0.5,0.75) circle (1pt);
    \fill (7.9+0.5,0.75) circle (1pt);
    \fill (8.1+0.5,0.75) circle (1pt);
    \draw[-,thick] (0.1,0.85)--(-0.1,0.65);
    \draw[-,thick] (0.1,0.65)--(-0.1,0.85);
    
    \draw[-,thick] (10.6+2,0.85)--(10.4+2,0.65);
    \draw[-,thick] (10.6+2,0.65)--(10.4+2,0.85);

    \node[right] at (10.6+2,0.75) {$-i\alpha$};
    \node[left] at (-0.1,0.75) {$i\alpha$};
    \end{tikzpicture}
    \label{Graphical Representation of the Quasimomentum}
}
    \caption{(a)
    The transfer matrix $\mathcal{T}$ (\ref{TM_END}) evaluated at $u_0=i\gamma/2$, where many $R$-matrices become proportional to the permutation operator.
    Using the unitarity condition (\ref{uni}) for the vertices in the second and third row, then between the first and fourth row one obtains the identity operator in the bulk.
    For the right boundary, one notes that the weights of the $K_-$-matrices differ by a minus sign, leading to the identity operator if one takes the explicit form of $K_-$ in (\ref{Kminus}) into account. The left boundary will simplify to a loop-diagram, giving also the identity when evaluated using the explicit form of $K_+$ (\ref{Kplus}).
    (b) Graphical representation of ${\tau\left(u-\frac{i\alpha}{2}\right)}/{\tau\left(u+\frac{i\alpha}{2}\right)}$ evaluated at $u_0$, Eq. (\ref{Quotient_of_double-TM}). The tiny crosses stand for the operator insertion of $\exp\left[{\pm i\alpha\sigma^z}/{2}\right]$.}
\end{figure}
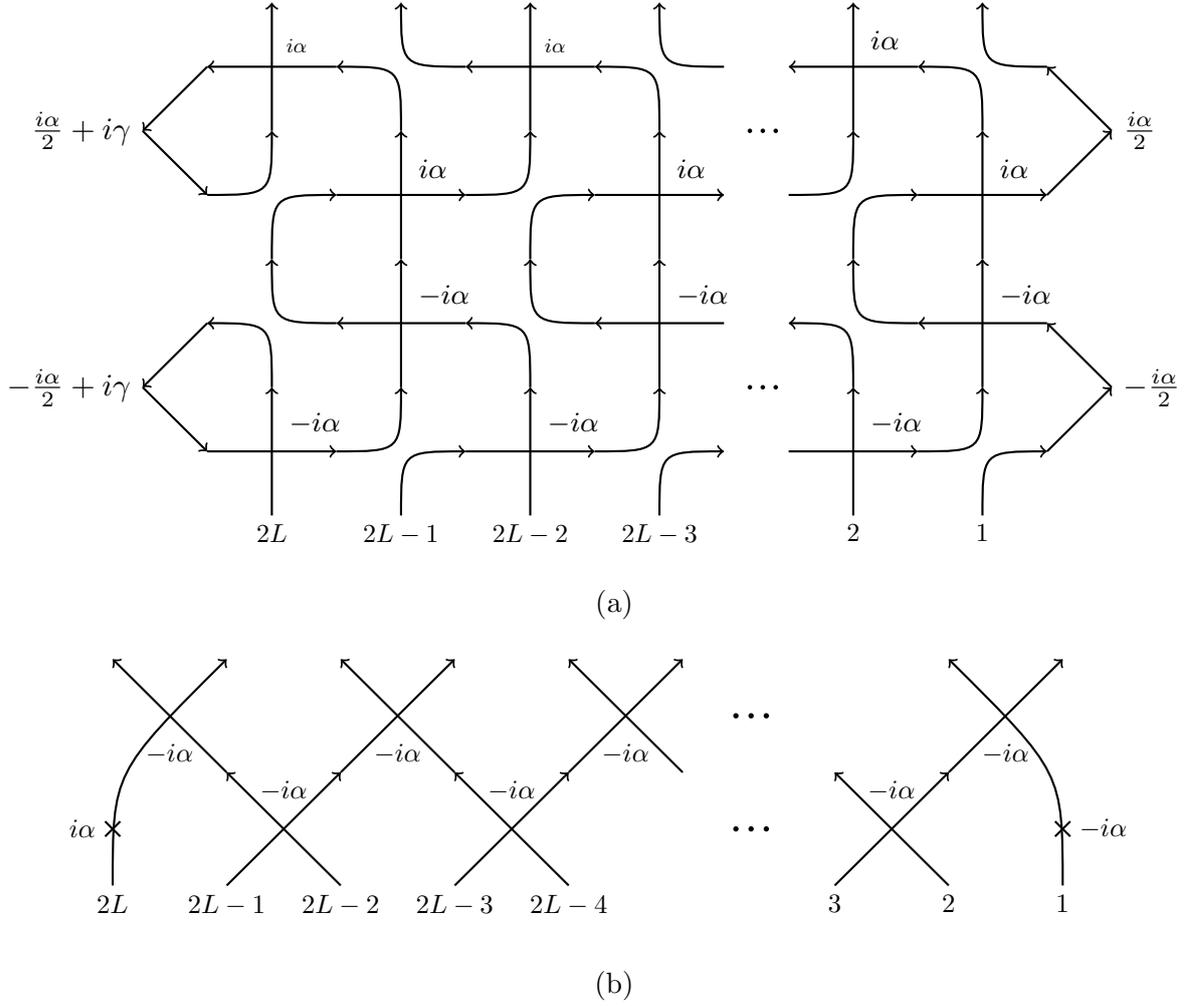

Using the unitarity relation (\ref{uni}), it is straightforward to show that the four-row transfer matrix (\ref{TM_END}) is proportional to the identity operator for the spectral parameter $u=u_0:=i\gamma/2$, see Figure \ref{transfermatrix_at_u0}. 
Note that the expansion of $\mathcal{T}(u)$ around $u_0$ yields a family of commuting operators including the Hamiltonian (\ref{End_H-Op}) constructed below which are even under the duality transformation (\ref{duality}).  This family can be complemented by considering the quotient of two double-row transfer matrices, $\tau(u-i\alpha/2)/\tau(u+i\alpha/2)$.
For $u=u_0$ and using the explicit form of $\tau(u)$, we can express this operator quotient as a product of the  operators $c_{i,j}(-\alpha)$ defined in (\ref{c-Op}):
\begin{equation}
\label{Quotient_of_double-TM}
\begin{aligned}
           \left.\frac{\tau\left(u-\frac{i\alpha}{2}\right)}{\tau\left(u+\frac{i\alpha}{2}\right)}\right|_{u=\frac{i\gamma}{2}}=&\Bigg[ \left(\prod^L_{i=1}c_{2i-1,2i}(-\alpha)\right)K_{-,1}\left(-\frac{i\alpha}{2}\right)\left(\prod^{L-1}_{i=1}c_{2i,2i+1}(-\alpha)\right)\\
       &\times \tr_0\left(c_{0,2L}(-\alpha)K_{+,0}\left(i\gamma-\frac{i\alpha}{2}\right) \right) \frac{1}{\rho(i\alpha)^L}\Bigg]^2\frac{\cos(2\gamma)-\cos(2\alpha)}{\cos(4\gamma)-\cos(2\alpha)}\\
       =&\left[ \left(\prod^L_{i=1}c_{2i-1,2i}(-\alpha)\right)\exp\left\{-\frac{i\alpha}{2}\sigma^z_1\right\}\left(\prod^{L-1}_{i=1}c_{2i,2i+1}(-\alpha)\right)\exp\left\{\frac{i\alpha}{2}\sigma^z_{2L}\right\}\right]^2\\&\times\frac{i^2\sin^2(\alpha-2\gamma)}{\rho(i\alpha)^{2L}}\frac{\cos(2\gamma)-\cos(2\alpha)}{\cos(4\gamma)-\cos(2\alpha)}\,.
\end{aligned}
\end{equation}
Compared to corresponding operator in the periodic model \cite{IkJS12} we see that the quantum group invariant boundaries are adding phase shifts on the quantum spaces related to the boundary, similar as in the presence of a twist in the periodic case \cite{CaIk13}. The operator (\ref{Quotient_of_double-TM}) can be represented graphically in Figure \ref{Graphical Representation of the Quasimomentum}.

\section{The Hamiltonian limit
}
\label{sec:HamilLimit}
By expanding $\mathcal{T}$ around $u_0$, we can define a local Hamiltonian  \cite{Sk88}:
\begin{align}
    \frac{\text{d}}{\text{d}u}\mathcal{T}(u)|_{u=u_0}&=a H+b, 
\end{align}
where $a$ and $b$ are constants given by
\begin{align*}
    a=&-\frac{2 i (\cos (4 \gamma )-\cos (2 \alpha ))}{\cos (2 \alpha )-\cos (2 \gamma )}\,,\\
    b=&-\frac{2 i \cot (\gamma ) (\cos (2 \alpha )+4 \cos (2 \gamma )-3 \cos (4 \gamma )-2)}{\cos (2 \alpha )-\cos (2 \gamma )}\notag\\&-\frac{2 i \cot (\gamma ) (\cos (4 \gamma )-\cos (2 \alpha )) ((1-2 L) \cos (2 \alpha )+(4 L-3) \cos (2 \gamma )-2 L+2)}{(\cos (2 \alpha )-\cos (2 \gamma ))^2}\,.
\end{align*}
The defined Hamiltonian reads in terms of the Pauli-matrices $\sigma_j^\alpha$:
\begin{equation}
\label{End_H-Op}
\begin{aligned}
    H=-\frac{1}{2\sin (\gamma )\rho(i\alpha) )}&\bigg\{\quad\;\,\color{black} -2\sin^2(\gamma)\sum^{2L-1}_{j=1}\cos(\gamma) \sigma^{z}_{j}\sigma^{z}_{j+1}+2\cos(\alpha)(\sigma^{+}_{j}\sigma^{-}_{j+1}+\sigma^{-}_{j}\sigma^{+}_{j+1})\\ 
\qquad&+\cos(\gamma)\sin^2(\alpha) \sum^{2L-2}_{j=1} \sigma^{z}_{j}\sigma^{z}_{j+2}+2(\sigma^{+}_{j}\sigma^{-}_{j+2}+\sigma^{-}_{j}\sigma^{+}_{j+2})\\
\qquad &+\sin(\alpha)\sin(2\gamma)\sum^{2L-2}_{j=1}(-1)^{j+1}\sigma^z_j\sigma^+_{j+1}\sigma^-_{j+2}+(-1)^{j}\sigma^z_j\sigma^-_{j+1}\sigma^+_{j+2}\\
&+\sin(\alpha)\sin(2\gamma)\sum^{2L-2}_{j=1}(-1)^{j+1}\sigma^+_j\sigma^-_{j+1}\sigma^z_{j+2}+(-1)^{j}\sigma^-_j\sigma^+_{j+1}\sigma^z_{j+2}\\
&+\sin(\gamma)\sin(2\alpha)\sum^{2L-2}_{j=1}(-1)^{j+1}\sigma^-_j\sigma^z_{j+1}\sigma^+_{j+2} +(-1)^j\sigma^+_j\sigma^z_{j+1}\sigma^-_{j+2}\\
&+\cos(\gamma)\sin^2(\alpha)(\sigma^z_1\sigma^z_2+\sigma^z_{2L-1}\sigma^z_{2L})\\
&+i(\sin(\alpha)\cos(2\gamma)-\sin(\alpha)e^{2i\alpha})(\sigma^+_1\sigma^-_2+\sigma^+_{2L-1}\sigma^-_{2L})\\
&-i(\sin(\alpha)\cos(2\gamma)-\sin(\alpha)e^{-2i\alpha})(\sigma^-_1\sigma^+_2+\sigma^-_{2L-1}\sigma^+_{2L})\\
&+2\rho(i\alpha)\sinh(i\gamma)(\sigma^z_1-\sigma^z_{2L})\\
&+\cos (\gamma ) (L \cos (2 \alpha )+(1-2 L) \cos (2 \gamma )+L-1)\bigg\}\,.
\end{aligned}
\end{equation}
We see that the Hamiltonian is indeed local since its maximal interaction range amounts to three neighboring lattice sites. 
Due to the particular choice of the constants $a,b$ the energy takes a simply form in terms of the Bethe roots $\{v_j\}$ 
\begin{align}
     E=\sum^{M}_{j=1}  \epsilon_0(v_j) \label{End_Energy_Formula},
\end{align}
where the bare energies $\epsilon_0$ are given as in the periodic case \cite{FrSe14}:
\begin{align}
            \epsilon_0(x)&=-\frac{2\sin(\alpha-\gamma)}{\cosh(2x)-\cos(\alpha-\gamma)}+\frac{2\sin(\alpha+\gamma)}{\cosh(2x)-\cos(\alpha+\gamma)}\,.
\end{align}

The Hamiltonian (\ref{End_H-Op}) takes a particularly simple form when written in terms of the generators  $e_{i,i+1}$ of the Temperley-Lieb (TL) algebra
\begin{equation}
\label{TL Algebra}
\begin{aligned}
    e_{i,i+1}^2&=-2\cos(\gamma)e_{i,i+1}\,,\\
    e_{i,i+1}e_{i+1,i+2}e_{i,i+1}&=e_{i,i+1}\,,\\
    e_{i+1,i+2}e_{i,i+1}e_{i+1,i+2}&=e_{i+1,i+2}\,,\\
    e_{i,i+1}e_{j,j+1}&=e_{j,j+1}e_{i,i+1}\,, \qquad |i-j|>1\,.
\end{aligned}    
\end{equation}
Employing the vertex representation of the Temperley-Lieb generators
\begin{align}
        e_{j,j+1}=\left(\mathbbm{1}_{\mathbb{C}^2}\right)^{\otimes j-1}\otimes\begin{pmatrix}0&0&0&0\\0&-e^{-i\gamma}&1&0\\0&1&-e^{i\gamma}&0\\0&0&0&0\end{pmatrix}\otimes \left(\mathbbm{1}_{\mathbb{C}^2}\right)^{\otimes 2L-j-1}\,,
\end{align}
one can express the Hamiltonian (\ref{End_H-Op}) as
\begin{equation}
\label{H-TL}
\begin{aligned}
    U^{odd}_{\mathcal{H}}(i\alpha)HU^{odd}_{\mathcal{H}}(-i\alpha)=-\frac{1}{\sin(\gamma)\rho(i\alpha)}\Bigg(&\sum^{2L-1}_{j=1} 2\rho(i\alpha) e_{j,j+1}\\+\sin(\alpha)&\sum^{2L-1}_{j=2} \sin(\alpha+(-1)^{j+1}\gamma)e_{j,j+1}e_{j-1,j}\\+\sin(\alpha)&\sum^{2L-1}_{j=2}\sin(\alpha+(-1)^j\gamma)e_{j-1,j}e_{j,j+1}\Bigg)\,,
\end{aligned}
\end{equation}
where $U^{odd}_{\mathcal{H}}(i\alpha)$  is the rotation of the spin variables on the odd lattice sites by $-i\alpha$ in the $x,y$-spin plane:
\begin{align}
  U^{odd}_{\mathcal{H}}(i\alpha)\sigma^\pm_{2j-1}U^{odd}_{\mathcal{H}}(-i\alpha)=e^{\mp i\alpha}\sigma^\pm_{2j-1}\,, \qquad j=1,...,L.
\end{align}

Using (\ref{eq:tau_eigenval}) the eigenvalues of the other conserved quantities of the model can be expressed in terms of the Bethe roots. One particular important quantity is the so-called quasi-momentum, which is related to the quotient of transfer matrices (\ref{Quotient_of_double-TM}) by:
\begin{equation}
\label{def:quasimom}
\begin{aligned}
    K=&\log\left(    \left.\frac{\tau\left(u-\frac{i\alpha}{2}\right)}{\tau\left(u+\frac{i\alpha}{2}\right)}\right)\right|_{u=\frac{i\gamma}{2}}\\=&2\log\left[\frac{1}{\rho(i\alpha)^L} \left(\prod^L_{i=1}c_{2i-1,2i}(-\alpha)\right)e^{-i\alpha/2\sigma^z_1}\left(\prod^{L-1}_{i=1}c_{2i,2i+1}(-\alpha)\right)e^{i\alpha/2\sigma^z_{2L}} \right]\\&+2\log\big[i\sin(\alpha-2\gamma)\big]+\log\left[\frac{\cos(2\gamma)-\cos(2\alpha)}{\cos(4\gamma)-\cos(2\alpha)}\right]
\end{aligned}
\end{equation}
The eigenvalues $\mathcal{K}$ of the quasi-momentum operator $K$ can be expressed by the Bethe roots
\begin{equation}
    \begin{aligned}
        \mathcal{K}=&\left.\log\left(\frac{\Lambda(u-\frac{i\alpha}{2})}{\Lambda(u+\frac{i\alpha}{2})} \right)\right|_{u=\frac{i\gamma}{2}}\\
    =&\sum^M_{i=1}k_0(v_i)+\log\left[\frac{\sin(2\gamma-\alpha)}{\sin(2\gamma+\alpha)}\right]+\left(2L-1\right)\log\left[\frac{\sin(\gamma-\alpha)}{\sin(\gamma+\alpha)}\right],
\end{aligned}    
\end{equation}where the bare quasi-momentum $k_0(u)$ takes the form: 
\begin{align}
    k_0(u)=2\log\left[\frac{\cosh(2u)-\cos(\alpha+\gamma)}{\cosh(2u)-\cos(\alpha-\gamma)}\right]
\end{align}
Note that the energy $E$ (\ref{End_Energy_Formula}) is invariant under the duality transformation (\ref{BAE_Dual})
while the quasi-momentum changes the sign $\mathcal{K}\to-\mathcal{K}$.

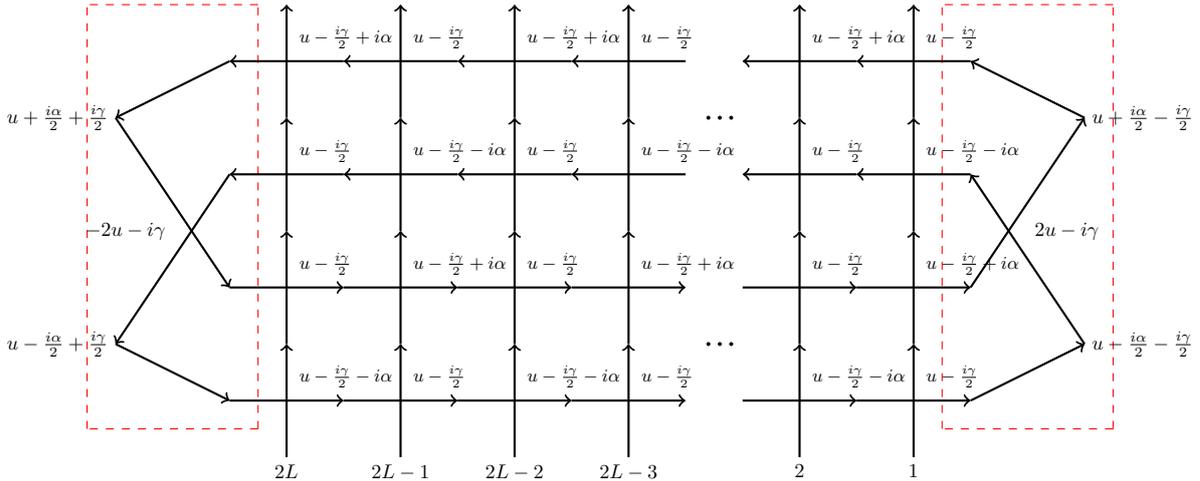
\begin{figure}[t]
    \centering
\begin{tikzpicture}[scale=0.75, transform shape]
\draw[red,dashed] (-0.5,-0.5)--(-0.5,7);
\draw[red,dashed] (-0.5,7)--(-3.5,7);
\draw[red,dashed] (-0.5,-0.5)--(-3.5,-0.5);
\draw[red,dashed] (-3.5,-0.5)--(-3.5,7);

\draw[red,dashed] (-0.5+15,-0.5)--(-0.5+15,7);
\draw[red,dashed] (-0.5+15,7)--(-3.5+15,7);
\draw[red,dashed] (-0.5+15,-0.5)--(-3.5+15,-0.5);
\draw[red,dashed] (-3.5+15,-0.5)--(-3.5+15,7);

\ecellr{0}{0}{$u-\frac{i\gamma}{2}-i\alpha$}
\ecellr{2}{0}{$u-\frac{i\gamma}{2}$}
\ecellr{4}{0}{$u-\frac{i\gamma}{2}-i\alpha$}
\ecellr{6}{0}{$u-\frac{i\gamma}{2}$}
\draw[->,thick] (-1,4)--(-3,1) node[left] {$u-\frac{i\alpha}{2}+\frac{i\gamma}{2}$};
\draw[->,thick] (-3,1)--(-1,0);
\ecelll{0}{4}{$u-\frac{i\gamma}{2}$}
\ecelll{2}{4}{$u-\frac{i\gamma}{2}-i\alpha$}
\ecelll{4}{4}{$u-\frac{i\gamma}{2}$}
\ecelll{6}{4}{$u-\frac{i\gamma}{2}-i\alpha$}
\fill (7.4,1) circle (1pt);
\fill (7.6,1) circle (1pt);
\fill (7.8,1) circle (1pt);
\ecellr{9}{0}{$u-\frac{i\gamma}{2}-i\alpha$}
\ecellr{11}{0}{$u-\frac{i\gamma}{2}$}
\ecelll{9}{4}{$u-\frac{i\gamma}{2}$}
\ecelll{11}{4}{$u-\frac{i\gamma}{2}-i\alpha$}
\draw[<-,thick] (12,4)--(14,1) node[right] {$u-\frac{i\alpha}{2}-\frac{i\gamma}{2}$};
\draw[<-,thick] (14,1)--(12,0);

\ecellr{0}{2}{$u-\frac{i\gamma}{2}$}
\ecellr{2}{2}{$u-\frac{i\gamma}{2}+i\alpha$}
\ecellr{4}{2}{$u-\frac{i\gamma}{2}$}
\ecellr{6}{2}{$u-\frac{i\gamma}{2}+i\alpha$}
\draw[->,thick] (-1,6)--(-3,5) node[left] {$u+\frac{i\alpha}{2}+\frac{i\gamma}{2}$};
\draw[->,thick] (-3,5)--(-1,2);
\ecelll{0}{6}{$u-\frac{i\gamma}{2}+i\alpha$}
\ecelll{2}{6}{$u-\frac{i\gamma}{2}$}
\ecelll{4}{6}{$u-\frac{i\gamma}{2}+i\alpha$}
\ecelll{6}{6}{$u-\frac{i\gamma}{2}$}
\fill (7.4,5) circle (1pt);
\fill (7.6,5) circle (1pt);
\fill (7.8,5) circle (1pt);
\ecellr{9}{2}{$u-\frac{i\gamma}{2}$}
\ecellr{11}{2}{$u-\frac{i\gamma}{2}+i\alpha$}
\ecelll{9}{6}{$u-\frac{i\gamma}{2}+i\alpha$}
\ecelll{11}{6}{$u-\frac{i\gamma}{2}$}
\draw[<-,thick] (12,6)--(14,5) node[right] {$u+\frac{i\alpha}{2}-\frac{i\gamma}{2}$};
\draw[<-,thick] (14,5)--(12,2);

\node[right] at (13,3) {$2u-i\gamma$};
\node[left] at (-2,3) {$-2u-i\gamma$};

\node[below] at (0,-1) {$2L$};
\node[below] at (2,-1) {$2L-1$};
\node[below] at (4,-1) {$2L-2$};
\node[below] at (6,-1) {$2L-3$};
\node[below] at (9,-1) {$2$};
\node[below] at (11,-1) {$1$};

\end{tikzpicture}

    \caption{Graphical representation of the swapped four-row transfer matrix for the relation of the $D^{(2)}_2$-model. The resulting new $K^{New}$-matrices (red boxes) acting on the enlarged unit cells can be related to the ones of the $D^{(2)}_2$-model via the spectral shift $u \to -u+\frac{i\gamma}{2}$ and some gauge transformations see (\ref{Robertson_K}).}
    \label{Enlarged_Four_Row_Transfermatrix_Graphic}
\end{figure}

We end this section by discussing some limiting cases of the Hamiltonian. It has recently been shown by Robertson \emph{et al.} \cite{RPJS20} that the transfer matrices of the staggered six-vertex model at the self-dual point $\alpha=\pi/2$ can be mapped to that of a spin chain constructed from the twisted quantum algebra $D^{(2)}_2$ with suitably chosen integrable boundary conditions.\footnote{Such a relation has been observed in Ref.~\cite{FrMa12} before within the analysis of the spectra of these models with periodic boundary conditions} 
In fact the transfer matrices of the $D_2^{(2)}$ model with both closed and open boundary conditions have been shown to factorize into the product of transfer matrices based on the $R$-matrix (\ref{R-Matrix}) considered in the present paper \cite{NeRe21a}.
For $\alpha=\pi/2$ the Hamiltonian (\ref{H-TL}) 
\begin{equation}
    U^{odd}_{\mathcal{H}}(\frac{i\pi}{2})HU^{odd}_{\mathcal{H}}(-\frac{i\pi}{2})=-\frac{2}{\sin(2\gamma)}\left(\sum^{2L-1}_{j=1} 2\cos(\gamma) e_{j,j+1}+\sum^{2L-1}_{j=2} e_{j,j+1}e_{j-1,j}+e_{j-1,j}e_{j,j+1}\right),
\end{equation}
(or (\ref{End_H-Op}) in the spin-formulation) reduces via a suitable parameter identification to the open integrable $D^{(2)}_2$ spin chain with boundary matrices from Refs.~\cite{MaGu00,NePi18}.  This model has recently been studied in \cite{RoJS21,NeRe21a}. 
Inserting a crossing relation into the four-row transfer matrix (\ref{TM_END}) and using the YBE to rearrange $R$-matrices as in Figure \ref{Enlarged_Four_Row_Transfermatrix_Graphic}, the $D^{(2)}_2$ $K$-matrices in the vertex representation can be expressed in terms of those of the staggered six-vertex model for the self-dual case $\alpha=\pi/2$ by a suitable shift in the spectral parameter,\ i.e.:
\begin{equation}
\begin{aligned}
   K^-_{D^{(2)}_2}(u)=&G_{\bar{0}}\Big(\frac{i\pi}{2}\Big)B(u)K_{-,\bar{0}}(-u+\frac{i\pi}{4})R_{\bar{0}0}(-2u)K_{-,0}(-u-\frac{i\pi}{4})P_{\bar{0}0}B(u)G_{\bar{0}}\Big(-\frac{i\pi}{2}\Big)\\&
  \times \text{csch}(2u+i\gamma),\label{Robertson_K}
\end{aligned}    
\end{equation}
where we used the transformations 
\begin{align}
 G(u)=\text{diag}(1,e^{u}) \,, \qquad B(u)=\text{diag}(e^{u},1,1,e^{-u})\,.
\end{align}
Note that the transformation $B(u)$ has recently been used to prove the $U_q(B_{n-p})\times U_q(B_{p})$ invariance of models based on the $D^{(2)}_{n+1}$ $K$-matrices \cite{NePi18}.

Finally, we note that the staggered model can be related to another well-known model in the limit $\gamma \to 0$. In this case, the Hamiltonian becomes, up to an overall scale factor unitary equivalent to the one of the \textit{ferromagnetic} XXX Heisenberg chain with periodic boundary conditions for all staggering parameters $\alpha$ (see \cite{RoJS21} for the self-dual case):
\begin{equation}
\label{H-OP-XXX}
\begin{aligned}
 \lim_{\gamma \to 0}\sin(\gamma)\,U^{odd}_{\mathcal{H}}(i\alpha)HU^{odd}_{\mathcal{H}}(-i\alpha)=&-\frac{1}{2}\sum^{2L-2}_{j=1}\sigma^{z}_j\sigma^z_{j+2}+2(\sigma^{+}_j\sigma^-_{j+2}+\sigma^{-}_j\sigma^+_{j+2})\\
&-\frac{1}{2}\left(\sigma^{z}_1\sigma^{z}_2+2\left(\sigma^+_1\sigma^-_2+\sigma^-_1\sigma^+_2\right)\right)\\
&-\frac12 \left(\sigma^{z}_{2L-1}\sigma^{z}_{2L} +2 \left(\sigma^+_{2L-1}\sigma^-_{2L}+\sigma^-_{2L-1}\sigma^+_{2L}\right)\right)\\
&+L\,.
\end{aligned}
\end{equation}

\section{Numerical Study of small number of lattice sites}
As a basis for our study of the  finite size spectrum of the staggered six-vertex model using its Bethe ansatz solution we have numerically diagonalized the Hamiltonian (\ref{End_H-Op}) and the quasi-momentum $K$ (\ref{def:quasimom}) for small lattices with $L$ unit cells.  The quantum group symmetry of the system with our choice of boundary conditions allows to decompose its Hilbert space $\left(\mathbb{C}^{2}\right)^{\otimes 2L}$ into a direct sum of sectors spanned by $U_q(\mathfrak{sl}(2))$ states with total spin $S=0,1,\dots,L$.

\begin{figure}[t]
    \centering
    \subfigure[]{\includegraphics[width=0.485\textwidth]{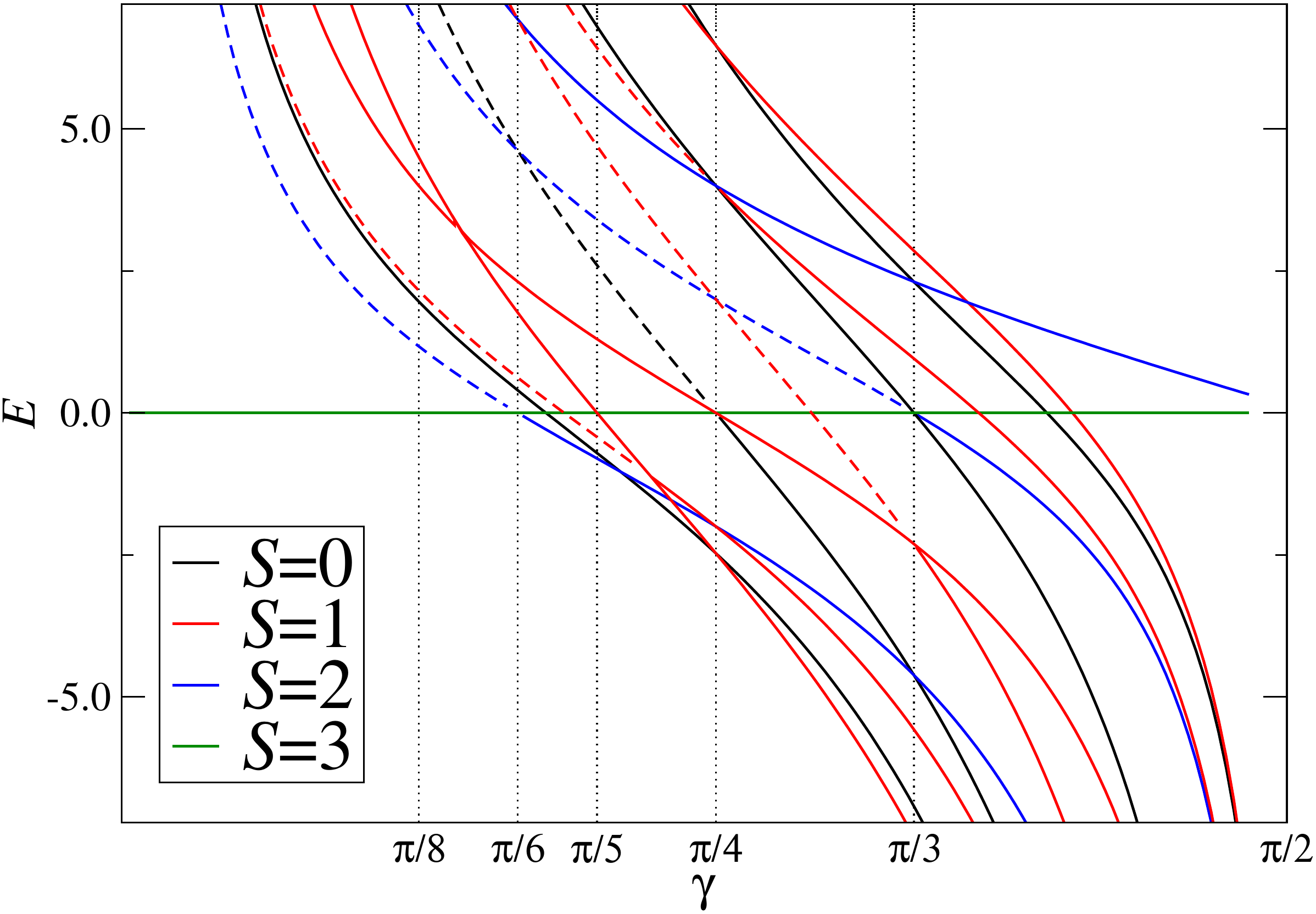}    \label{L3SpektrumKlein}}
    \subfigure[]{\includegraphics[width=0.485\textwidth]{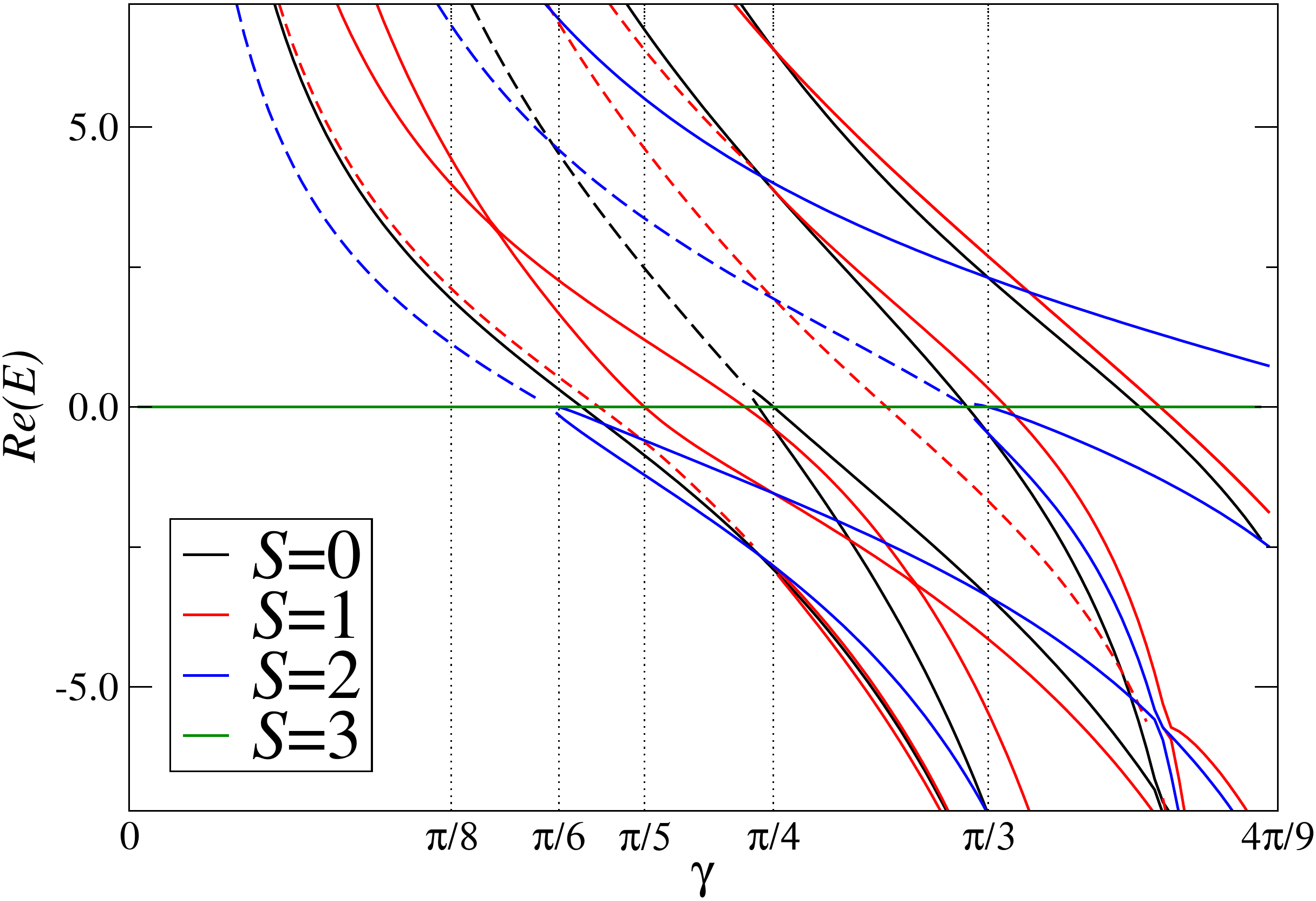}
    \label{L3SpektrumAlpha}}\\
    \subfigure[]{\includegraphics[width=0.7\textwidth]{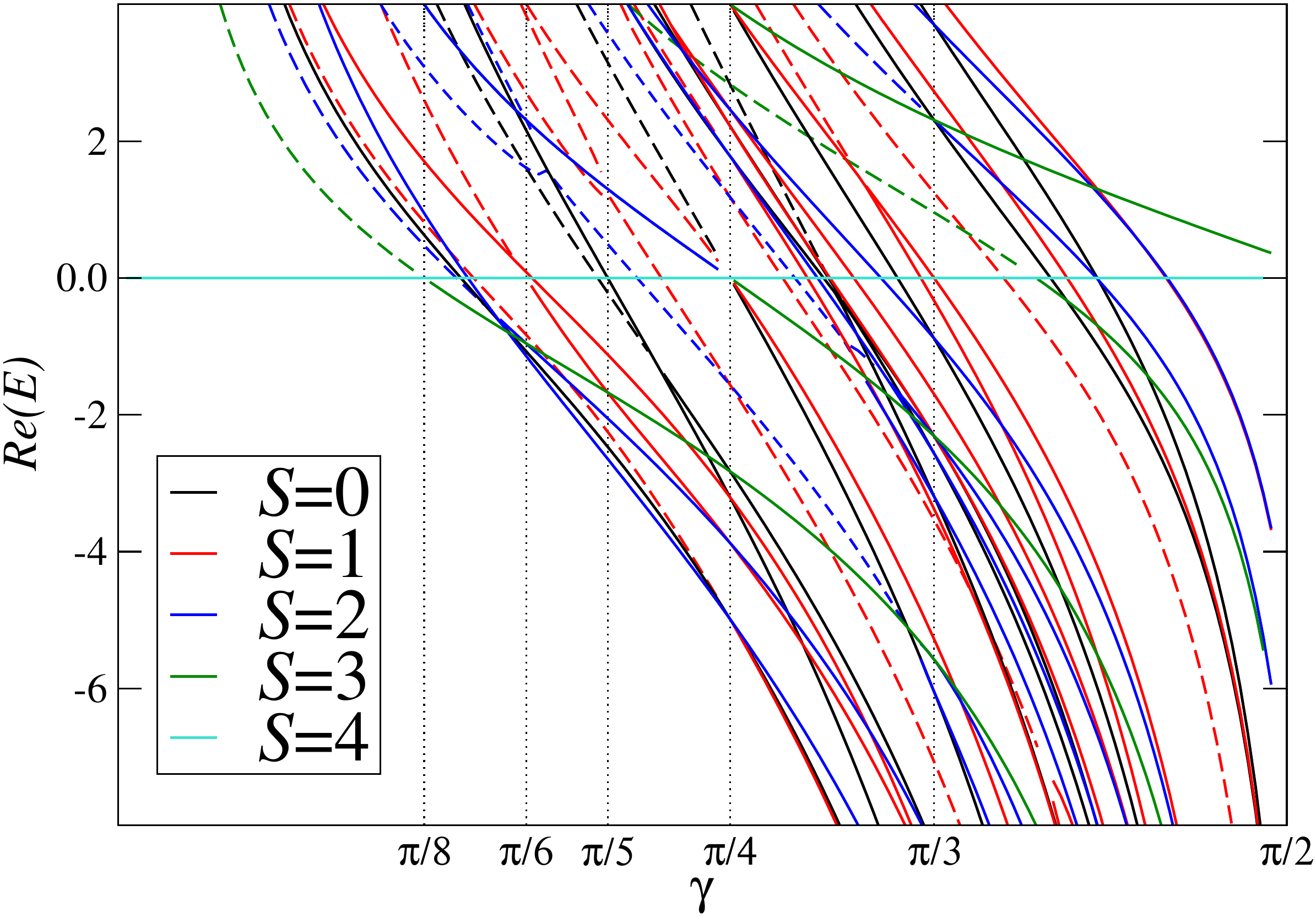}
    \label{L4SpektrumKlein}}
    \caption{ (a) Spectrum of the Hamiltonian for $L=3$ and $\alpha=\pi/2$, which is completely real for this choice of parameters. The solid lines indicate levels with real quasi-momentum, while the dashed ones represent a purely imaginary quasi-momentum. The colours indicate the quantum number $S$, i.e. the highest $S_z$ of the multiplet leading to the displayed energy level. (b)  Real part of the spectrum of the model with staggering $\alpha=9\pi/4$ and $L=3$.  Notation as used in Figure~\ref{L3SpektrumKlein}. (c) Real part of the spectrum of the self-dual model $\alpha=\pi/2$ for $L=4$. Same Notation as in \ref{L3SpektrumKlein}. We see a new ground state crossing appearing at $\gamma=\pi/8$ in comparison to the $L=3$ case.}
\end{figure}

From our numerics we find that majority of the spectrum has real energies although we also find complex ones. Since the latter come with large real parts, we expect that they do not play a role at low energies.  In Figures \ref{L3SpektrumKlein}-\ref{L4SpektrumKlein} we present the real parts of the eigenenergies as a function of the anisotropy $\gamma$ for $L=3$ and $4$ and different values of the staggering parameter. 
Surprisingly, we find the spin of the model's ground state (GS) depends on the anisotropy parameter: ground state crossings are observed at certain rational fractions of $\gamma/\pi$.  For the small lattices which are accessible to the numerical diagonalization we find as an approximate rule that the ground state has $U_q(\mathfrak{sl}(2))$ spin $S^{GS}=1,2,\dots,L-1$ for anisotropies
\begin{equation}
\label{GS Crossing Formula}
    \frac{\pi}{2(S^{GS}+1)}\lesssim\gamma\lesssim\frac{\pi}{2S^{GS}}\,,
\end{equation}
(we will confirm that these inequalities become exact for larger systems through our Bethe ansatz analysis below).
For $0\le\gamma\lesssim\pi/{2L}$ the ground state is in the sector with maximum $S^{GS}=L$, matching our observation that the Hamiltonian (\ref{End_H-Op}) becomes that of the \emph{ferromagnetic} Heisenberg chain in the limit $\gamma\to0$.   Hence the $U_q(\mathfrak{sl}(2))$ is maximal spontaneously broken in this range of $\gamma$.
This is different from the periodic model, where the ground state is a unique state with total $S^z=0$ for all anisotropies \cite{IkJS08}. 

Furthermore, we find that the eigenvalues of the quasi-momentum operator may transmute from real into purely imaginary ones when the anisotropy $\gamma$ is lowered. The lower $\gamma$, the more energy levels acquire a purely imaginary quasi-momentum.  Such a transmutation has been found to be related to the appearance of discrete states in the spectrum of conformal weights of a staggered superspin chain based on a deformation of the algebra $sl(2|1)$ \cite{FrHo17}.

The above analysis has been carried out for small lattice sizes only. However, we will see how the same results appear when using the root density formalism valid for all lattice sizes.

\section{The root density approach for the ground state 
\label{Root_Density_Chapter}}
To proceed with our studies, we have to identify the Bethe root configurations describing the low energy states. We have solved the Bethe equations (\ref{BAE}) for small lattice sizes and compared the resulting energies (\ref{End_Energy_Formula}) and quasi-momenta with the results from the direct diagonalization of the last chapter. We find that the Bethe states parameterized by $M=L-S$ Bethe roots always realize the highest-weight state in a given spin-$S$ multiplet. \\
Moreover, this allows to identify the configurations of the Bethe roots, which correspond to the low-energy regime of the spin-chain. For this regime, renormalization trajectories can be defined by fixing the deviation of the pattern of Bethe roots from that for the ground state.\\
Regarding the ground states, we find in the entire regime of anisotropies $0<\gamma<\pi/2$ that the root configurations consist of two types of Bethe roots, either completely real or with an imaginary part $\pi/2$:
\begin{equation}
\label{RootTypes}
\begin{aligned}
    v^0_m&=x_m\,,\qquad \qquad m=1,2,3,\dots,M^0\,,
    \\
    v^{\frac{\pi}{2}}_n&=y_n+\frac{i\pi}{2}\,, \qquad n=1,2,3,\dots,M^{\frac{\pi}{2}}
    \,. 
\end{aligned}
\end{equation}
This observation enables us to study the model further in the root density formalism \cite{YaYa69}: plugging (\ref{RootTypes}) into the Bethe equations (\ref{BAE}) we obtain equations for their yet undetermined real parts, $x_m$ and $y_n$. By taking the logarithm, we obtain the following coupled equations: 
\begin{equation}
\label{Log BAE Equations}
\begin{aligned}
   2\pi I^x_m&=-2L\phi\left(x_m,\frac{\gamma-\alpha}{2}\right)-2L\phi\left(x_m,\frac{\alpha+\gamma}{2}\right)+\sum_{k=1,\neq m}^{M^0}\phi\left(x_m-x_k,\gamma\right)\\&+\sum_{k=1,\neq m}^{M^{0}} \phi\left(x_m+x_k,\gamma\right)
    -\sum_{k=1}^{M^{\frac{\pi}{2}}}\psi\left(x_m-y_k,\gamma\right)-\sum_{k=1}^{M^{\frac{\pi}{2}}}\psi\left(x_m+y_k,\gamma\right)\,,
    \quad m=1,\dots,M^0\,,\\
    2\pi  I^y_n&=2L\psi\left(y_n,\frac{\gamma-\alpha}{2}\right)+2L\psi\left(y_n,\frac{\alpha+\gamma}{2}\right)-\sum_{k=1}^{M^0}\psi\left(y_n-x_k,\gamma\right)\\&-\sum_{k=1}^{M^{0}} \psi\left(y_n+x_k,\gamma\right)
    +\sum_{k=1,\neq m}^{M^{\frac{\pi}{2}}}\phi\left(y_n-y_k,\gamma\right)+\sum_{k=1,\neq m}^{M^{\frac{\pi}{2}}}\phi\left(y_n+y_k,\gamma\right)\,,
    \quad n=1,\dots,M^{\frac\pi2}\,.
\end{aligned}    
\end{equation}
Here we have introduced the quantum numbers $I^{x,y}_m \in \mathbb{N}$ characterizing the different branches of the logarithm and further have defined
\begin{align*}
    \phi(x,y)&=2\arctan\left(\tanh(x)\cot(y)\right)\\
    \psi(x,y)&=2\arctan\left(\tanh(x)\tan(y)\right)
\end{align*}
The solutions of these equations become dense on the whole real lines in the thermodynamic limit $L\to \infty$ with $M^{0,\frac\pi2}/L$ fixed. This allows to describe the distributions of the Bethe roots for the ground state by two densities. The coupled linear integral equations fixing these densities can be derived by the doubling procedure of the Bethe roots (see e.g.\ \cite{AsSu96}) and are given by: 
\begin{equation}
    \label{Density Eq}
    \begin{aligned}
    \rho^x(x)&=\sigma^x_0(x)+\frac{\tau^x_0(x)}{L}+\int_{-\infty}^{\infty}\text{d}x'K_0(x-x')\rho^x(x')+\int_{-\infty}^{\infty}\text{d}x'K_1(x-x')\rho^y(x')+\mathcal{O}\Big(\frac{1}{L^2}\Big)
    \,,
    \\\rho^y(x)&=\sigma^y_0(x)+\frac{\tau^y_0(x)}{L}+\int_{-\infty}^{\infty}\text{d}x'K_1(x-x')\rho^x(x')+\int_{-\infty}^{\infty}\text{d}x'K_0(x-x')\rho^y(x')+\mathcal{O}\Big(\frac{1}{L^2}\Big)
    \,.
    \end{aligned}
\end{equation}
The driving terms and the integral kernels are given by the following expressions, where the prime denotes the derivative in respect to the first argument:
\begin{equation}
\label{Bar Root Densities}
\begin{aligned}
    \sigma^x_0(x)=&-\frac{1}{\pi}\phi'\left(x,\frac{\gamma-\alpha}{2}\right)-\frac{1}{\pi}\phi'\left(x,\frac{\alpha+\gamma}{2}\right)\,,\\
    \sigma^y_0(x)=&\frac{1}{\pi}\psi'\left(y,\frac{\gamma-\alpha}{2}\right)+\frac{1}{\pi}\psi'\left(y,\frac{\alpha+\gamma}{2}\right)\,,\\
    \tau^x_0(x)=&-\frac{1}{\pi}\phi'(2x,\gamma)-\frac{1}{2\pi}\phi'(x,\gamma)+\frac{1}{2\pi}\psi'(x,\gamma)\,,\\
    \tau^y_0(x)=&-\frac{1}{\pi}\phi'(2x,\gamma)-\frac{1}{2\pi}\phi'(x,\gamma)+\frac{1}{2\pi}\psi'(x,\gamma)\,,\\
     K_0(x)=&\frac{1}{2\pi}\phi'(x,\gamma)\,,\quad
     K_1(x)=-\frac{1}{2\pi}\psi'(x,\gamma)
\end{aligned}
\end{equation}
Note that for $\alpha=\pi/2$, the driving terms coincide, reflecting the self-duality of the model for this value of the staggering parameter. 
The integral equations can be solved order by order in $1/L$ by Fourier transformation. We obtain the results for the first two orders:
\begin{equation}
\label{Root Densities}
\begin{aligned}
    \sigma^{x}(x)&=\frac{2\sin\left(\frac{\pi(\alpha-\gamma)}{\pi-2\gamma}\right)}{\pi-2\gamma}\frac{1}{\cosh\left(\frac{2\pi x}{\pi-2\gamma}\right)-\cos\left(\frac{\pi (\alpha-\gamma)}{\pi-2\gamma}\right)}\,,\\
    \sigma^y(x)&= \frac{2\sin\left(\frac{\pi(\alpha-\gamma)}{\pi-2\gamma}\right)}{\pi-2\gamma}\frac{1}{\cosh\left(\frac{2\pi x}{\pi-2\gamma}\right)+\cos\left(\frac{\pi (\alpha-\gamma)}{\pi-2\gamma}\right)}\,,\\
    \tau^x(x)&=
\tau^y(x)=\frac{1}{4\pi} \int^{\infty}_{-\infty} \text{d}\omega e^{i\omega x}  \frac{\sinh\left(\frac{3\gamma-\pi}{4}\omega\right)}{\sinh\left(\frac{\gamma \omega}{4}\right) \cosh \left(\frac{2\gamma-\pi}{4}\omega\right)}\,.
\end{aligned}
\end{equation}
Note that $\sigma^x(x)\leftrightarrow\sigma^y(x)$ under the duality transformation $\alpha \to \pi-\alpha$, cf. (\ref{BAE_Dual}). Furthermore, the staggering has to be restricted to values $\gamma<\alpha<\pi-\gamma$ for the bulk parts of the root densities to be positive \cite{FrMa12}.
From these densities we compute the number of Bethe roots describing the ground state and obtain:
\begin{equation}
\label{Number_Roots}
    \begin{aligned}
    \frac{2M^0_{GS}+1}{L}&=2\cdot\frac{\pi-\alpha-\gamma}{\pi-2\gamma}+\frac{1}{L}\left(\frac{3}{2}-\frac{\pi}{2\gamma}\right)+\mathcal{O}\Big(\frac{1}{L^2}\Big)\,,\\
    \frac{2M^{\frac{\pi}{2}}_{GS}+1}{L}&=2\cdot\frac{\alpha-\gamma}{\pi-2\gamma}+\frac{1}{L}\left(\frac{3}{2}-\frac{\pi}{2\gamma}\right)+\mathcal{O}\Big(\frac{1}{L^2}\Big)\,.
\end{aligned}    
\end{equation}
The individual numbers of Bethe roots are $\alpha$ and $\gamma$ dependent. With the above expression and the fact that all Bethe states are highest weight states, we can compute the sector $S$ in which the ground state is realized. The surface contribution would imply a non-zero spin $-\frac{1}{2}+\frac{\pi}{2\gamma}$ of the ground state which is independent of the staggering $\alpha$ but is non-integer due to the explicit $\gamma$ dependence. This can be resolved by rounding the number of Bethe roots (\ref{Number_Roots}) and the resulting ground state spin $S^{GS}$ to the nearest integer number, giving
\begin{align}
    S^{GS}=\left[-\frac{1}{2}+\frac{\pi}{2\gamma}\right]\label{SzGS_Rounded},
\end{align}
where the brackets indicate the rounding. Inverting this relation we obtain a range of anisotropies $\gamma$ for which the ground state is realized in the sector with spin $S^{GS}$:
\begin{align}
    \frac{\pi}{2S^{GS}+2}<\gamma<\frac{\pi}{2S^{GS}}. \label{GS Crossings Theo}
\end{align}
This formula refines the approximate rule (\ref{GS Crossing Formula}), which we have conjectured based on our numerical investigations of small systems above. The minor differences between (\ref{GS Crossings Theo}) and the numerically observations for small lattices can be interpreted as a result of the excessive influence of the boundary terms for small $L$. Note that (\ref{SzGS_Rounded}) tends to infinity as $\gamma\to0$ reflecting the relation to the ferromagnetic XXX Heisenberg chain in this limit. This generalizes our findings regarding the spontaneously broken $U_q(\mathfrak{sl}(2))$ symmetry from small system sizes to general $L$. 

While the spin $S=L-(M^0_{GS}+M_{GS}^{\frac\pi2})$ of the ground state is independent of $\alpha$, the difference or the ratio of the numbers $M^{0,\frac\pi2}_{GS}$ of the corresponding Bethe roots does depend on the staggering parameter: from the bulk contributions to (\ref{Number_Roots}) we obtain
\begin{align}
    dN_{GS}=&M^{0}_{GS}-M^{\frac{\pi}{2}}_{GS}
    =L\frac{\pi-2\alpha}{\pi-2\gamma}\,,\label{dN_GS}\\
    \frac{2M^0_{GS}+1}{2M^{\frac{\pi}{2}}_{GS}+1}=&\frac{\pi-\alpha-\gamma}{\alpha-\gamma}+\mathcal{O}\Big(\frac{1}{L}\Big)\,.
\end{align}
Hence, by varying $\alpha$, crossings between different spin-$S$ states may be  induced. To realize the corresponding root configurations on a given lattice the numbers of roots should be commensurate with $L$, i.e. have a rational ratio and a simple scaling of the difference of the number of Bethe roots for the bulk contribution.  This is achieved by fixing $\alpha$ as 
\begin{align}
    \alpha=\frac{p\gamma+q(\pi-\gamma)}{p+q},\label{Condition_alpha_Scaling}
\end{align}
where $p$ and $q$ are positive integers and relatively prime to each other \cite{FrSe14}. With that condition, one would obtain the following expression for the ratio and difference between the two types of roots
\begin{equation}
\label{Diff_RatofBRoots}
\begin{aligned}
    \frac{2M^0_{GS}+1}{2M^{\frac{\pi}{2}}_{GS}+1}&=\frac{p}{q}+\mathcal{O}(\frac{1}{L})\\
    M^0_{GS}-M^{\frac{\pi}{2}}_{GS}&=L\frac{p-q}{p+q}.
\end{aligned}    
\end{equation}
Note that by setting $p=q=1$, which corresponds to the self-dual case $\alpha=\pi/2$, the numbers of the two types of Bethe roots become the same for all $\gamma$, corresponding to the additional degeneracy of the spectrum. 

Knowing the ground state densities (\ref{Root Densities}) the bulk and boundary contributions to the expectation values of the conserved quantities in the thermodynamic limit can be calculated: being bulk quantities, the energy density and the Fermi velocity agree with those of the periodic model \cite{FrSe14}
\begin{align}
        e_\infty&=-2\int^{\infty}_{-\infty}\text{d}\omega  \frac{\sinh(\frac{\gamma \omega}{2})\left(\sinh \left(\frac{\pi\omega}{2}-\frac{\omega \gamma}{2}   \right)\cosh(\frac{\omega\pi}{2}-\alpha \omega)-\sinh(\frac{\gamma \omega}{2})\right)}{\sinh(\frac{\omega \pi}{2})\sinh((\frac{\pi-2\gamma}{2})\omega)}\,,
       \qquad v_F=\frac{2\pi}{\pi-2\gamma}\,,
\end{align}
while the surface contribution to the energy reads
\begin{align}
f_\infty=-\int^{\infty}_{-\infty}\text{d}\omega \frac{\cosh \left(\frac{1}{4} (\pi -2 \alpha ) \omega \right) \sinh \left(\frac{1}{4} (3 \gamma -\pi ) \omega \right) \cosh \left(\frac{\gamma  \omega }{4}\right) }{\cosh\left(\frac{1}{4} (\pi -2 \gamma ) \omega \right) \sinh\left(\frac{\pi  \omega }{4}\right)}-\frac{4 \sin (2 \gamma )}{\cos (2 \alpha )-\cos (2 \gamma )}.
\end{align}
Similarly, we obtain an expression for the value of quasi-momentum of the ground state in the thermodynamic limit
\begin{align}
    \mathcal{K}_{thermo}=Lk_{\infty}+k_s+\log\Big[\frac{\sin(2\gamma-\alpha)}{\sin(2\gamma+\alpha)}\Big]+\left(2L-1\right)\log\Big[\frac{\sin(\gamma-\alpha)}{\sin(\gamma+\alpha)}\Big]+\mathcal{O}\big(\frac{1}{L}\big),
\end{align}
where the bulk $k_\infty$ and surface $k_{s}$ contributions read
\begin{align}
    k_{\infty}=&4\int^{\infty}_{-\infty}\text{d}\omega \frac{\sinh(\frac{\omega \gamma}{2})\sinh(\frac{\pi\omega}{2}-\alpha\omega)\sinh(\frac{\pi-\gamma}{2}\omega)}{\omega \sinh(\frac{\omega \pi}{2})\sinh(\frac{\pi-2\gamma}{2}\omega)}\,,\\
    k_{s}=&2\int^{\infty}_{-\infty}\text{d}\omega \frac{\sinh(\frac{3\gamma-\pi}{4}\omega)\cosh(\frac{\gamma\omega}{4})\sinh(\frac{\pi-2\alpha}{4}\omega)}{\omega \sinh(\frac{\omega \pi}{4})\cosh(\frac{2\gamma-\pi}{4}\omega)}\notag\\&-\log \left(\frac{\cos (\alpha +\gamma )-1}{\cos (\alpha -\gamma )-1}\right)-\log \left(\frac{\cos (\alpha +\gamma )+1}{\cos (\alpha -\gamma )+1}\right)\,.
\end{align}
Having these explicit expressions, we can identify the effective field theory describing the low-energy regime of the spin-chain for large system sizes by studying the finite size spectrum. Due to the criticality of the model and the open boundary condition, we expect that the field theory is a boundary conformal field theory (BCFT). The scaling dimensions of the BCFT in the finite strip geometry formulation of the BCFT can be accessed by calculating the energy $E_{(n,d)}(L)$ of a state in the finite lattice. The explicit relationship is given by \cite{BlCN86,Cardy84a}
\begin{align}
    E_{(n,d)}(L)=Le_\infty+f_\infty+\frac{\pi v_F}{L}\left(-\frac{c}{24}+h_n+d\right),\label{CFT_Formula}
\end{align}
where $c$ is the central charge of the CFT, $h_n$ the conformal weight of the corresponding primary field and $d$ is the level of the descendants. Note here, since the true central charge and the true scaling dimensions appear in (\ref{CFT_Formula}) as a sum, the spectrum of the lattice model provides only the \textit{effective} central charge and \textit{effective} conformal weights by the formulae
\begin{equation}
\label{Finite_Size_Formula}
\begin{aligned}
     c_{\text{eff}}&=-\frac{24L}{\pi v_F}\Delta E_{0}\,,\\
     h^n_{\text{eff}}&=\frac{L}{\pi v_F}\Delta E_n\,,
\end{aligned}    
\end{equation}
where $\Delta E_n$ is the energy gap of a state with energy $E_n$ in respect to thermodynamic ground state: 
\begin{align}
    \Delta E_n=E_{(n,0)}(L)-L e_\infty-f_\infty\,.
\end{align}

\section{Analysis of the finite size spectrum}
\subsection{Continuous part}
To determine the (effective) scaling dimensions, one needs to take low lying excitations above the ground state into account. We find that most root configurations describing the low energy regime are still given by (\ref{RootTypes}) but with different quantum numbers $S$ and $dN$ as compared to the ground state (\ref{GS Crossings Theo}), (\ref{dN_GS}). Hence, the low-lying excited states can be described in the framework of the root density formalism analogously as already done for the ground state. However, the integral boundaries of the linear integral equation for the excited states will differ from those of the ground state representing the different values of quantum numbers $S$, $dN$. 

\begin{figure}[t]   
\centering
\subfigure[]{\includegraphics[width=0.48\textwidth]{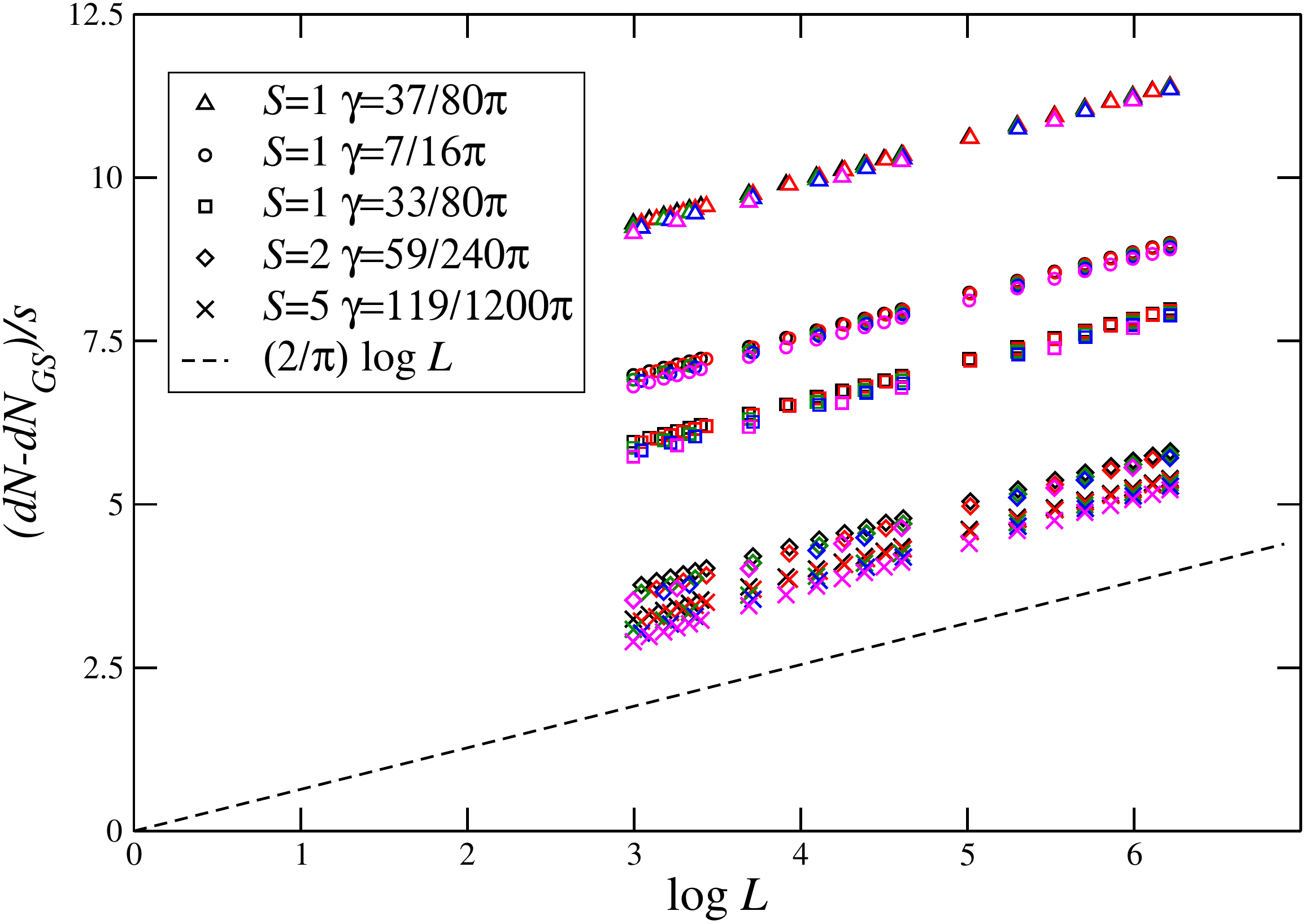}
\label{QI-Geraden}
}    
\subfigure[]{\includegraphics[width=0.48\textwidth]{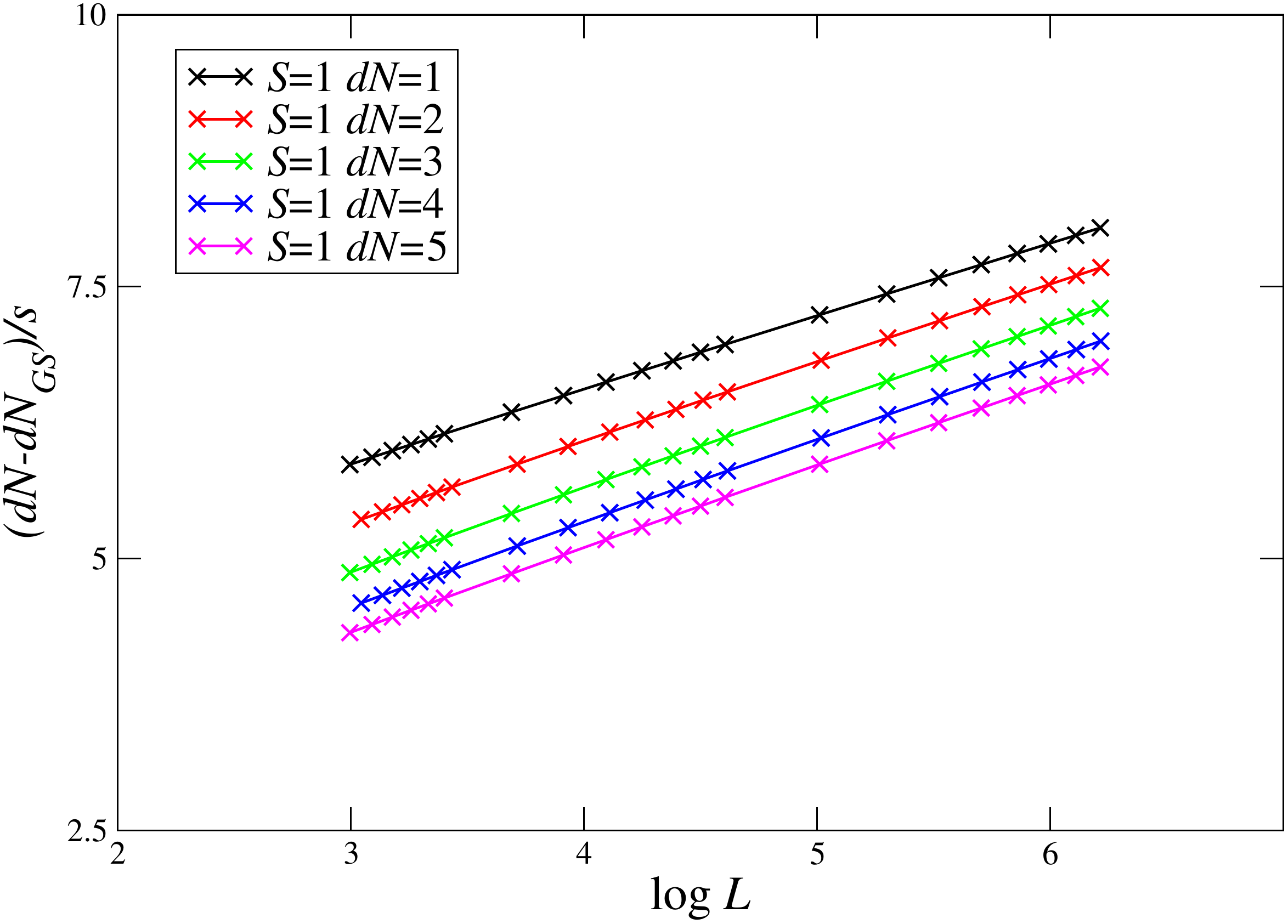}
  \label{QI-Spread}
}
\caption{ $(dN-dN_{GS})/s$ vs.\ $\log(L)$ on the self-dual line $\alpha=\pi/2$. Different symbols indicate different combinations of $S$ and $\gamma$ as labeled in the legend. 
(a) Collapse of data for spin-$S$ states and anisotropies $\gamma\lesssim \pi/2S$ corresponding to the continuous part of the conformal spectrum with different $dN$ (encoded by coloring: black $dN=1$, red $dN=2$, green $dN=3$, blue $dN=4$ and magenta $dN=5$).
(b) Lifting of this degeneracy w.r.t.\ $dN$ for the states in the $S=1$-continuum for anisotropy $\gamma=23\pi/80 \gtrsim \pi/4$, where the ground state crosses into the $S=2$ sector.
}.
\end{figure}

For the model considered in this paper with two branches of excitations and the same Fermi velocity the resulting finite size energies can be expressed in terms of these quantum numbers as  \cite{VeWo85,BoIK86,Suzu88,FrYu90} (for the particular case of open boundary conditions see \cite{AsSu96,HUBBARD})
\begin{align}
    E(L) = Le_\infty + f_\infty + \frac{\pi v_F}{L}\left\{ \frac12 \Delta \vec{M}^T \left( Z Z^\top\right)^{-1} \Delta \vec{M}  \right\} + o\left(L^{-1}\right)\,,\label{Standard Formula E}
\end{align}
where 
\begin{align*}
    \Delta \vec{M}=\begin{pmatrix}M^0-M^0_{GS}\\M^{\frac{\pi}{2}}-M^{\frac{\pi}{2}}_{GS} \end{pmatrix}\,,\qquad Z=\lim_{x\to\infty}\begin{pmatrix}\xi_{11}(x)&&\xi_{12}(x)\\\xi_{21}(x)&&\xi_{22}(x)\end{pmatrix}\,,
\end{align*}
We recall that $S=L-M^{0}-M^{\frac{\pi}{2}}$ is the $S_z$-quantum number of the  Bethe state, i.e.\ highest weight state in the corresponding spin-$S$ multiplet, and $dN=M^0-M^{\frac{\pi}{2}}$ is the difference in the number of Bethe roots of the two different types (\ref{RootTypes}).  $\mathbf{\xi}$ is the so-called dressed charge matrix defined by linear integral equations similar to (\ref{Density Eq}):
\begin{equation}
   \label{eq:dressedcharge}
    \begin{aligned}
    \xi_{11}(x)&=1+&\int^{\infty}_{-\infty}\text{d}x'\,K_{0}(x-x')\xi_{11}(x')
                +\int^{\infty}_{-\infty}\text{d}x'\,K_{1}(x-x')\xi_{21}(x')\,,\\
    \xi_{21}(x)&=&\int^{\infty}_{-\infty}\text{d}x'\,K_{1}(x-x')\xi_{11}(x')
               +\int^{\infty}_{-\infty}\text{d}x'\,K_{0}(x-x')\xi_{21}(x')\,,\\
    \xi_{12}(x)&=&\int^{\infty}_{-\infty}\text{d}x'\,K_{0}(x-x')\xi_{12}(x')
               +\int^{\infty}_{-\infty}\text{d}x'\,K_{1}(x-x')\xi_{22}(x')\,,\\
    \xi_{22}(x)&=1+&\int^{\infty}_{-\infty}\text{d}x'\,K_{1}(x-x')\xi_{12}(x')
                 +\int^{\infty}_{-\infty}\text{d}x'\,K_{0}(x-x')\xi_{22}(x')\,.
\end{aligned}
\end{equation}
By means of the Wiener-Hopf method one finds that \cite{FrYu90}
\begin{align*}
  \left(Z Z^\top\right) = \begin{pmatrix} 1-\int_{-\infty}^\infty \text{d}x\,K_0(x) && -\int_{-\infty}^\infty \text{d}x\,K_1(x) \\ -\int_{-\infty}^\infty \text{d}x\,K_1(x) && 1-\int_{-\infty}^\infty \text{d}x\,K_0(x) \end{pmatrix}^{-1}
\end{align*}
giving
\begin{align}
    E(L)=Le_\infty+f_{\infty}+\frac{\pi v_F}{L}\left(-\frac{1}{12}+\frac{\gamma}{4\pi} \left(2S+1-\frac{\pi}{\gamma}\right)^2+\frac{1}{4}\frac{(dN-dN_{GS})^2}{\tilde{Z}_D^2}+n_{ph} \right),\label{Main_Formula}
\end{align}
where $dN_{GS}$ has been defined in (\ref{dN_GS}), $n_{ph}$ is the number of particle-hole excitations in the vicinity of the Fermi points and
\begin{align*}
  \tilde{Z}_D = \lim_{\omega\to0}\left(1-\int_{-\infty}^{\infty}\text{d}x\,\text{e}^{i\omega x}\,(K_0(x) - K_1(x))\right)^{-1}\,.
\end{align*}
We note that $\tilde{Z}_D$ diverges in the limit $\omega\to0$ as a consequence of the degeneracy of the integral kernel in (\ref{eq:dressedcharge}). This is a characteristic feature in several lattice models with a continuous spectrum of scaling dimensions emerging in the continuum limit, see e.g.\  \cite{EsFS05,IkJS08,FrMa11}:  as a consequence of this singularity the penultimate term in (\ref{Main_Formula}) does not contribute to the finite size scaling for any finite $dN-dN_{GS}$ in the limit $L\to\infty$. For large but finite $L$ one finds that the energy gaps between states with different $dN-dN_{GS}$ vanish as $1/L(\log L)^2$, forming a continuum in the thermodynamic limit.  As in the periodic model we find that these logarithmic corrections are determined in terms of  the eigenvalues of the quasi-momentum operator: to bring the logarithmic corrections in the scaling dimensions under control we introduce the continuous quantum number $s$ based on the difference between the quasi-momentum of the renormalization trajectory and the one of the thermodynamic ground state
\begin{align}
    s&=\frac{\pi-2\gamma}{4\pi\gamma}\Big( \mathcal{K}-\mathcal{K}_{thermo} \Big)\notag\\
    &=\frac{\pi-2\gamma}{4\pi\gamma}\Big( \sum^M_{i=1}k_0(v_i)-Lk_{\infty}-k_s \Big)\,.\label{Def_s}
\end{align}
As in the periodic model the variable $s$ can be related to the deviation of the quantum number $dN$ from the one of the ground state $dN_{GS}$ (note that $s$ is real for root configurations (\ref{RootTypes})):
\begin{align}
    dN-dN_{GS}=\frac{2s}{\pi}\left( \log\left(\frac{L}{L_0}\right) +B(s)\right),\label{Quotient_DeltaDn_s}
\end{align}
where $L_0$ is a non-universal length, which depends on the anisotropy $\gamma$, while $B(s)$ is the density of states in the continuum. 
For the quantum group invariant open boundary case we have checked this identification numerically for levels in the spin sectors with several $S=S^{GS}$ and corresponding anisotropies $\gamma$ from Eq.~(\ref{GS Crossings Theo}) on the self-dual line.  For $\gamma\lesssim \pi/2S$, i.e.\ close to the \emph{right} boundary of these intervals, we find that the quotient $(dN-dN_{GS})/s$ for different $dN$ (and $L$) collapse to a single line $\propto\log L$ with slope $2/\pi$ independent of $\gamma$, as predicted by (\ref{Quotient_DeltaDn_s}), see Figure~\ref{QI-Geraden}.
On the other hand, for values of $\gamma\gtrsim\pi/(2S+2)$, i.e.\ close to the transition $S^{GS}\to S^{GS}+1$, we observe a splitting into lines for different values of $dN$, although still with slope $2/\pi$ for sufficient large system sizes $L$, see Figure~\ref{QI-Spread}.  In Section~\ref{Dis Part} below we will see that the resulting modification of the density of states $B(s)$ in (\ref{Quotient_DeltaDn_s}) can be attributed to the transmutation of levels near the bottom of the continuous part of the spectrum into discrete states.  We emphasize, however, that this does not affect the quality of the parameterization of the logarithmic corrections to scaling in the continuum part of the spectrum, see Figures \ref{S_z_Small},\ref{S_z_Large} and \ref{alpha ungleich pi2} below.

The subleading corrections in (\ref{Quotient_DeltaDn_s}), in particular the density of states $B(s)$, provide additional information for the description of the low energy behavior of the system by means of an effective field theory. A reliable determination of $B(s)$, however, requires the use of methods allowing the analysis of systems which are significantly larger than what is possible in the approach used here, e.g.\ based on the formulation of the spectral problem in terms of non-linear integral equations as in \cite{CaIk13,FrSe14}.  While we do not address this problem here let us note that the identification (\ref{Def_s}) of the continuous quantum number $s$ with the eigenvalues of the quasi-momentum operator is an essential condition for further progress in this direction.

\begin{figure}[t]
    \centering
    \includegraphics[width=0.8\textwidth]{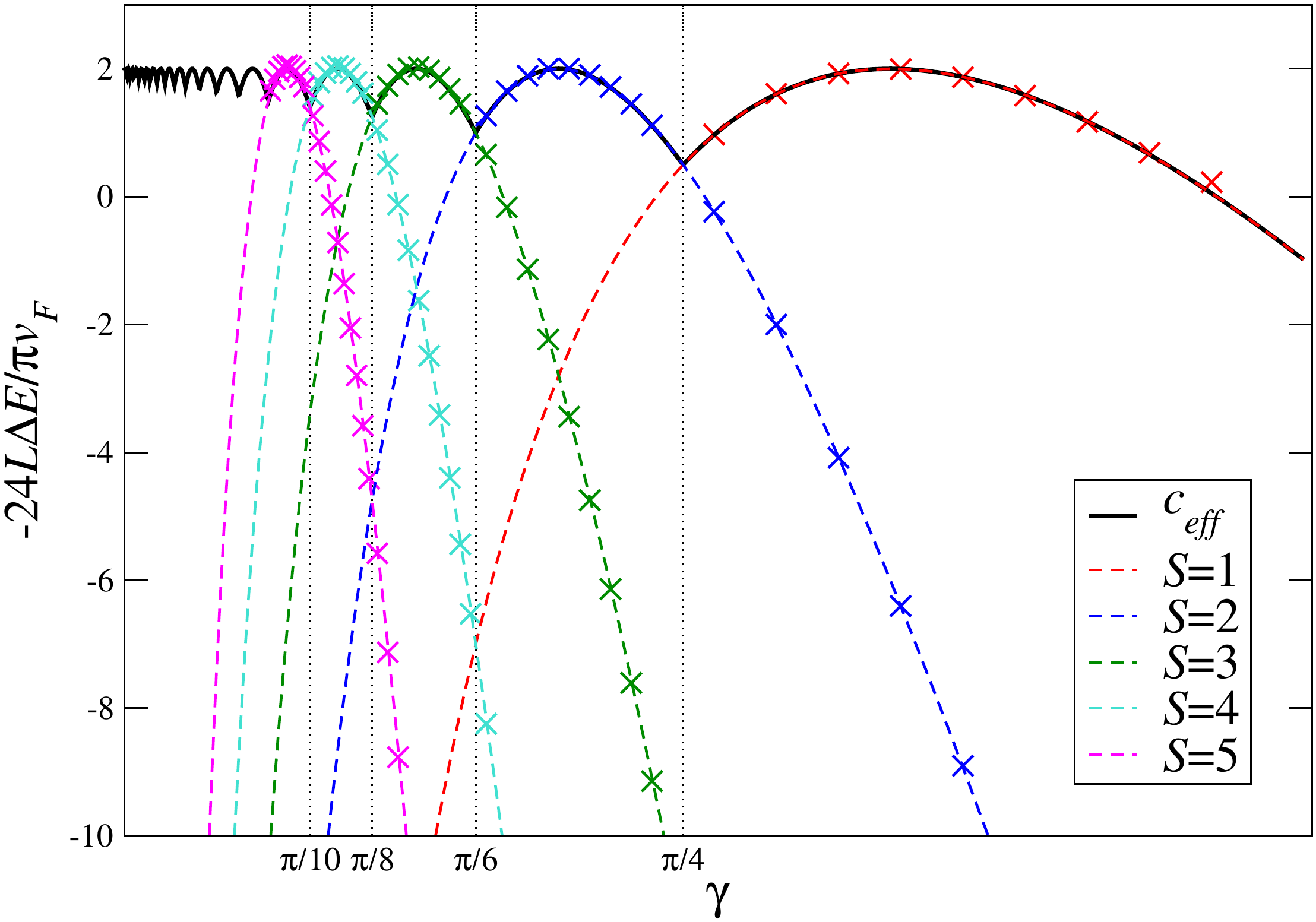}
    \caption{The effective central charge (\ref{ceff}) of the staggered spin chain as a function of the anisotropy is represented as the solid black line. Dashed colored lines are plots of (\ref{ceff}) in sectors with given spin $S$, crosses represent the effective central charges $c_{\text{eff}}$ obtained from the Bethe ansatz solutions for large $L$. Vertical lines  represent the ground state crossings.}
    \label{Ceff_Plot}
\end{figure}

Comparing (\ref{Main_Formula}), written in terms of quantum number $s$, with the prediction (\ref{CFT_Formula}) we find the effective conformal weights corresponding to levels in the sector with spin $S$ to be
\begin{align}
    h_{\text{eff}}&=-\frac{1}{12}+\frac{\gamma}{4\pi} \left(2S+1-\frac{\pi}{\gamma}\right)^2+\frac{\gamma s^2}{\pi-2\gamma}+n_{ph}, \qquad s\in \mathbb{R}\,.\label{h_eff}
\end{align}
The ground state of the lattice model is realized in the sector with spin $S=S^{GS}(\gamma)$, Eq. (\ref{SzGS_Rounded}), and $s=0$.  This leads to the effective central charge \cite{RoJS21}
\begin{align}
    c_{\text{eff}}&=2-\frac{6\gamma}{\pi}\left(2S^{GS}(\gamma)+1-\frac{\pi}{\gamma}\right)^2=
    2-\frac{24\gamma}{\pi}\, \left(\mathrm{frac}\left(\frac{\pi}{2\gamma}\right)-\frac{1}{2}\right)^2\,,
    \label{ceff}
\end{align}
where $\mathrm{frac}\left(\frac{\pi}{2\gamma}\right)$ denotes the fractional part of $\frac{\pi}{2\gamma}$. The cusps due to the fractional part $(\mathrm{frac}(\frac{\pi}{2\gamma})-\frac{1}{2})$ of the effective central charge are a consequence of the ground state crossings occurring at integer values of $\pi/2\gamma$ in the staggered six-vertex model, see Figure \ref{Ceff_Plot}.
%-------------------------------------------
%-------------------------------------------
The emergence of a continuous spectrum, parameterized by the quantum number $s$, in the thermodynamic limit is shown in Figures \ref{S_z_Small}, \ref{S_z_Large} and \ref{alpha ungleich pi2} where we have computed the effective conformal weights $h_{\mathrm{eff}}$ from the finite size energies using (\ref{Finite_Size_Formula}) and, using the Bethe ansatz results for $s$, from (\ref{h_eff}) for various anisotropies $\gamma$ and staggering $\alpha$ in the spin sector containing the ground state. Extrapolation of the finite size data to $L\to\infty$ by means of a rational function of $1/\log L$ shows that various levels with $dN\neq dN_{GS}$ converge to the bottom of the corresponding continuum (given by the Bethe state with $dN=dN_{GS}$).
%-------------------------------------------
\begin{figure}[H]
    \centering
    \subfigure{\includegraphics[width=0.49\textwidth]{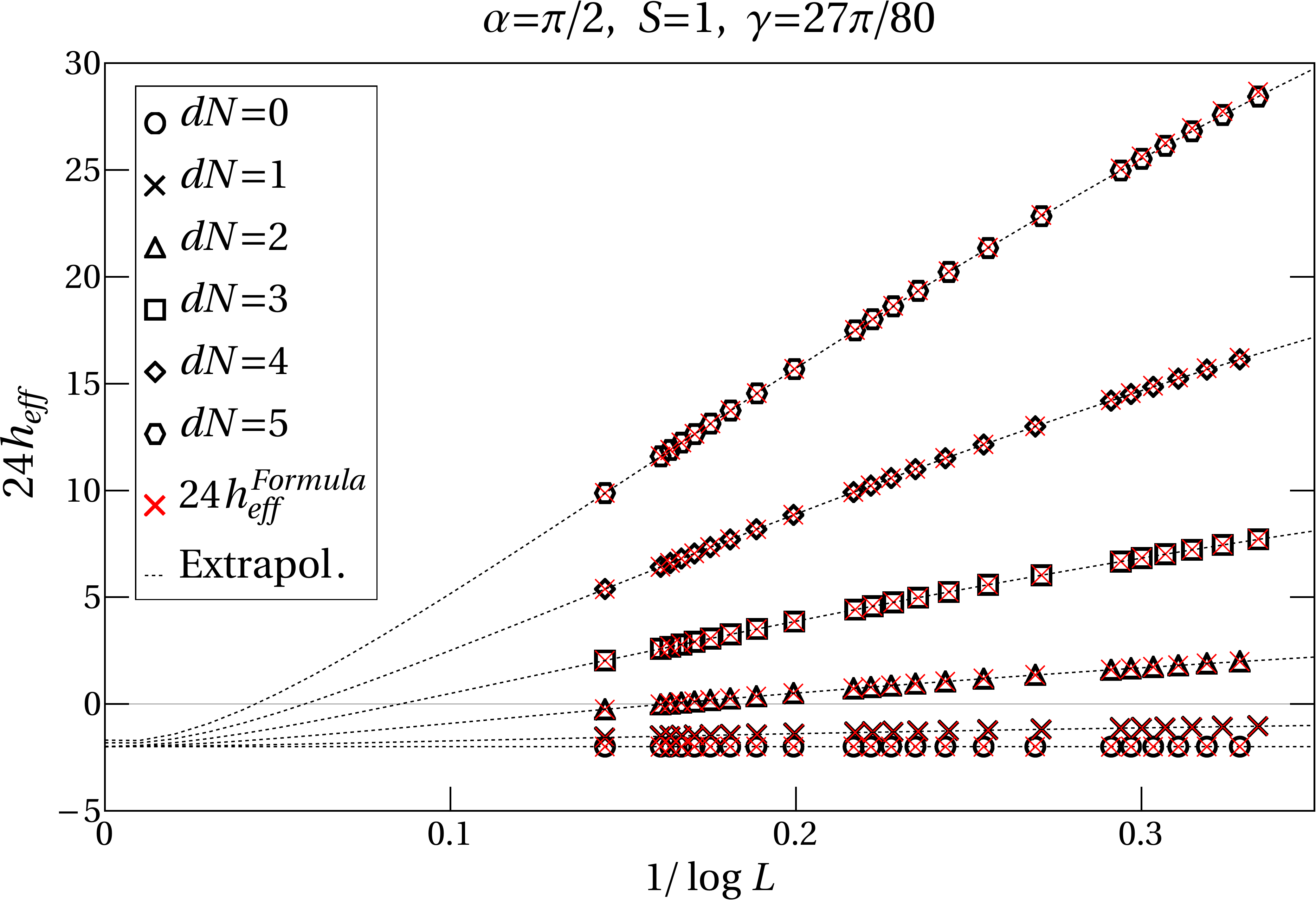}}
    \subfigure{\includegraphics[width=0.49\textwidth]{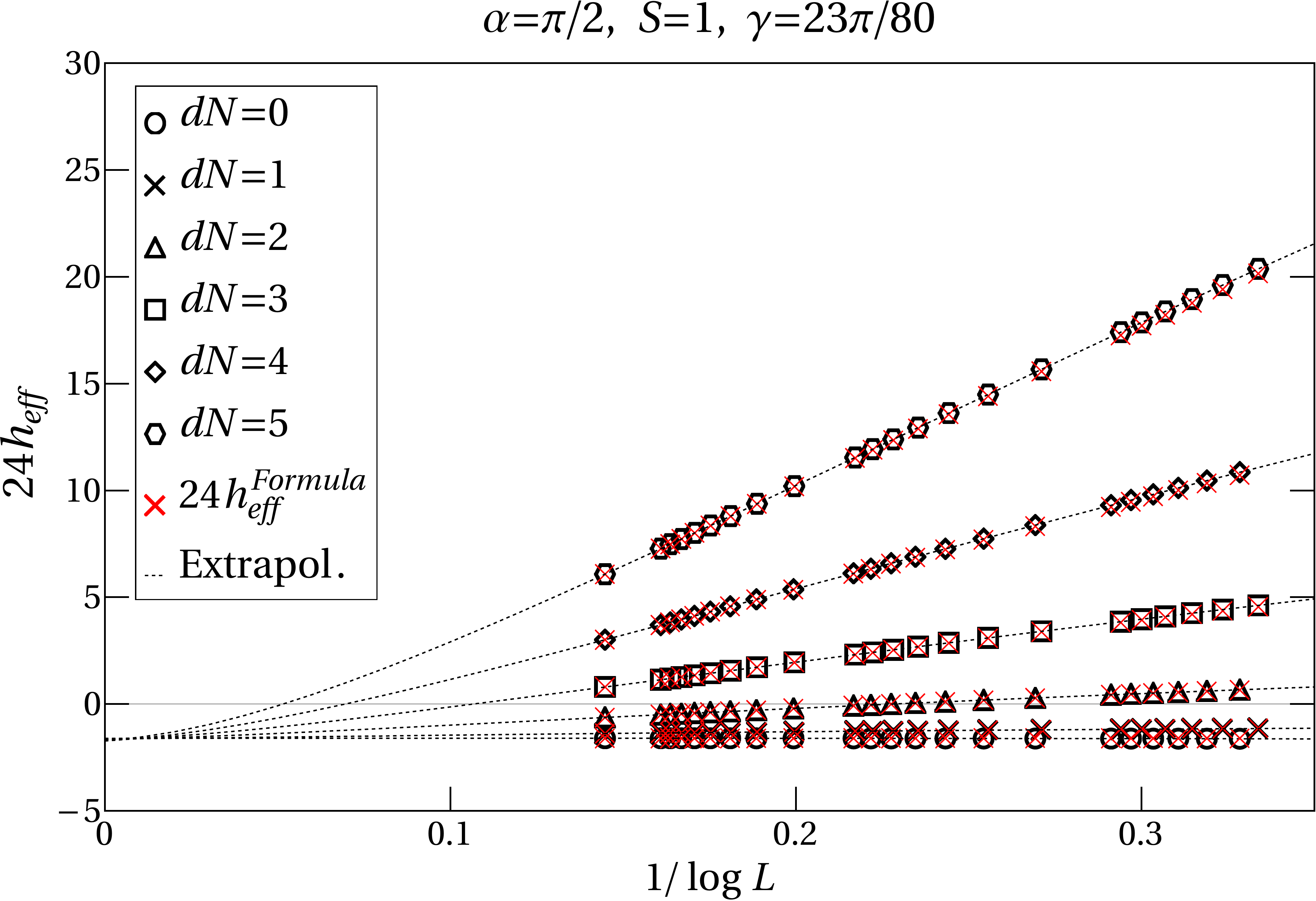}}\\
    \subfigure{\includegraphics[width=0.49\textwidth]{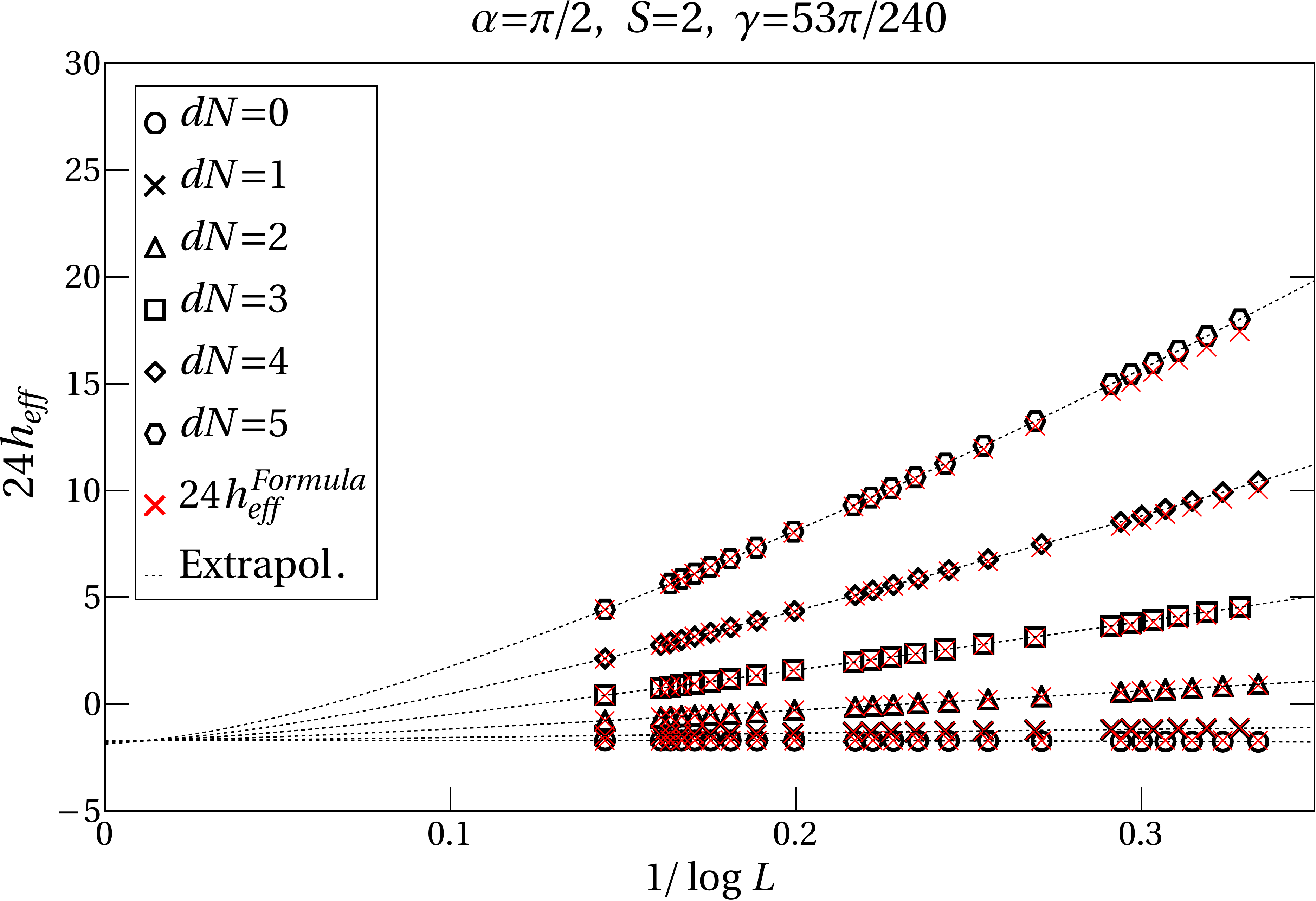}}
    \subfigure{\includegraphics[width=0.49\textwidth]{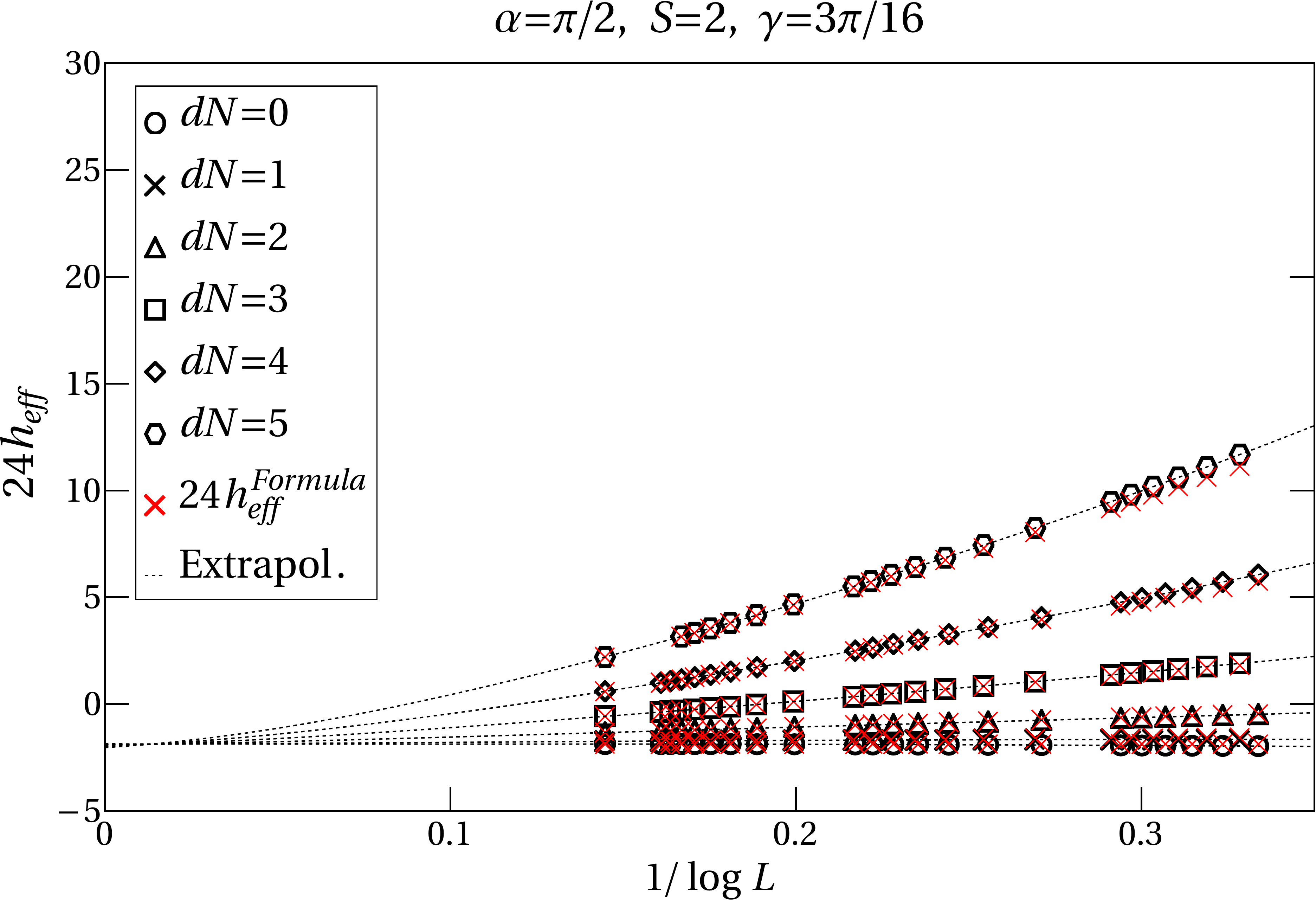}}\\
    \subfigure{\includegraphics[width=0.49\textwidth]{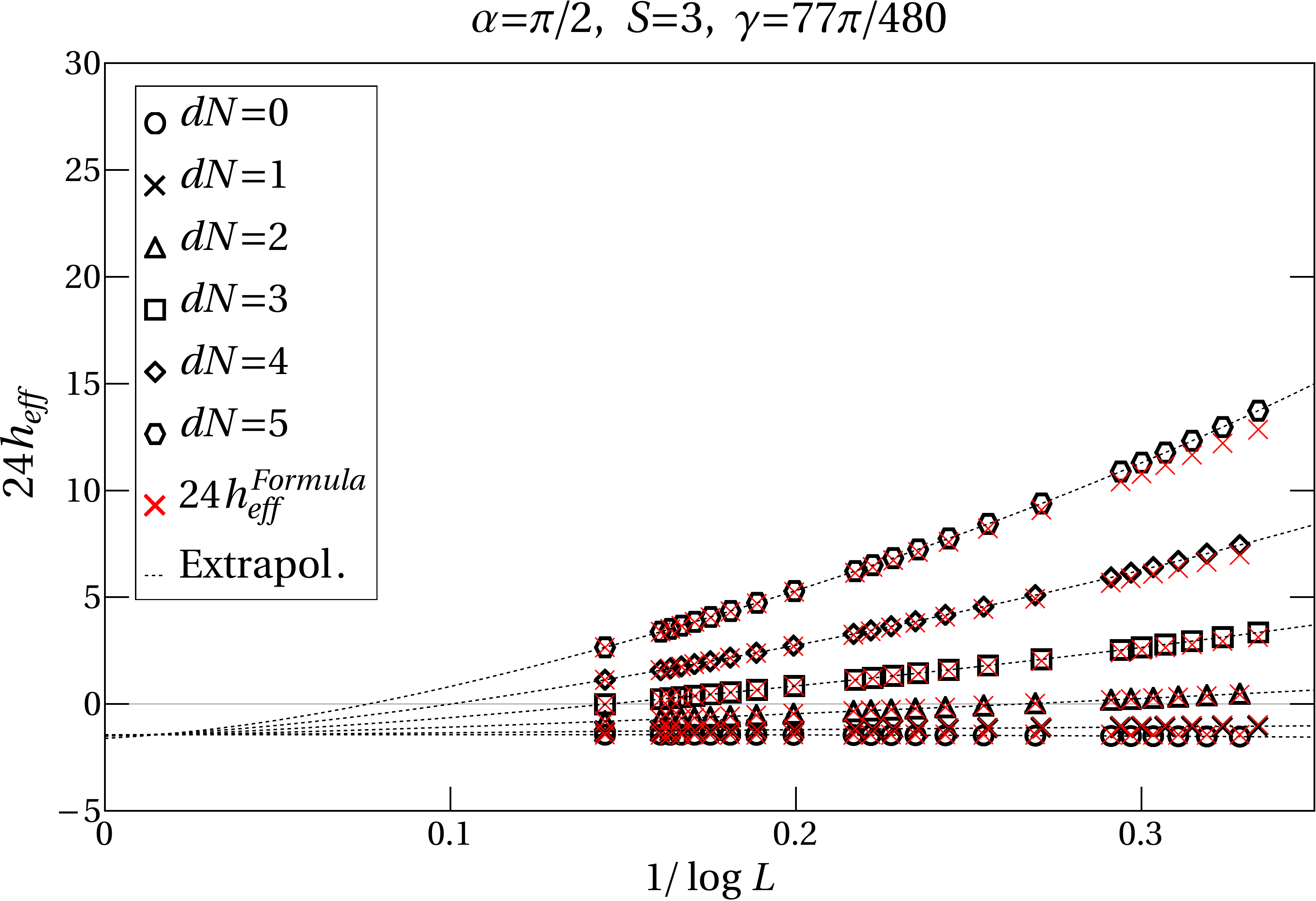}}
    \subfigure{\includegraphics[width=0.49\textwidth]{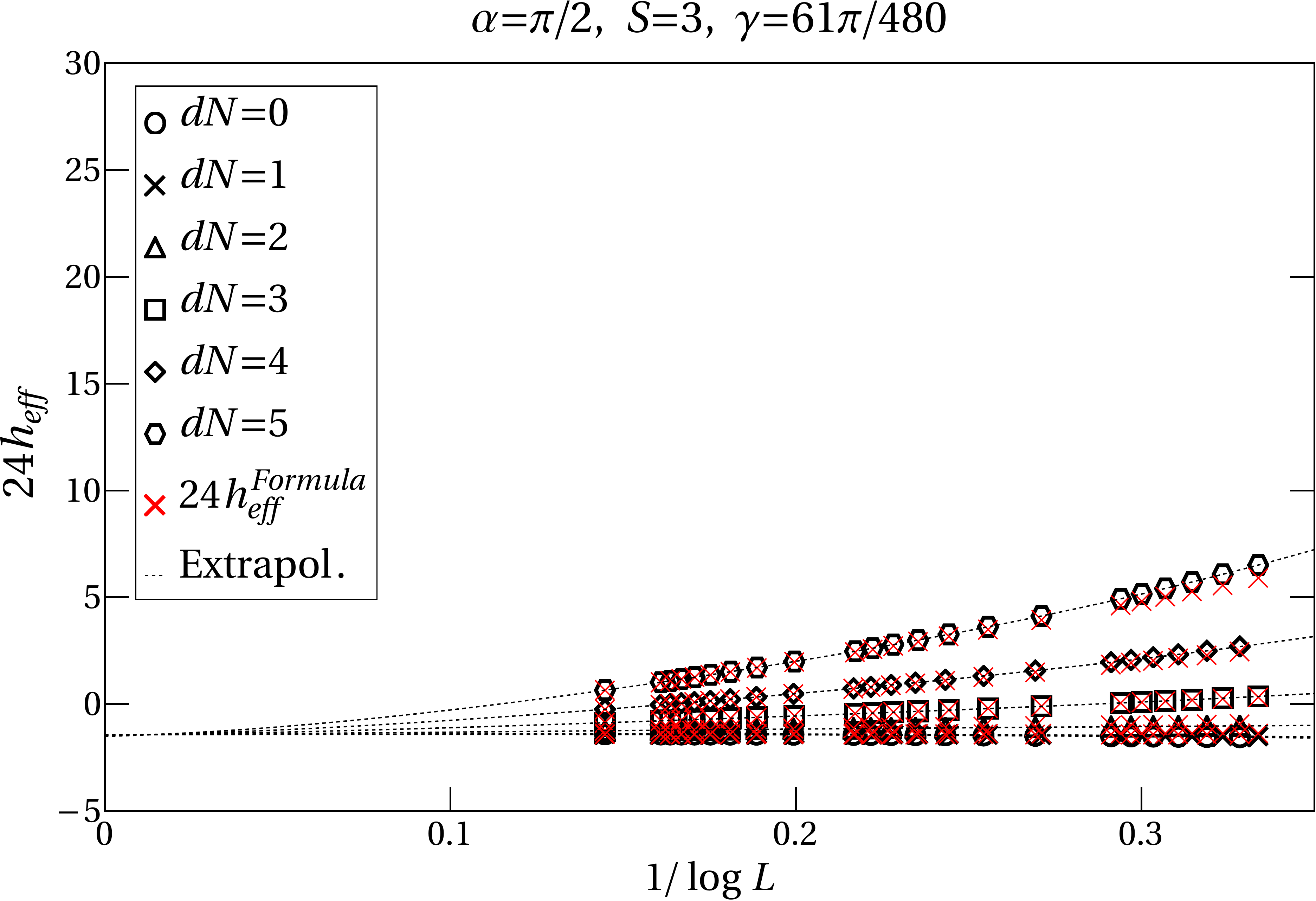}}
    \caption{Spectrum of conformal weights for the self-dual model (i.e.\ $dN_{GS}=0$):  for each $S=1,2,3$ we have chosen two values of the anisotropy in the intervals (\ref{GS Crossings Theo}) $S$ is the spin of the ground state. Black symbols are the effective scaling dimensions obtained from Eq.~(\ref{Finite_Size_Formula}) using the finite size energies obtained from the solution of the Bethe equations. Red symbols represent the scaling dimensions obtained from (\ref{h_eff}) using the Bethe ansatz results for the quasi-momentum $s$. Note that the latter provide an excellent parameterization of the logarithmic corrections via the continuous variable $s$ for all $\gamma$ values considered (e.g. \ in the top-right figure for the parameters used in Figure \ref{QI-Spread} showing the lifting of the degeneracy of (\ref{Quotient_DeltaDn_s}) w.r.t.\ $dN$). }
    \label{S_z_Small}
\end{figure}
%-------------------------------------------
\begin{figure}[H]
    \centering
    \subfigure{\includegraphics[width=0.49\textwidth]{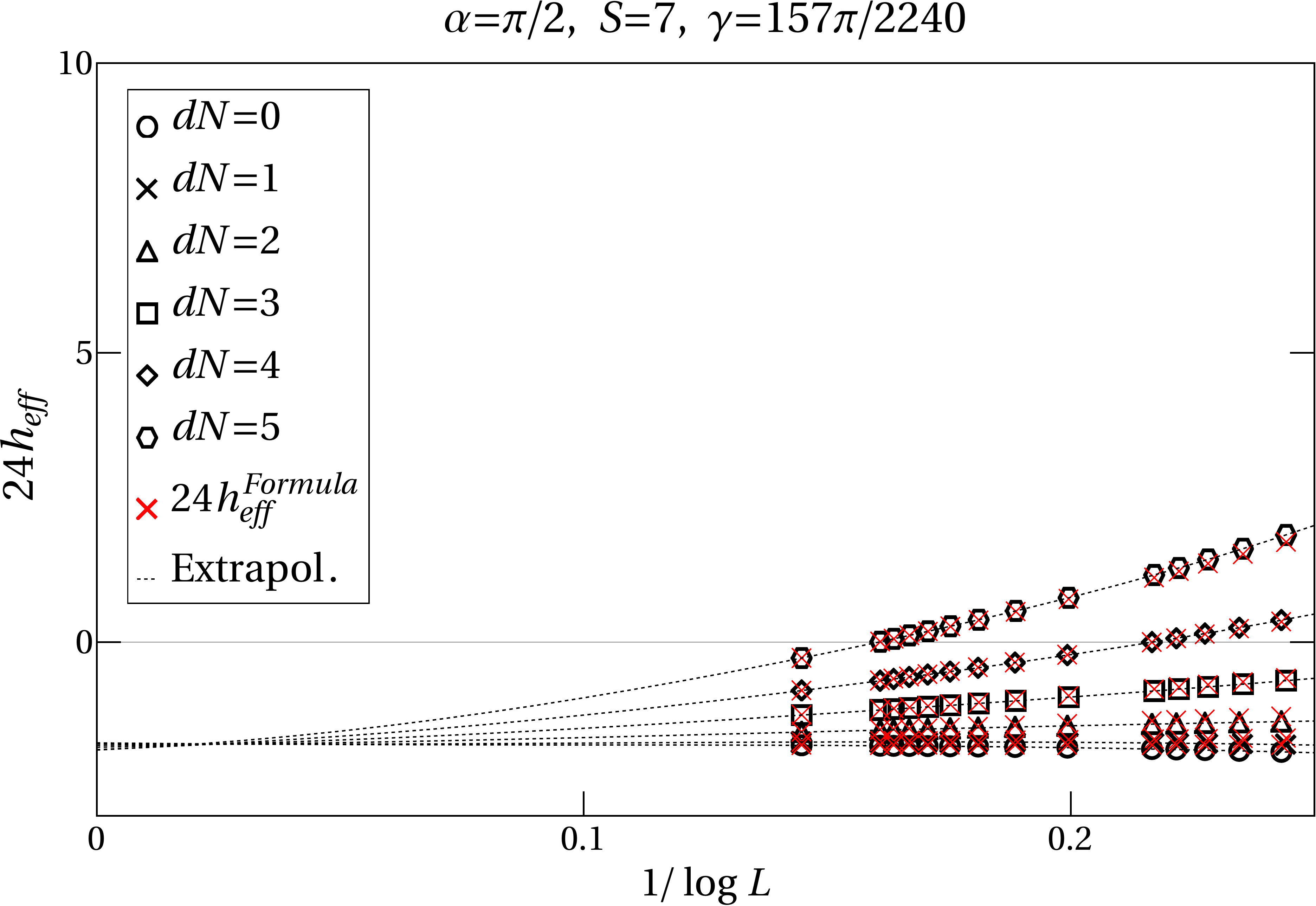}}
    \subfigure{\includegraphics[width=0.49\textwidth]{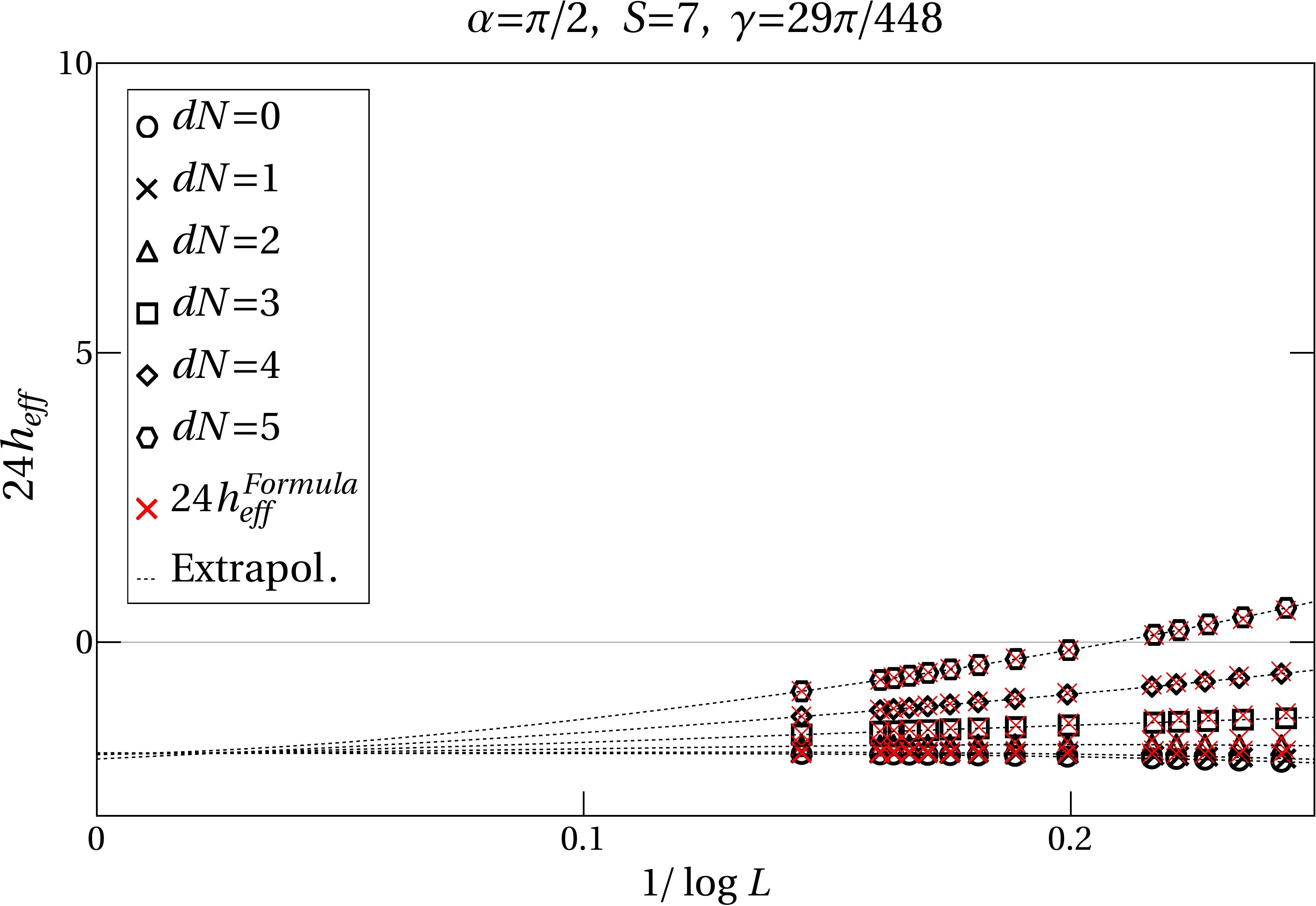}}\\
    \subfigure{\includegraphics[width=0.49\textwidth]{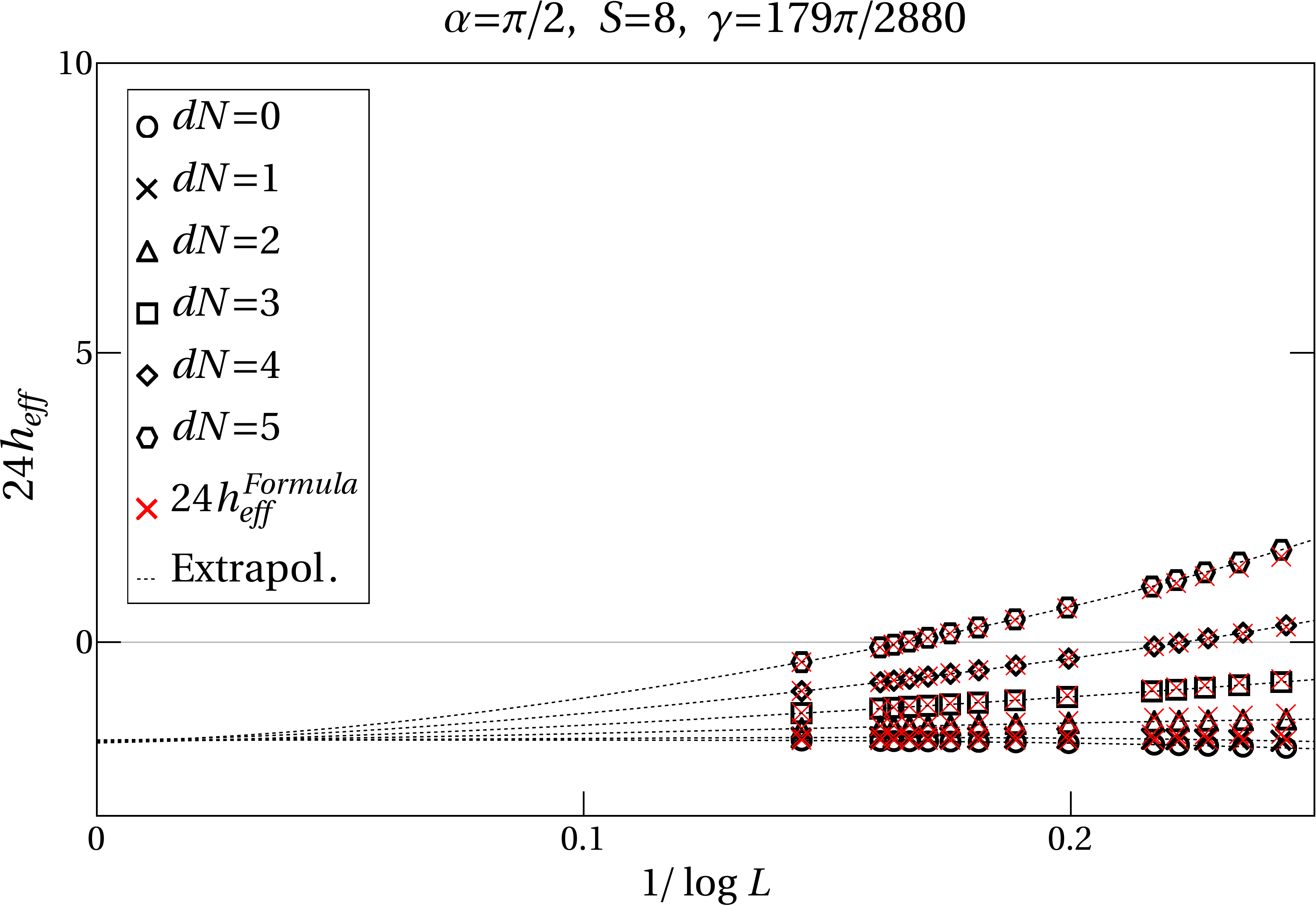}}
    \subfigure{\includegraphics[width=0.49\textwidth]{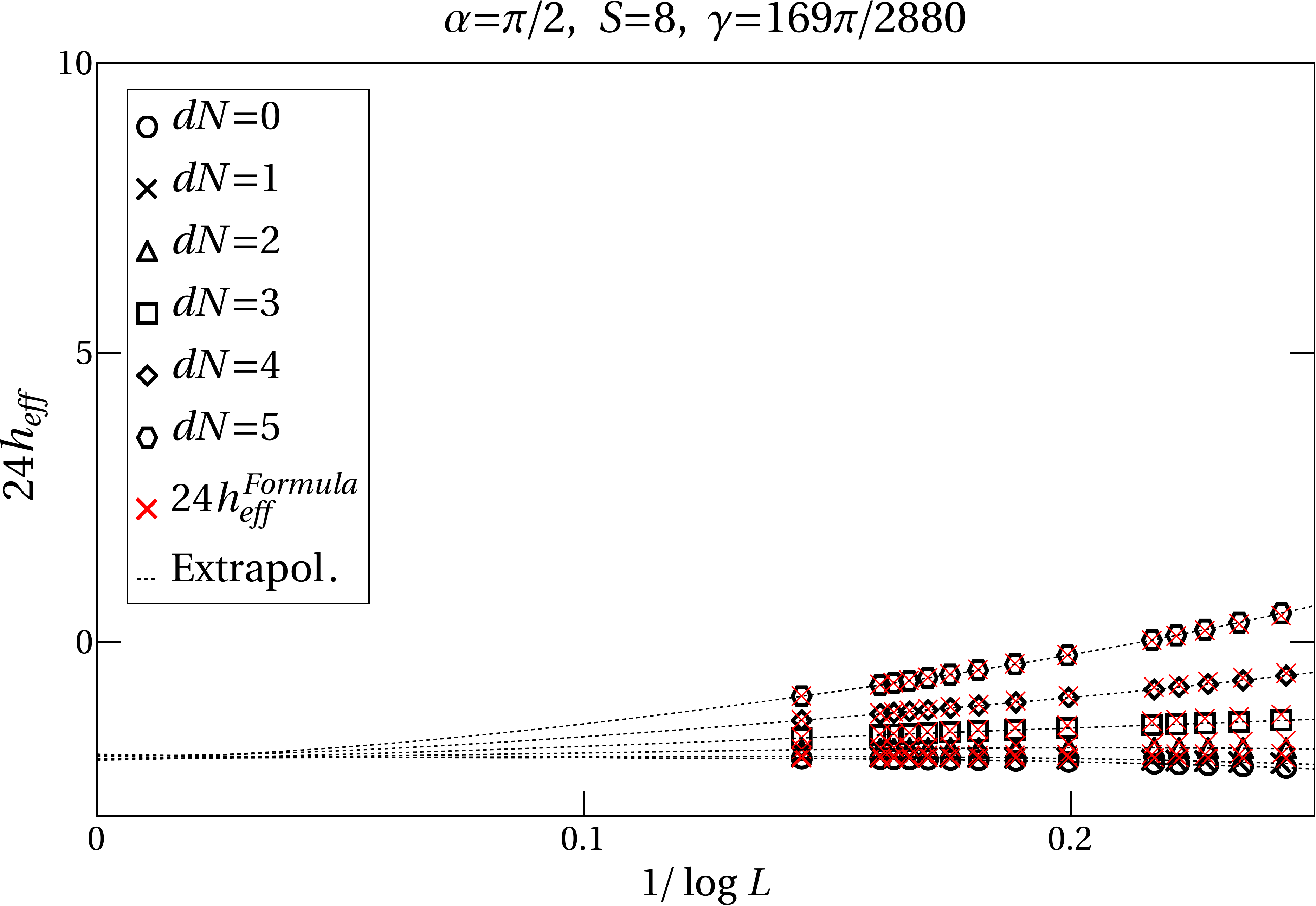}}\\
    \subfigure{\includegraphics[width=0.49\textwidth]{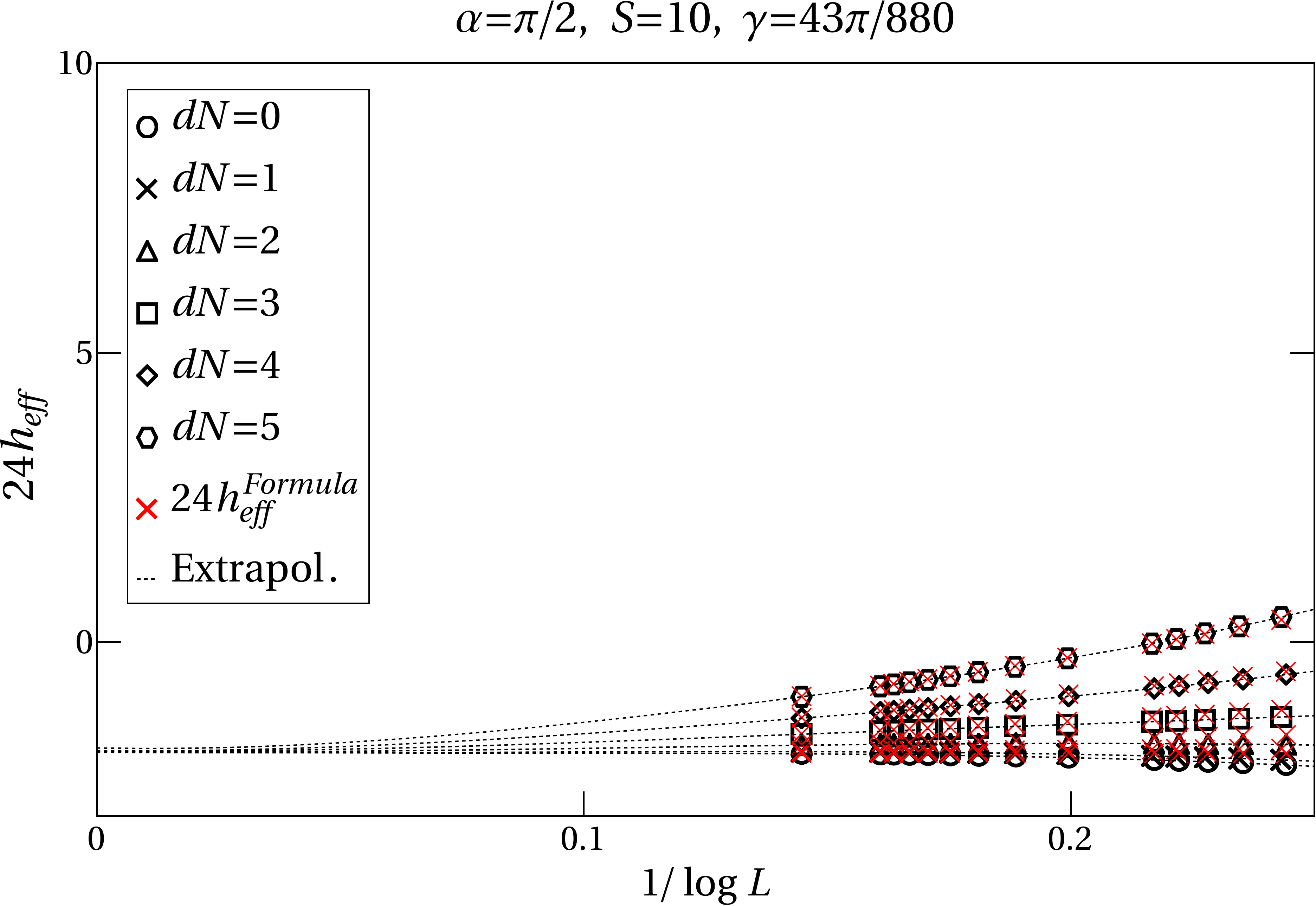}}
     \subfigure{\includegraphics[width=0.49\textwidth]{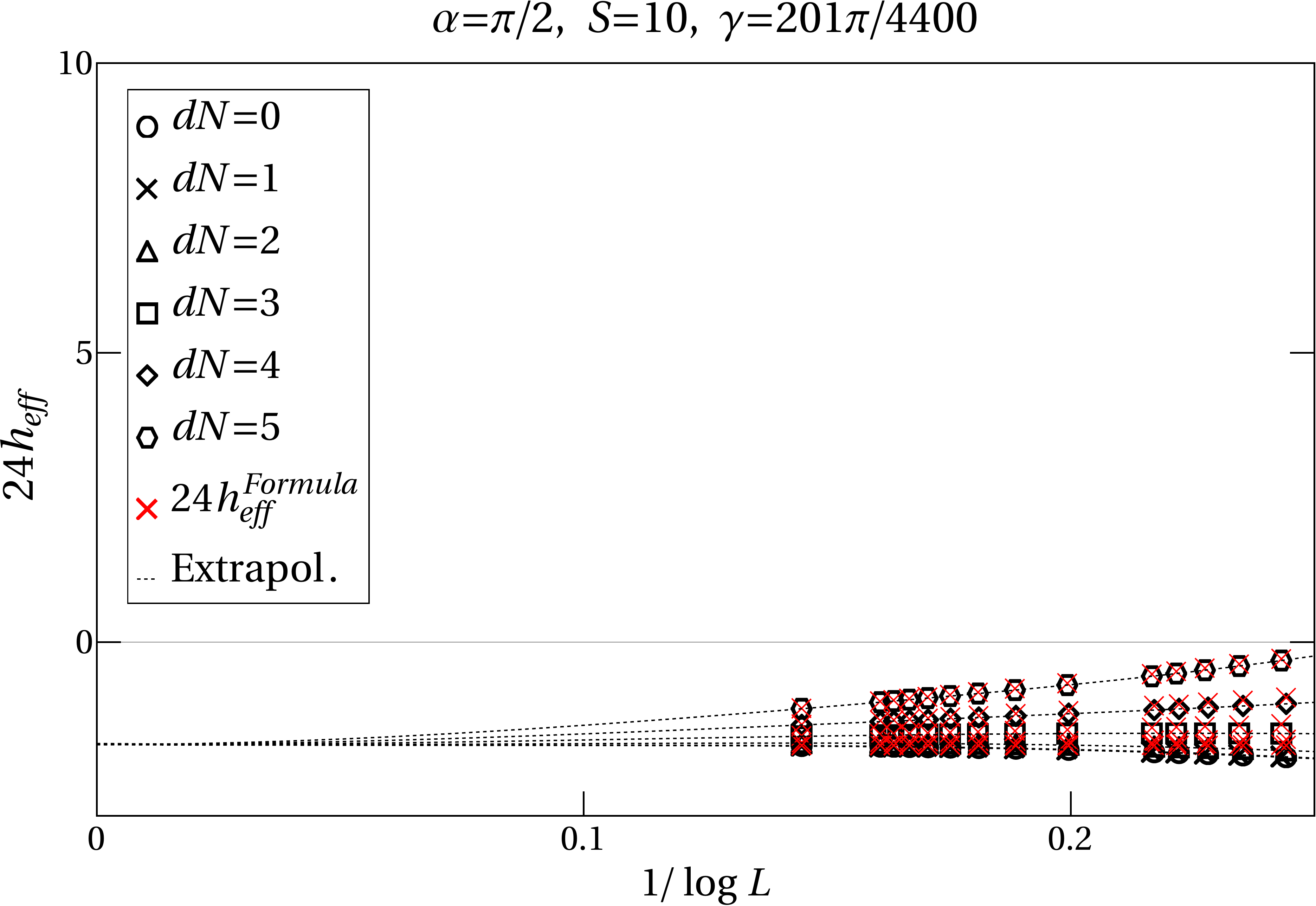}}
    \caption{Similar as Fig.~\ref{S_z_Small}, but for anisotropies where the ground state is found in the sectors $S=7, 8, 10$.}
    \label{S_z_Large}
\end{figure}
%-------------------------------------------
\begin{figure}[H]
    \centering
    \subfigure{\includegraphics[width=0.8\textwidth]{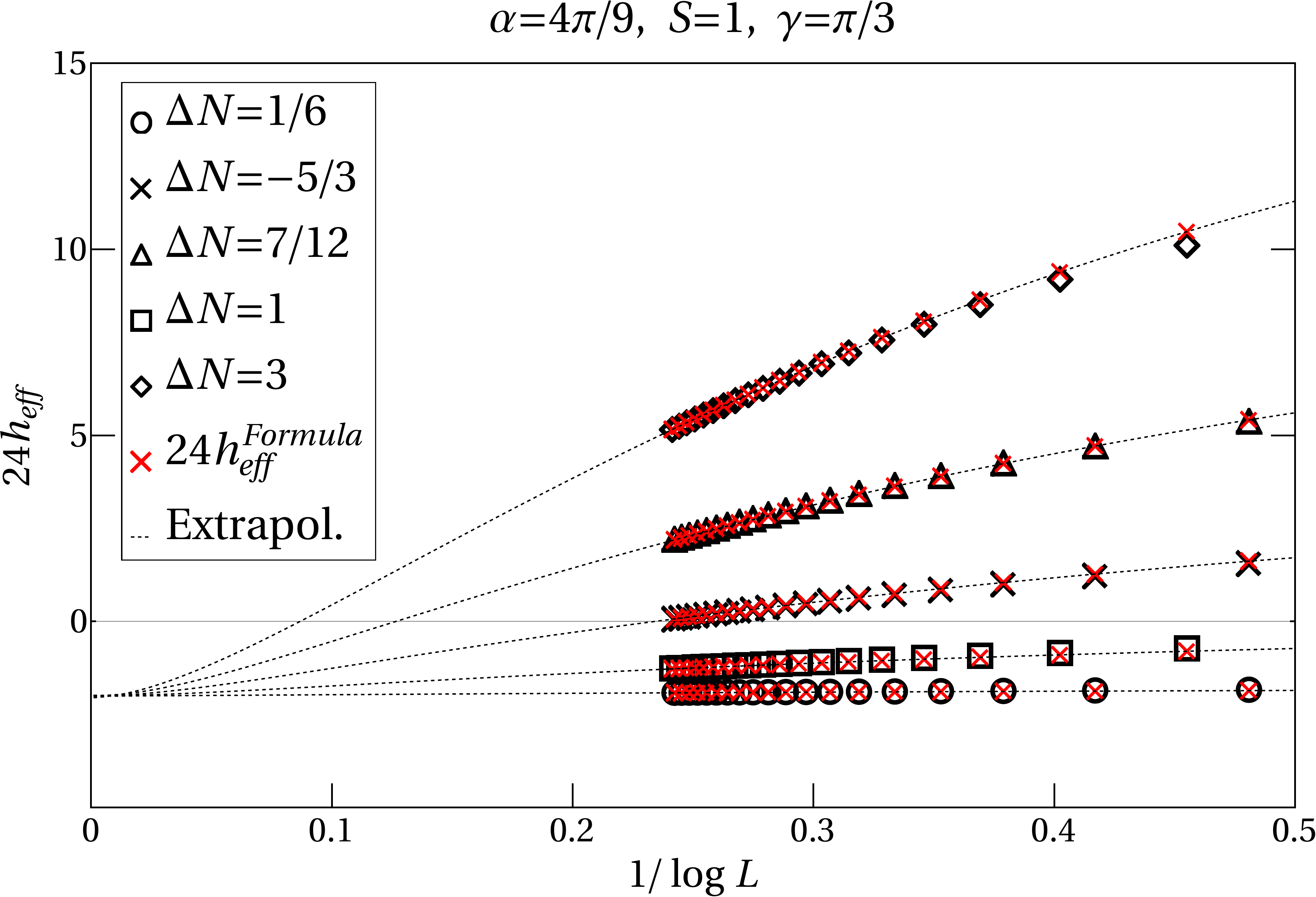}}\\
    \subfigure{\includegraphics[width=0.49\textwidth]{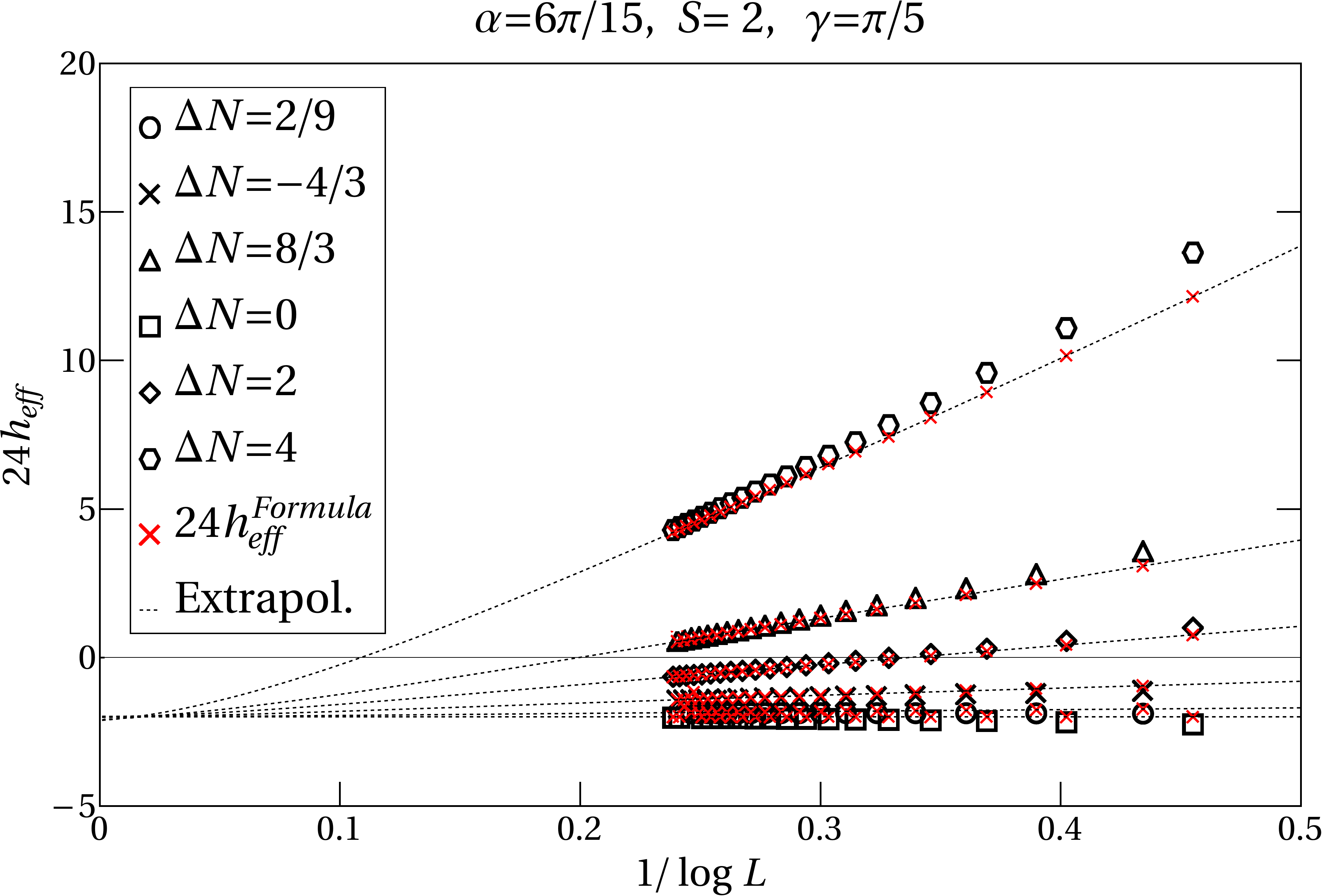}}\subfigure{\includegraphics[width=0.49\textwidth]{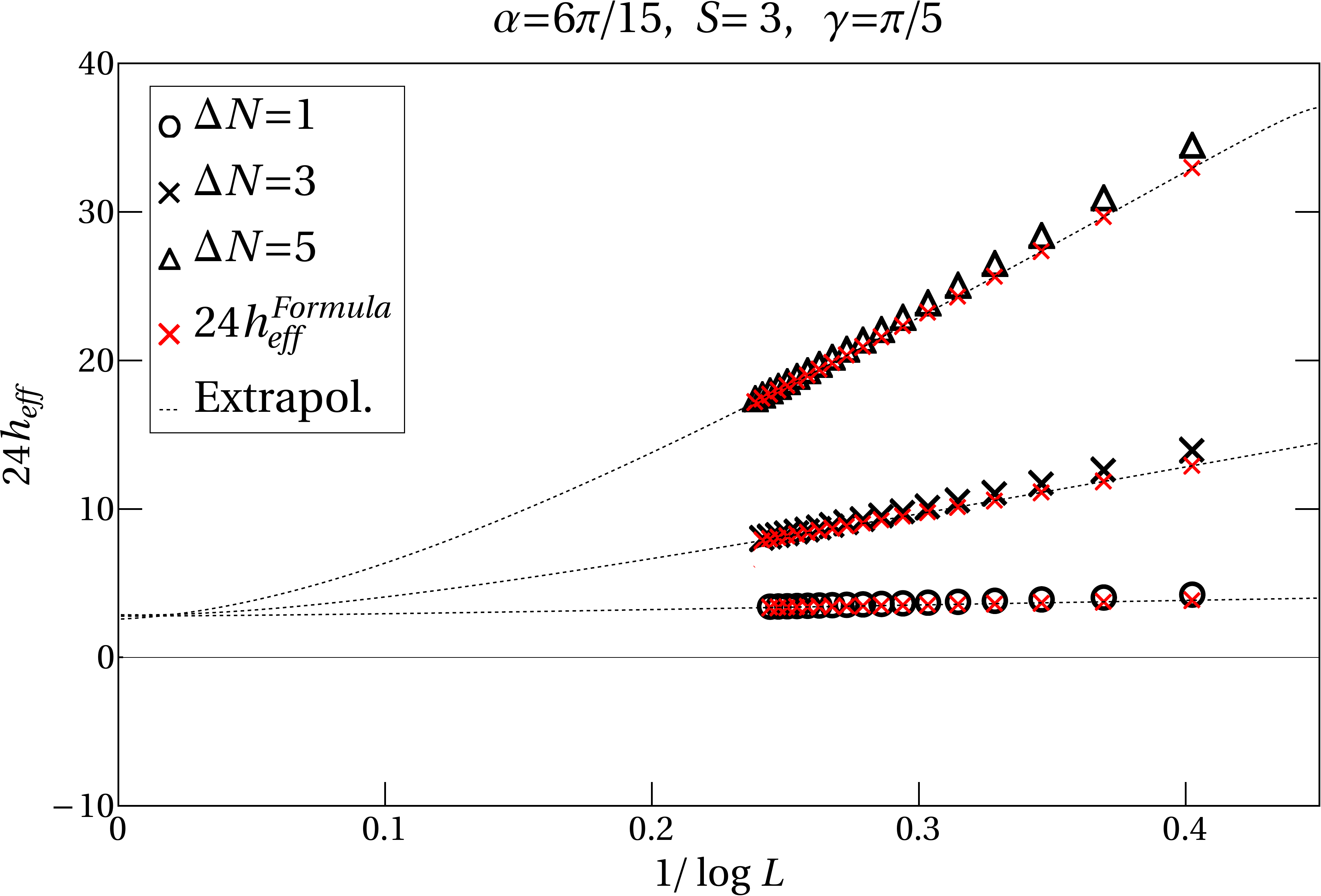}}
    \caption{Similar as Fig.~\ref{S_z_Small} but for values of $\alpha$ away from the self-dual point ($\Delta N=dN-dN_{GS}$). 
 \label{alpha ungleich pi2}}
   
\end{figure}
%-------------------------------------------

Motivated by previous studies we interpret these results in the context of the $SL(2,\mathbb{R})/U(1)$ black hole CFT at level $k$: this model has been found to describe the continuum limit of the staggered six-vertex model for periodic boundary conditions \cite{IkJS12,CaIk13,FrSe14} and, at the self-dual point, for open boundary conditions in the $D^{(2)}_2$-formulation \cite{RoJS21}.  The central charge $c_{BH}$ and and conformal weights $h_{BH}$ of the primary fields in the black hole CFT are given by \cite{HaPT02,RiSc04}
\begin{align}
    c_{BH}=2+\frac{6}{(k-2)}\,,\quad h_{BH}=\frac{(n + w k)^2}{4k}-\frac{J(J-1)}{k-2}\label{c,h_BH-CFT} \quad\text{with}\quad J=\frac{1}{2}+i\tilde{s},\quad \tilde{s}\in \mathbb{R}^+_0\,,
\end{align}
where the integers $n$ and $w$ label the momentum and winding in the compact direction of the semi-infinite cigar-shaped target space of the CFT while $i\tilde{s}=J-1/2$ is the momentum along the uncompactified direction.  The resulting effective conformal weights are
\begin{align}
    h^{BH}_{\text{eff}}=-\frac1{12}+\frac{(n + w k)^2}{4k}
    -\frac1{k-2}\,\left(J-\frac12\right)^2+d\,,\label{h_eff_BH}
\end{align}
where $d$ describes the level of descendant. The identification of the studied spin chain as a lattice regularization of the black hole CFT can be made by comparing (\ref{h_eff_BH}) with (\ref{h_eff}) leading to the following correspondence\footnote{Note $s$ as defined in (\ref{Def_s}) can have either sign and its relation to $J-1/2$ can be fixed up to this sign only. Similarly, we could replace $(n,w)\to(-n,-w)$. Here  we follow the convention used in Ref.~\cite{RoJS21}.}
\begin{align}
  k=\frac{\pi}{\gamma},\qquad  n=-2S-1,\qquad w=1, \qquad \left(J-\frac{1}{2}\right)^2=(is)^2, \qquad d=n_{ph}. \label{CFT_Identification_Conti}
\end{align}
The appearance of a non-vanishing winding number $w$ in the open boundary model is a consequence of the boundary conditions (\ref{Kminus}), (\ref{Kplus}) which lead to an effective twist for the propagating modes in the spin chain.
Note that, as expected from the presence of the continuous quantum number $s$, this is a non-rational CFT. The Virasoro vacuum is not normalizable which is reflected by the fact that the identity field with $h_{BH}=0$ appears neither in the spectrum of conformal weights (\ref{c,h_BH-CFT}) nor in the discrete part of the spectrum discussed in the following section.

\subsection{Discrete part\label{Dis Part}}
\begin{figure}[t]
    \centering
    \subfigure{\includegraphics[width=0.49\textwidth]{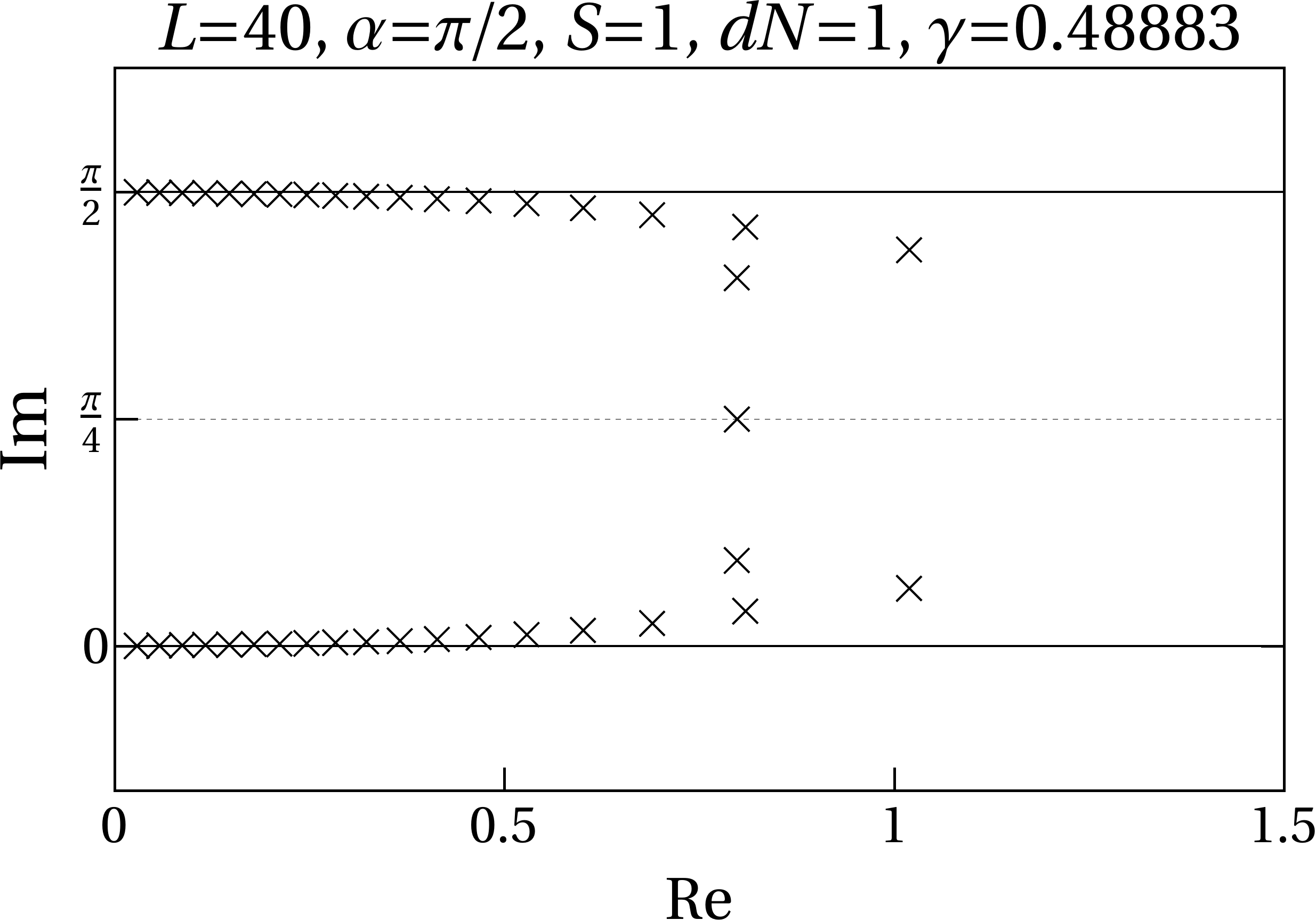}}
    \subfigure{\includegraphics[width=0.49\textwidth]{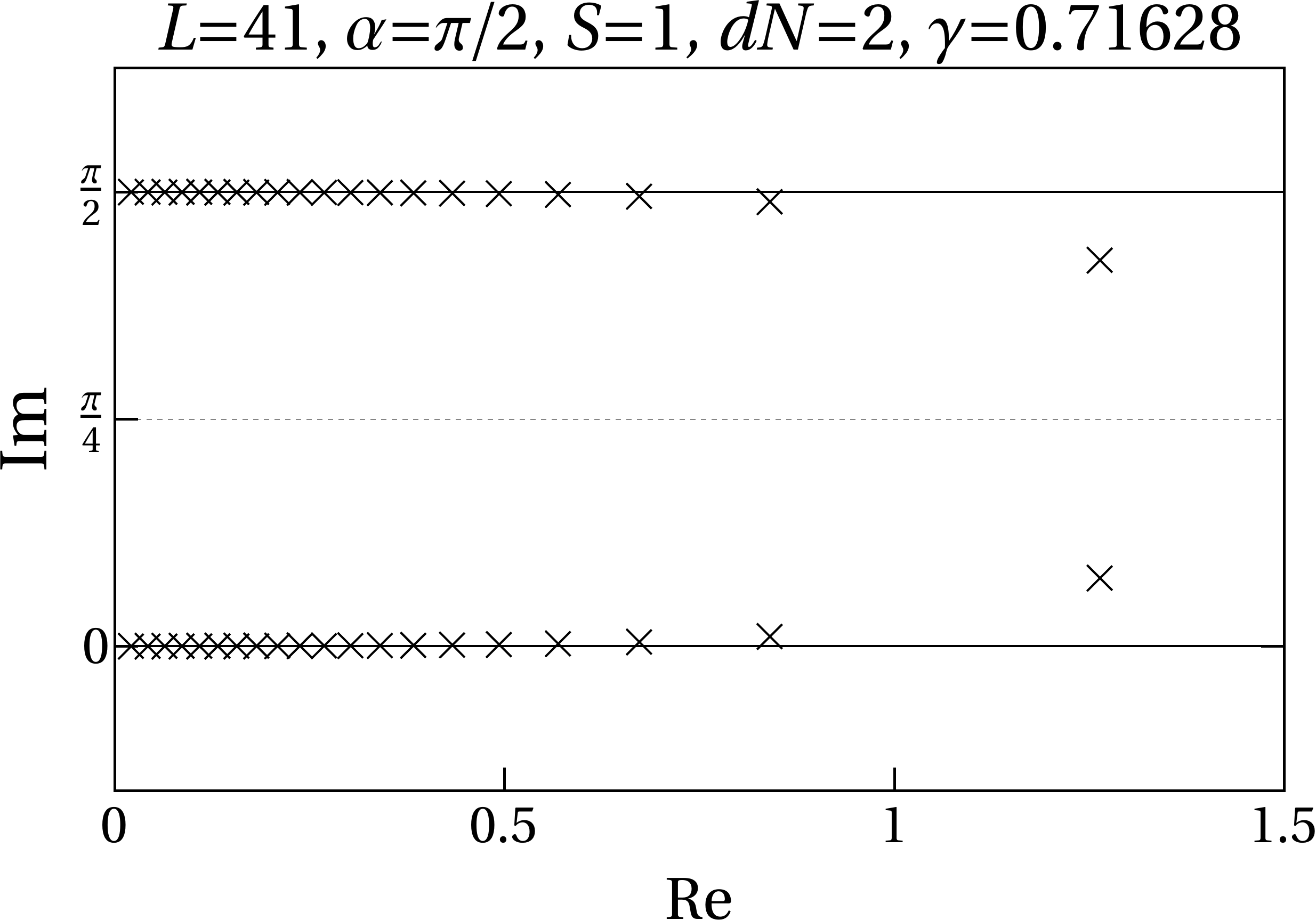}}
    \caption{Pattern of Bethe roots leading to an imaginary quasi-momentum. The configurations are obtained by following configurations given as (\ref{RootTypes}) with $dN$ as indicated by lowering the anisotropy. One the left, we have an odd number of Bethe roots and so one root lies on the line $i\pi/4$, while the rest is paired. On the right, we have an even number of roots, hence, the pairing works.}
    \label{BR_Plot}
\end{figure}

In the black hole CFT, besides the continuous scaling dimensions (\ref{c,h_BH-CFT}) with  $J=\frac12+i\tilde{s}$ for real $\tilde{s}$ considered above, there exists also a set of conformal weights with discrete values $J$, related to bound states localized near the tip of the cigar-shaped target space of the model \cite{MaOo01,HaPT02,RiSc04}. For the corresponding states to be normalizable and to guarantee non-negative conformal weights the possible values of $J$ are restricted to obey the following two conditions \cite{RiSc04,HaPT02}: 
\begin{align}
    \frac{1}{2}<J<\frac{(k-1)}{2}\,, \quad J=\frac{|kw|-|n|}{2}-\ell, \qquad \ell=0,1,2\dots\,. \label{Dis_CFT_State_Condition}
\end{align}
From the first of these conditions we expect that a discrete state with fixed $J$ is realized in the spin-chain spectrum for anisotropies $\gamma$ ranging from $0$ to some root of unity. Specifically, a discrete state with given $J$ may be realized for anisotropies 
\begin{align}
    \label{eq:disc-threshold}
    0<\gamma<\frac{\pi}{2J+1}\,.
\end{align}
The discrete CFT states and their realization conditions can also be related to the quasi-momentum of the Bethe state. In our study of the continuous part of the spectrum above we found that the Bethe states are parameterized by root configurations (\ref{RootTypes}) resulting in real eigenvalues for the quasi-momentum. However, as seen in the study of small lattice sizes, the quasi-momentum can also change from real values to purely imaginary ones when the anisotropy $\gamma$ is lowered. On the level of Bethe roots, this translates into a root pattern changing from (\ref{RootTypes}) to more complicated complex configurations: 
the real parts of one or more of the roots diverge as the anisotropy approaches from above certain rational multiples of $\pi$. Reducing the anisotropy further these roots reappear in the finite domain with different imaginary parts.  Depending on the state considered this process may be repeated several times until the root configuration acquires the following remarkable pattern:
%
%In more detail the real parts of one or more Bethe roots of the form (\ref{RootTypes}), typically those with largest real part, diverge when the anisotropy approaches from above certain rational multiples of $\pi$ allowing the root pattern to change drastically.  Specifically, we find one particular remarkable pattern of Bethe roots: 
for states parameterized by an even number $M$ of Bethe roots they come in pairs $v_j$, $\bar{v}_j$ mirrored at the line $i\pi/4$ 
\begin{equation}
\label{Dis_BR1}
    \begin{aligned}
    &v_{j}=x_j+iy_j\,, \qquad \bar{v}_{j}=x_j+i\left(\frac{\pi}{2}-y_j\right)\,\\ &\text{with}\quad x_j,y_j\ge 0\, \qquad j=1,2\dots\frac{M}{2}\,.
\end{aligned}
\end{equation}
If the number $M$ of Bethe roots is odd there appears an additional root with imaginary part $\pi/4$, i.e.\
\begin{equation}
\label{Dis_BR2}
    \begin{aligned}
    &v_{j}=x_j+iy_j \,, \quad \bar{v}_{j}=x_j+i\left(\frac{\pi}{2}-y_j\right)\,,\quad v_M=x+\frac{i\pi}{4}\\  &\text{with}\quad x_j,y_j,x\ge 0\,, \qquad j=1,2\dots\frac{M-1}{2}\,.
\end{aligned}
\end{equation}
Examples of such root configurations for spin-$1$ states on lattices with $L=40$ ($41$) sites, i.e.\ $M=39$, $M=40$, evolving from the (\ref{RootTypes}) with $dN=M^0-M^{\frac\pi2}=1$ and $2$, respectively, as $\gamma$ is lowered are shown in Figure~\ref{BR_Plot}.

\begin{figure}[t]
    \centering
    \includegraphics[width=0.8\textwidth]{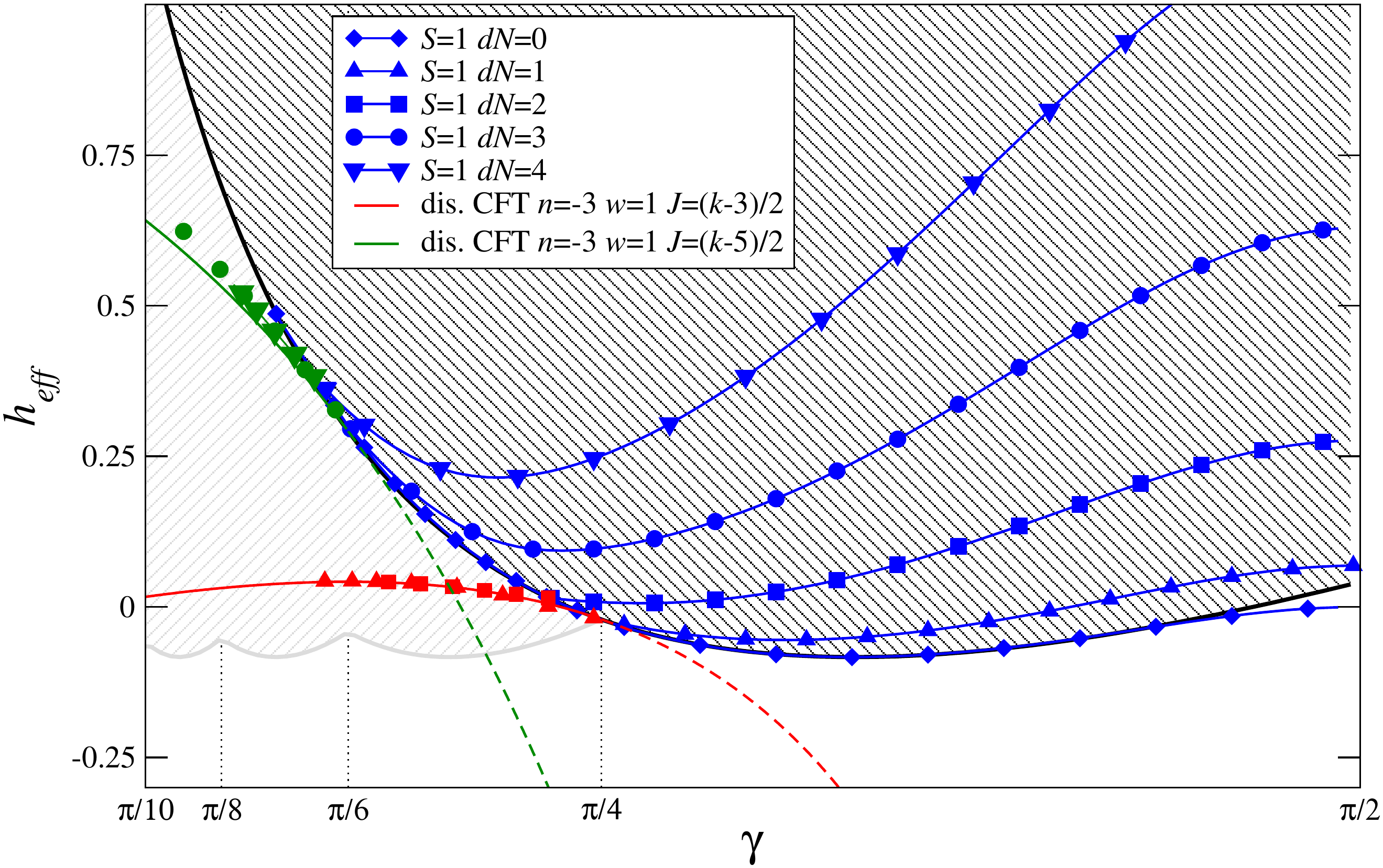}
    \caption{Effective scaling dimensions vs.\ anisotropy $\gamma$ for the self-dual model, $\alpha=\pi/2$, of size $L=40,41$ derived for the lowest states in the $S=1$ sector, showing the transmutations of continuous states into discrete ones. The black shaded area represents the continuum of levels starting at the spin-$S^{GS}=1$ ground state. 
    Grey shading indicates the continua in the spin $S^{GS}>1$ sectors, which overlap with each other and the spin-$S^{GS}=1$ one. The lower edge of the continua corresponds to $c_{eff}$ (\ref{ceff}) up to the factor of $-24$.
    Blue symbols denote the effective scaling dimensions obtained from finite size data (\ref{Finite_Size_Formula}) corresponding to weights from the continuous part with $S=1$ of the CFT spectrum given in terms of root configurations (\ref{RootTypes}) with different $dN$. Red and green symbols depict the continuation of the corresponding states (same symbol shape) to anisotropies where the Bethe root patterns change to (\ref{Dis_BR1}) or (\ref{Dis_BR2}) and the quasi-momentum $s$ becomes imaginary. 
    Red (green) solid lines are the effective scaling dimensions (\ref{h_eff_BH}) of the primary fields from the discrete part of the CFT spectrum with $n=-3$, $w=1$ and $J=(k-3)/2$ ($(k-5)/2$) for $k=\pi/\gamma$. Dashed lines are their continuation to anisotropies $\gamma>\pi/(2J+1)$ where the corresponding operators in the CFT become non-normalizable.
    }
    \label{Dis_States_E}
\end{figure}

Following the transmutation of configurations (\ref{RootTypes}) to (\ref{Dis_BR1}), (\ref{Dis_BR2}) under the variation of $\gamma$ for small values of $dN$ we are able to observe what happens to the scaling dimensions when the quasi-momentum changes from from real to imaginary: as discussed above the scaling dimensions corresponding to spin-$1$ states are in the continuous part of the spectrum (\ref{h_eff}) for anisotropies $\pi/4<\gamma<\pi/2$ where they are separated by finite size gaps $\sim (dN/\log L)^2$.\footnote{%
In the limit $\gamma\to\pi/2$ one observes a crossover to a linear dependence on $1/\log L$.  This can be understood from the fact that the staggered six-vertex model at $\gamma=\pi/2$ coincides with the integrable $OSp(2|2)$ model which is in a different universality class \cite{MaNR98,JaRS03}.}
As $\gamma$ is reduced further the lowest levels approach the lower bound of the $S=1$ continuum, leaving it when the quasi-momentum becomes purely imaginary, see Figure~\ref{Dis_States_E} for the lowest states in this spin sector.  Specifically we find that the finite size energies of the states with $dN=1$, $2$ ($dN=3$, $4$), realized for even and odd lattice sizes $L$ respectively, lead to the conformal weights (\ref{h_eff_BH}) of the black hole CFT with $J=(k-3)/2$ ($(k-5)/2$) in the regime where $s\in i\mathbb{R}$.
%------------------------------------
%------------------------------------

\begin{figure}[t]
    \centering
    \includegraphics[width=0.8\textwidth]{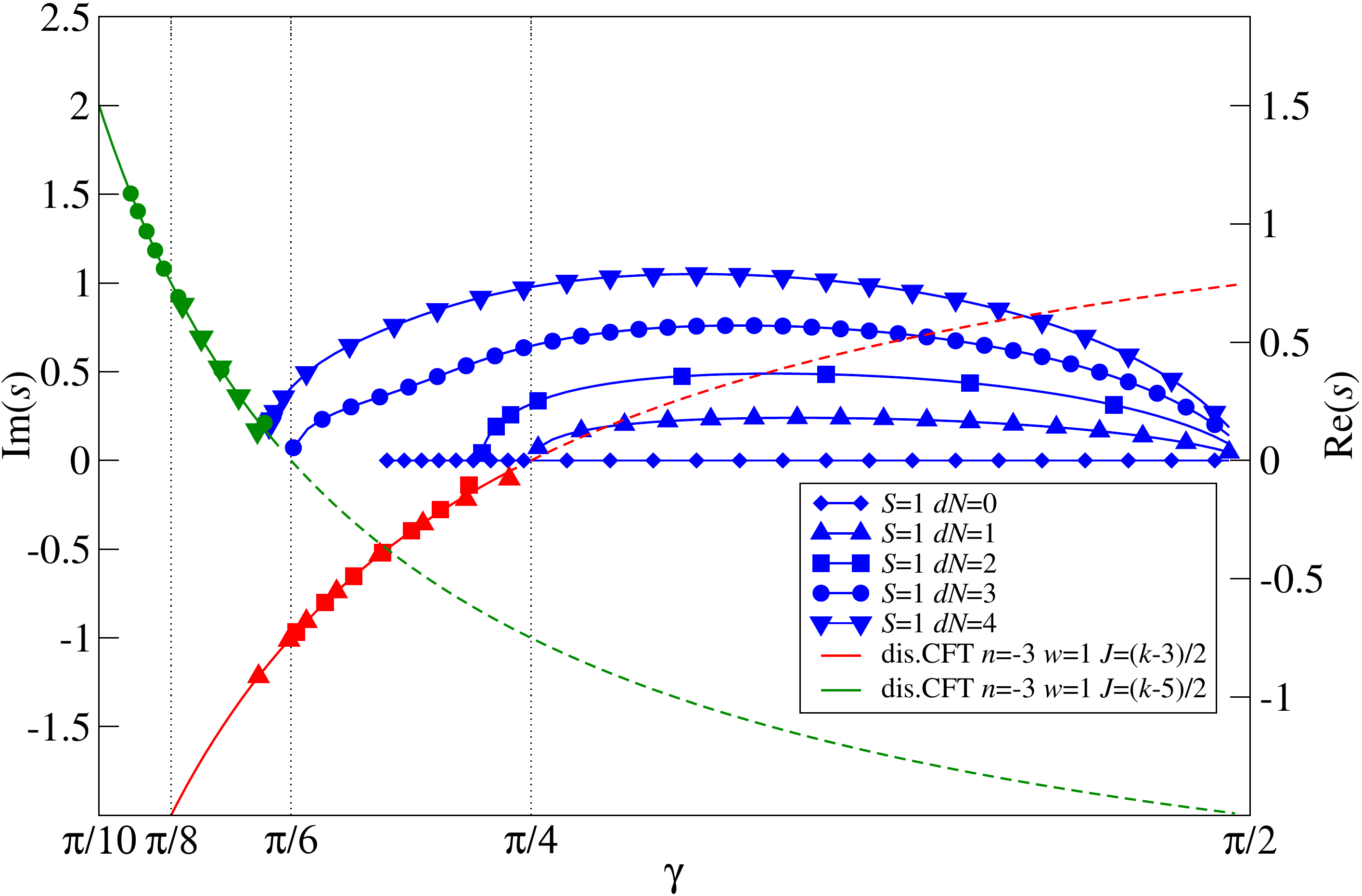}
    \caption{Real (blue symbols) and imaginary (red and green symbols) part of the quasi-momentum $s$ vs.\ anisotropy $\gamma$ for the spin $S=1$ states considered in Figure \ref{Dis_States_E}. The solid red and green lines depict the quasi-momentum given by the CFT predictions via (\ref{Im(s)_Dependence}). Dashed lines are obtained by continuation of the CFT data into  the region where the unitarity  condition (\ref{Dis_CFT_State_Condition}) is violated. Note that $s\equiv0$ for the $dN=0$ state.}
    \label{QI_Plot}
\end{figure}

It turns out that the finite size formula (\ref{h_eff}) continues to hold for purely imaginary $s$: in this case $J=\frac12+is$ has to be an element of the discrete set (\ref{Dis_CFT_State_Condition}). In the CFT sector with quantum numbers $(n,w)=(-2S-1,1)$ for the compact degree of freedom this leads to the following condition on the allowed imaginary values of $s$ in the lattice model (as mentioned above the sign of $s$ is not fixed by (\ref{CFT_Identification_Conti})):
\begin{align}
    s_\ell=\pm i\left(\frac{\pi}{2\gamma}-S-1-\ell\right)\,, \qquad \ell=0,1,2,\dots < \frac{\pi}{2\gamma}-S-1\,.
    \label{CFT_Prediction_s}
\end{align}
Note that this implies that the thresholds for the appearance of discrete levels of the black hole CFT resulting from the unitarity condition (\ref{Dis_CFT_State_Condition}) coincide with the anisotropies (\ref{GS Crossings Theo}) where the spin of the ground state in the lattice model changes.
This prediction can be compared with our numerically findings for the $S=1$ states with $J=(k-3)/2 - \ell$, $\ell=0,1$, considered in Figure \ref{Dis_States_E}: their quasi-momentum, once it becomes imaginary, is expected to match the condition above for $\ell=0,1$, i.e.\
\begin{equation}
\label{Im(s)_Dependence}
    \begin{aligned}
        s_{\ell=0}=\pm i\left(\frac{\pi}{2\gamma}-2\right)\quad\text{for~}\gamma<\frac\pi4\,, \\
        s_{\ell=1}=\pm i\left(\frac{\pi}{2\gamma}-3\right)\quad\text{for~}\gamma<\frac\pi6\,. 
\end{aligned}
\end{equation}
Our numerical data show that the change from real to imaginary quasi-momentum takes place at or slightly below these values of $\gamma$. Overall, i.e.\ up to finite size corrections near these thresholds, Eqs.~(\ref{Im(s)_Dependence}) match our results for the quantum number $s$ obtained for system sizes $L=40$ and $41$, see Figure \ref{QI_Plot}. We presume that this transmutation of low-lying levels from the continuous part of the spectrum into discrete ones is the origin of the modification of the density of states $B(s)$ for small $s$ close to these anisotropies observed in the  previous section.
%------------------------------------
%------------------------------------

Therefore, the expression for the effective scaling dimensions (\ref{h_eff}) as obtained from the finite size analysis of the staggered vertex model correctly describes both the continuous and the discrete part of the spectrum of the $SL(2,\mathbb{R})/U(1)$ sigma model at level $k$ via the identifications (\ref{CFT_Identification_Conti}).
\section{Summary}
Starting from the six-vertex model with quantum group invariant boundary conditions we have constructed an integrable anisotropic $\mathbb{Z}_2$-staggered spin-$1/2$ chain. The $U_q(\mathfrak{sl}(2))$-symmetry of the model is spontaneously broken: in contrast to the periodic chain where the lowest state is always in the sector with $S_z=0$, the ground state of the open chain has non-zero spin depending on the anisotropy $\gamma$, becoming completely polarized for sufficiently small $\gamma$ ($\gamma\to0$ in the thermodynamic limit).

For the self-dual choice of the staggering, i.e.\ $\alpha=\pi/2$, this model is equivalent to the $D_2^{(2)}$-spin chain with integrable boundary conditions studied in Refs.~\cite{RoJS21,NeRe21a}.  This model has been identified as a lattice regularization of the $SL(2,\mathbb{R})_k/U(1)$ sigma model whose non-compact degrees of freedom lead to a continuous spectrum of conformal weights in the thermodynamic limit.  We find that, similar as in the staggered six-vertex model with periodic boundary conditions \cite{FrSe14}, this identification can be extended to the entire range of staggering parameters $\gamma<\alpha<\pi-\gamma$.

As in previous studies of the periodic model \cite{IkJS12,CaIk13,FrSe14} the formulation as a staggered model facilitates the definition of a conserved quasi-momentum operator $K$ (\ref{def:quasimom}). Unlike the Hamiltonian (\ref{End_H-Op}) and the other conserved quantities generated by the four-row transfer matrix (\ref{TM_END}) the quasi-momentum is odd under the duality transformation.  The eigenvalues of this operator allow to distinguish states corresponding to the levels in the continuous (discrete) part of the spectrum of the sigma model (see also \cite{FrHo17}): already for the very small systems that have been diagonalized numerically the corresponding quasi-momenta are real (imaginary), respectively. Moreover, the eigenvalues of the quasi-momentum can be related to the momentum along the non-compact direction of the semi-infinite cigar-shaped target space of the black hole CFT. As a consequence they determine the amplitudes of the strong logarithmic corrections to scaling of the states leading to the continua and reflect the appearance of the quantized discrete states in the spectrum of the lattice model.  An open problem is the calculation of the density of states in the continuous part of the CFT spectrum from the finite size data for the vertex model and in particular the effect of the ground state crossings between sectors with different $U_q(\mathfrak{sl}(2))$ spin. We expect that this calculation is facilitated by the relation (\ref{Def_s}) between the continuous quantum number $s$ in  the CFT and the quasi-momentum $\mathcal{K}$ for the Bethe states, see Refs.~\cite{CaIk13,FrSe14} for the periodic chain.

\begin{acknowledgments}
  Funding for this work has been provided by the Deutsche Forschungsgemeinschaft under grant No.\ Fr~737/9-2 as part of the research unit   \emph{Correlations in Integrable Quantum Many-Body Systems} (FOR2316).
\end{acknowledgments}

\appendix
%\bibliography{s6v.bib}

\end{document}